\documentclass[12pt]{iopart}

\usepackage{iopams}
\usepackage{amssymb,bm,graphicx}
 
\usepackage[usenames,dvipsnames]{color}
\usepackage[normalem]{ulem}

\begin{document}

\title[Effect of driving on coarsening dynamics in phase-separating systems]{Effect of driving on coarsening dynamics in phase-separating systems}

\author{D Tseluiko$^1$, M Alesemi$^2$, T-S Lin$^3$, and U Thiele$^{4,5}$}

\address{$^1$Department of Mathematical Sciences, Loughborough University, Loughborough LE11 3TU, UK}
\address{$^2$Department of Mathematics, Jazan University, Jazan, Saudi Arabia}
\address{$^3$Department of Applied Mathematics, National Chiao Tung University,\\ 1001 Ta Hsueh Road, Hsinchu 300, Taiwan}
\address{$^4$Institut f\"ur Theoretische Physik, Westf\"alische Wilhelms-Universit\"at M\"unster, Wilhelm Klemm Str. 9, 48149 M\"unster, Germany}
\address{$^5$Center of Nonlinear Science (CeNoS), Westf{\"a}lische Wilhelms-Universit\"at M\"unster, Corrensstr.\ 2, 48149 M\"unster, Germany}
\ead{d.tseluiko@lboro.ac.uk}
\vspace{10pt}
\begin{indented}
\item[]March 2020
\end{indented}

\submitto{\NL}

\begin{abstract}
We consider the Cahn-Hilliard (CH) equation with a Burgers-type convective term that is used as a model of coarsening dynamics in laterally driven phase-separating systems. In the absence of driving, it is known that solutions to the standard CH equation are characterized by an initial stage of phase separation into regions of one phase surrounded by the other phase (i.e., clusters or drops/holes or islands are obtained) followed by the coarsening process, where the average size of the structures grows in time and their number decreases. Moreover, two main coarsening modes have been identified in the literature, namely, coarsening due to volume transfer and due to translation. In the opposite limit of strong driving, the well-known Kuramoto-Sivashinsky (KS) equation is recovered, which may produce complicated chaotic spatio-temporal oscillations. The primary aim of the present work is to perform a detailed and systematic investigation of the transitions in the solutions of the convective CH (cCH) equation for a wide range of parameter values, and, in particular, to understand in detail how the coarsening dynamics is affected by an increase of the strength of the lateral driving force. Considering symmetric two-drop states, we find that one of the coarsening modes is stabilized at relatively weak driving, and the type of the remaining mode may change as driving increases. Furthermore, there exist intervals in the driving strength where coarsening is completely stabilized. In the intervals where the symmetric two-drop states are unstable they can evolve, for example, into one-drop states, two-drop states of broken symmetry or even time-periodic two-drop states that consist of two traveling drops that periodically exchange mass. We present detailed stability diagrams for symmetric two-drop states in various parameter planes and corroborate our findings by selected time simulations.
\end{abstract}

%
%
%
%
%

\section{Introduction}
In recent years, there has been a renewed interest in the convective Cahn-Hilliard (cCH) equation as a model of coarsening dynamics in driven phase-separating systems. In the present study, we consider the following one-dimensional cCH equation that contains an additional nonlinear driving term of Burgers type:
\begin {equation} \label{Eq.2}
u_t+D u u_x + (u-u^3+u_{xx})_{xx}=0.
\end{equation}
Here, $u(x,t)$ is the order parameter field, with $x$ and $t$ denoting the spatial coordinate and time, respectively, and $D$ is the driving strength. This equation was derived, for example, by Golovin et al.\ \cite{Ref.15,Ref.20} as a model for a kinetically controlled growing crystal surface with a strongly anisotropic surface tension. In such a context,  $u$ is the surface slope and $D$ is the growth driving force proportional to the difference between the bulk chemical potentials of the solid and fluid phases (see also Liu and Metiu~\cite{Ref.14} for modelling of growing crystal surfaces). Equation (\ref{Eq.2}) was also obtained by Watson \cite{Watson2004} as a small-slope approximation of the crystal-growth model obtained by Di~Calro et al.\ \cite{Carlo_etal_1992} and Gurtin~\cite{Ref.21}. Related models have also been derived, for instance, in the context of epitaxial growth (see, for example, \v Smilauer et~al.~\cite{Smilauer_etal_1999}) and liquid droplets on inclined planes (see, for example, Thiele and Knobloch \cite{Ref.10,Ref.11}, Thiele~\cite{Ref.50}).

In the absence of driving, the cCH equation reduces to the standard CH equation \cite{Bray1994ap,Novi1985jsp}, that was proposed as a model to describe phase separation (or spinodal decomposition) of two-component mixtures (see, for instance, Cahn \cite{Ref.31,Ref.35,Ref.72},  Cahn and Hilliard~\cite{Ref.25,Ref.73}). Note that the standard CH equation can be written in the following general gradient-dynamics form:
\begin{equation} \label {Eq.B64}
u_t = \left[Q(u) \left( \frac{\delta F[u]}{\delta u}\right)_x \right]_x, 
\end{equation}
where $\delta/\delta u$ denotes the variational derivative. The free energy $F[u]$ is given by
\begin{equation} \label {Eq.B65} 
F[u]= \int \varphi(u, u_x) dx,
\end{equation}
and $\varphi(u,u_x)= \frac{1}{2} u_{x}^{2}+f(u)$ is the energy density, with the first term being the square-gradient term that penalizes interfaces and with the double-well potential $f(u)=\frac{1}{4}u^4-\frac{1}{2}u^2$ as the local free energy.

The initial dynamics of the solutions of the standard CH equation from a perturbed homogeneous state is characterized by separation into regions corresponding to different components, i.e., clusters (drops/holes or islands) of one phase surrounded by the other phase, or labyrinthine patterns of the two phases. However, after this initial stage of evolution, these structures slowly grow in size and their number decreases, i.e., the structure coarsens. In the following we refer to the structures as ``drops''.

Two main modes of coarsening have been identified, namely, coarsening by volume transfer and by translation. In coarsening by the volume transfer mode (which is also known as Ostwald ripening \cite{Ostwald}), the centres of the drops remain fixed in space, while the sizes of the drops change -- some grow in time, while others decrease in size and, eventually, disappear. In coarsening by the translation mode, the centres of the drops do not remain fixed, and coarsening occurs due to motion and merging of the drops. The coarsening process continues until only a single large drop remains. For a more detailed discussion of coarsening for the CH and related equations see, for example, Onuki~\cite{Ref.60}, Desai~\cite{Ref.61}, Thiele~et~al.~\cite{Ref.66}, and Pototsky~et~al.~\cite{Pototsky2014}).

In the limit of strong driving, the cCH equation reduces to the well-known Kuramoto-Sivashinsky (KS) equation \cite{Kura1984ptp,Siva1977aa}. Indeed, substituting $u= \tilde u / D$ into (\ref{Eq.2}) and taking the limit $D\rightarrow\infty$, one obtains the KS equation for $\tilde u$ (see, for example, Golovin~et~al.~\cite{Ref.1}). In contrast to the solutions of the CH equation, the long-time dynamics of the solutions of the KS equation is characterized by complicated chaotic spatio-temporal oscillations \cite{HyNi1986pd,KeNS1990sjam,Smyrlis_Papageorgiou_1991}. Thus, as the driving force is increased from zero to large values, there must appear transitions leading from the coarsening dynamics typical of the standard CH equation to complicated chaotic oscillations typical of the KS equation. We note that coarsening dynamics for the cCH equation has been studied in the limit of a weak driving force numerically by Emmott and Bray \cite{Ref.6} and Golovin~et~al.~\cite{Ref.1} and analytically by Watson et~al.~\cite{Ref.3}, {and for moderately large driving force by  Podolny et al.\ \cite{PODOLNY2005291},} and scaling laws for the average separation between the successive phases as a function of time have been obtained. Zaks et al.~\cite{Ref.51} reported that driving can be used to stop coarsening for certain parameter values. Some stationary solutions of the cCH equation have been analysed by Korzec et al.\ \cite{Ref.52}. We also note that Eden and Kalantarov~\cite{EdenKalantarov2007} demonstrated the existence of a finite-dimensional inertial manifold for the cCH equation. The main aim of the present work is to perform a detailed and systematic investigation of the transitions in the solutions of the cCH equation for a wide range of parameter values as the driving force is increased and to construct detailed stability diagrams in the parameter planes. Finally, note that similar transitions with increasing lateral driving strength have been investigated for various thin-film equations \cite{Ref.11,ThVK2006jfm}. The place of the cCH and thin-film equations in a classification of one-field equations based on mass conservation and variational character is discussed in the introduction of \cite{EGUW2019springer}.

The rest of the present work is organized as follows. In Sect.~\ref{Sec:convCH}, we discuss basic background on the cCH equation. 
In Sect.~\ref{sec:single_double}, we discuss some theory behind single-interface (i.e., front) and double-interface (i.e., drop) solutions. We present the results of numerical continuation of periodic drop solutions in Sect.~\ref{sect:double_numerics}. First, we discuss the results of numerical continuation with respect to the domain size for different values of the mean concentration, and then we analyze how the driving force affects inhomogeneous solutions of the CH equation. In Sect.~\ref{Sect5}, we present a systematic study of the linear stability properties of various spatially periodic traveling solutions of the cCH equation, and analyze the effect of driving on the coarsening modes of symmetric two-drop states. We produce detailed bifurcation diagrams additionally including two-drop states of broken symmetry and time-periodic two-drop states that consist of two drops that periodically exchange mass. We present detailed stability diagrams for symmetric two-drop states in various parameter planes. In addition, we support the numerical continuation results by selected time simulations. Finally, in Sect.~\ref{Sect:conclusions} we present our conclusions.

\section{The convective Cahn-Hilliard equation}
\label{Sec:convCH}
As we focus on analyzing solutions that are stationary or time-periodic in a moving frame, it is convenient to rewrite  equation (\ref{Eq.2}) in a frame moving with velocity $v$, i.e., 
\begin {equation} \label{Eq.2mov}
u_t-vu_x+D u u_x + (u-u^3+u_{xx})_{xx}=0.
\end{equation}
We are primarily interested in analyzing solutions on a spatially periodic domain, say $x\in[0,L]$, and we note that $u(x,t)$ is a conserved quantity, i.e., the mean value  $\bar u=\frac{1}{2L}\int_{-L}^{L}u\, d x$ is constant. Note that due to the symmetry $(D,u)\rightarrow(-D,-u)$, it is sufficient to only consider nonnegative values of $D$. In addition, the symmetry $(x,u)\rightarrow(-x,-u)$ implies that it is sufficient to only consider nonnegative mean values $\bar u$. For the rest of the manuscript, we therefore assume that $D\geq 0$ and $\bar u\geq 0$.

To analyze the linear stability of a spatially uniform solution $\bar u$, we consider a small perturbation of the form $\propto \exp(ikx+\beta t)$ and, after linearizing equation~(\ref{Eq.2mov}), obtain the following dispersion relation:
\begin {equation} \label {Eq.10}
\beta(k) = i v k-i D  \bar u k +k^2 - 3 \bar u^{2} k^2 - k^4.
\end{equation}  
Thus, the growth rate $w(k)=\mathrm{Re}\,\beta(k)$ of a small-amplitude sinusoidal wave of wavenumber $k$ is
\begin {equation} \label {Eq.11}
w(k) = [(1-3 \bar u^2 ) - k^2 ] k^2,
\end{equation}
as for the standard CH equation, and the phase speed is $-\mathrm{Im}\beta(k)/k=D\bar u-v$. 

By solving equation $w(k_c)=0$, we find the cutoff wavenumber $k_c$:
\begin{equation} \label{Eq.13}
k_c =  \sqrt{1- 3 \bar u^2}.
\end{equation}
This solution exists only when  $1-3 \bar u^2 > 0 $, i.e., when $|\bar u |< \sqrt{1/3}$. In this case, there is a band of unstable wavenumbers, $k\in(0,\,k_c)$. Otherwise, if $|\bar u|\geq \sqrt{1/3}$, we find that $w(k)<0$ for all $k>0$, and we obtain the linearly stable case. Note that these uniform states may still be nonlinearly unstable.

\section{Front and one-drop solutions}
\label{sec:single_double}

In this section, we discuss single-interface solutions (i.e., kinks and anti-kinks, or fronts) and double-interface solutions (i.e., one-drop solutions) of the standard and convective CH equations.
For this purpose, we consider the cCH equation on an infinite domain. A front solution is a solution that approaches two different constants as $x\rightarrow\pm\infty$. Let us denote these constants by $u_\mathrm{a}$ and $u_\mathrm{b}$ for $x\rightarrow-\infty$ and $x\rightarrow+\infty$, respectively. If $u_\mathrm{a}<u_\mathrm{b}$ we obtain a so-called kink solution. If $u_\mathrm{a}>u_\mathrm{b}$, we obtain an anti-kink solution. Here we call both ``front''. A double-interface (or one-drop) solution, is a solution that approaches the same constant (say $u_\mathrm{b}$) as $x\rightarrow\pm\infty$, but has a region where it approaches a different constant (say $u_\mathrm{a}$), {so that this region is macroscopic, i.e., sufficiently long compared to the lengths of the regions where the solution first transitions from $u_\mathrm{b}$ to $u_\mathrm{a}$ and then from $u_\mathrm{a}$ to $u_\mathrm{b}$. Such a solution may be considered as a superposition (with small correction) of well-separated kink and anti-kink solutions.} If $u_\mathrm{a}>u_\mathrm{b}$, we obtain a solution in the form of a drop, otherwise, we obtain a solution in the form of a hole. We note that our discussion of single- and double-interface solutions (i.e., front and one-drop states) partly follows the discussions of Emmott and Bray \cite{Ref.6}, Golovin et al.\ \cite{Ref.1}, Korzec et al.\ \cite{Ref.52}, Zaks et al.\ \cite{Ref.51}.

For the standard CH equation, it is well-known that front solutions have zero speed and are of the form (see Novick-Cohen and Segel \cite{NovickCohen_Segel1984})
\begin{equation}
u_0(x)=\pm\tanh\biggl(\frac{x}{\sqrt{2}}\biggr).
\end{equation}
There also exist periodic drop and hole solutions of drops/holes of arbitrarily large size. In the course of our work we consider domain sizes where one or two periods of a periodic solution fit. Note that in the latter case the solution has a discrete translation symmetry with respect to a shift of half the domain size. We refer to the respective solutions as ``one-period'' and ``two-period'' states. Alternatively we refer to them as ``one-drop'' and ``symmetric two-drop'' states.

For the cCH equation, a solution $u_0$ that is stationary in a frame moving at speed $v$  satisfies the equation
\begin {equation} \label {Eq.14a_steady1}
 -v  u_{0x} + D  u_0   u_{0 x} +( u_0 -  u_0^3 + u_{0x x})_{x x}=0,
\end{equation}
which, when integrated once, becomes
\begin {equation} \label {Eq.14a_steady2}
 -v  u_0 + \frac{D}{2}  u_0^2 + ( u_0 -  u_0^3 + u_{0x x})_{x}=C_0,
\end{equation}
where $C_0$ is a constant of integration that corresponds to the flux in the moving frame. Equation (\ref{Eq.14a_steady2}) can be rewritten as a three-dimensional dynamical system by introducing the functions $y_1=u_0$, $y_2=u_{0x}$ and $y_3=u_{0xx}$:
\begin{eqnarray}
y_1'&=&y_2,\label{eq:dynsyt1}\\
y_2'&=&y_3,\\
y_3'&=&C_0+v y_1-\frac{D}{2}y_1^2-y_2+3y_1^2y_2 \label{eq:dynsyt3}.
\end{eqnarray}
{We note that this dynamical system preserves phase space volume, since the divergence of the corresponding vector field (or, equivalently, the trace of the Jacobian matrix) is identically zero.}

The fixed points of (\ref{eq:dynsyt1})--(\ref{eq:dynsyt3}) satisfy $y_2=y_3=0$ and
\begin{equation}
\frac{D}{2}y_1^2-vy_1-C_0=0.
\end{equation}
Assuming that there exists a front solution that connects uniform solutions $u_\mathrm{a}$ and $u_\mathrm{b}$ 
we obtain that 
\begin{equation}
v=\frac{D}{2}(u_\mathrm{a}+u_\mathrm{b}),\qquad C_0=-\frac{D}{2}u_\mathrm{a}u_\mathrm{b}.
\label{eq:step_speed_flux}
\end{equation}
A front solution then corresponds to a heteroclinic orbit connecting the fixed point $(u_\mathrm{a},0,0)$ along the unstable manifold of $u_\mathrm{a}$, denoted by $W_{u}(u_\mathrm{a})$, to the fixed point $(u_\mathrm{b},0,0)$ along the stable manifold of $u_\mathrm{b}$, denoted by $W_{s}(u_\mathrm{b})$. 

In fact, it is known that equation (\ref{Eq.2}) has exact kink and anti-kink solutions which have $v=0$ and which are given by (see Golovin et al.~\cite{Ref.1})
\begin {equation} \label {Eq.7}
u_0^{\pm} (x) = \pm u^{\pm} \tanh \frac{u^{\pm}}{\sqrt{2}} x,   \qquad  \qquad  u^{\pm} = \sqrt{1 \mp {D}/{\sqrt{2}}},
\end{equation}
for $\pm$, respectively. Thus, for these solutions, $u_\mathrm{a}=-u^+$ and $u_\mathrm{b}=u^+$ for the case of the kink, and $u_\mathrm{a}=-u^-$ and $u_\mathrm{b}=u^-$ for the case of the anti-kink. Note that these solutions reduce to the front solutions of the standard CH equation when $D=0$. Note also that kink solutions exist only for $D<\bar D\equiv \sqrt{2}$. 

The eigenvalues for the fixed points $(u_\mathrm{a,b}, 0, 0)$ satisfy
\begin{equation}
\lambda^3+(1-3u_\mathrm{a,b}^2)\lambda+\frac{D}{2}(u_\mathrm{a,b}-u_\mathrm{b,a})=0.
\label{eq:lin_stab_uab}
\end{equation}

\begin{figure}
 \centering
 \includegraphics [width=7.5cm]{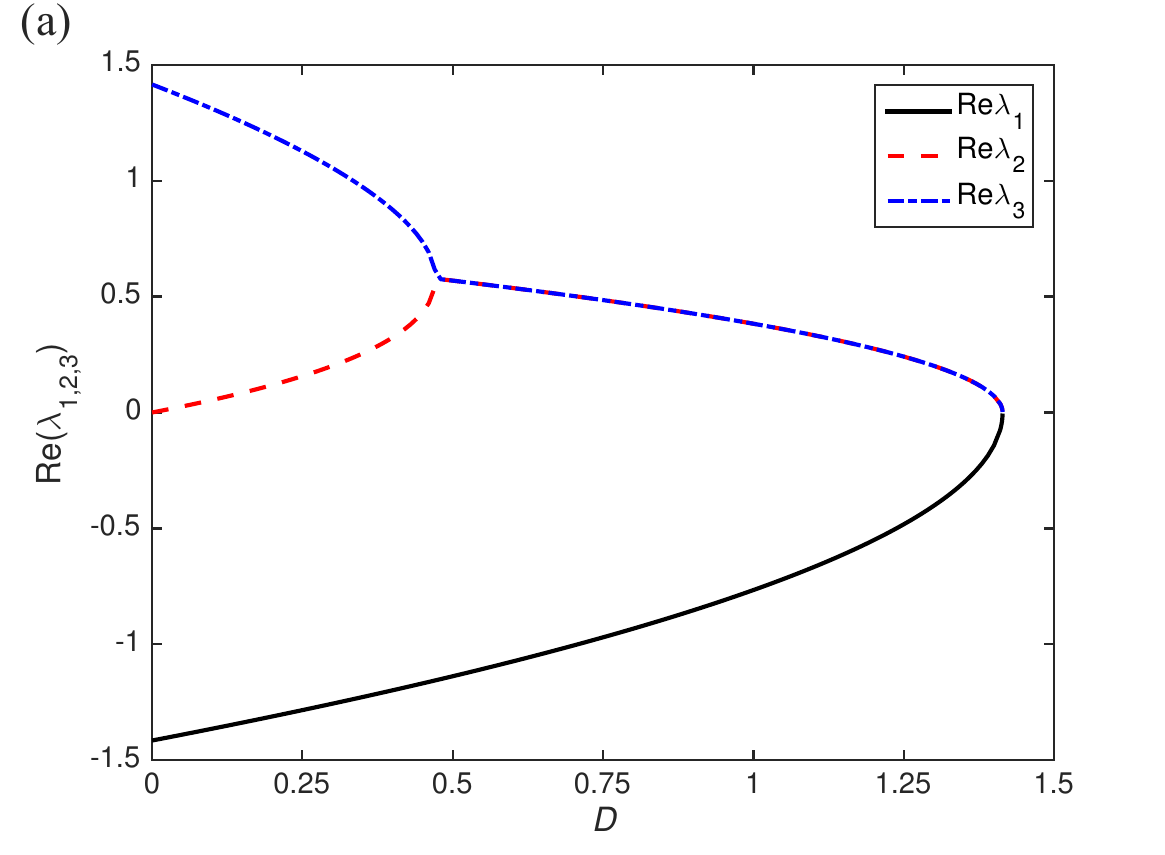}\hspace{-0.5cm}
  \includegraphics [width=7.5cm]{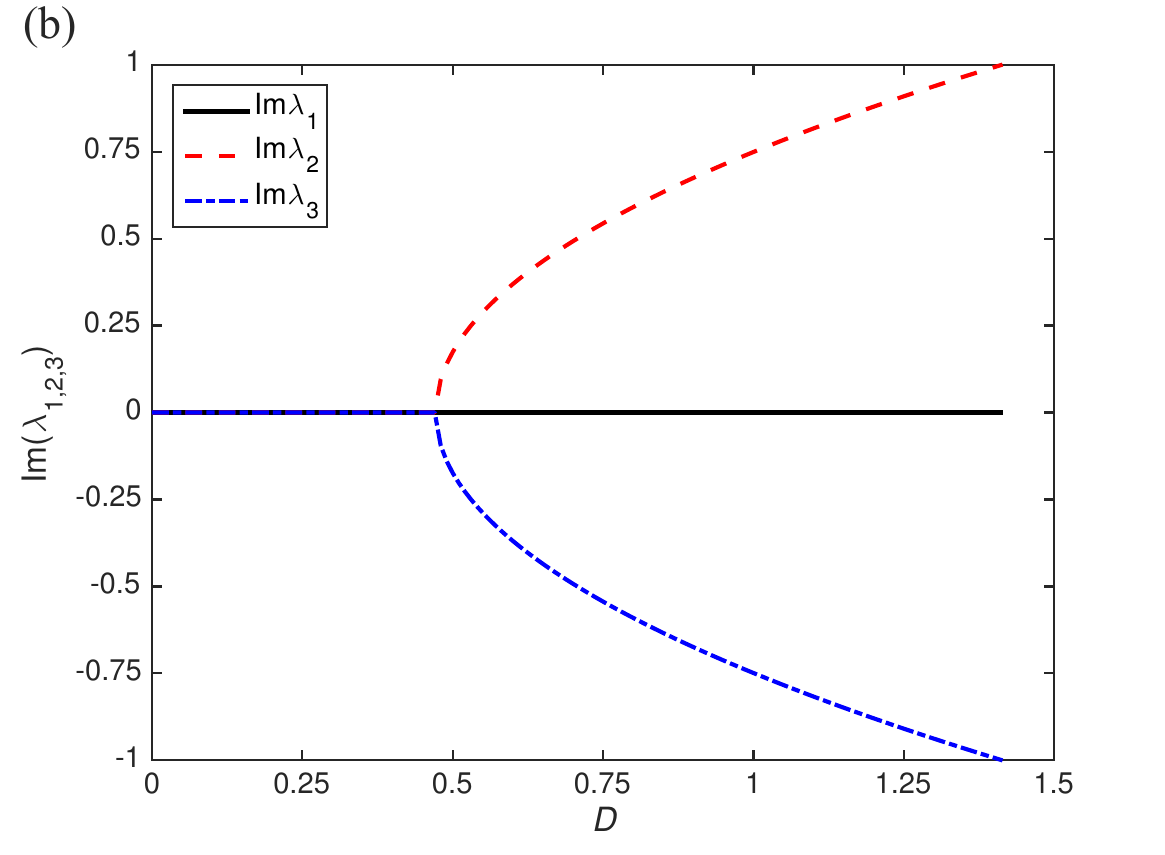}
  \vspace{-0.3cm}
 \caption{Shown is the dependence on $D$ of the (a) real and (b) imaginary parts of the eigenvalues
$\lambda_{1}, \lambda_{2}$ and $\lambda_{3}$ for $u^+ = \sqrt{1-D/\sqrt{2}}$ and $v= 0$ by solid, dashed and dotted lines, respectively.}
 \label{fig:Re_roots_v2}
 \vspace{-0.3cm}
\end{figure}

Figures~\ref{fig:Re_roots_v2}(a) and \ref{fig:Re_roots_v2}(b) show the dependence on $D$ of the real and imaginary parts, respectively, of the eigenvalues for $u^{+}$, which can be found analytically (see  Zaks et al.\ \cite{Ref.51}):
\begin{equation}
\lambda_1=-\sqrt{2-\sqrt{2}D},\quad
\lambda_{2,3}=\left(\sqrt{1-D/\sqrt{2}}\mp\sqrt{1-3D/\sqrt{2}}\right)\big/\sqrt{2},
\end{equation}
It can be seen that $\lambda_1$ is real and negative for all $D\in(0,\bar{D})$. The other two eigenvalues, $\lambda_{2,3}$, have positive real parts and are real for $D\in(0,\widehat{D})$ and complex conjugate for $D\in(\widehat D, \bar D)$, where $\widehat D=\sqrt{2}/3$. {This was first pointed out by Podolny et al.\ \cite{PODOLNY2005291}.} Note that as $D\rightarrow \bar D$, $u^+\rightarrow 0$, and $\lambda_1\rightarrow 0$, $\lambda_{2,3}\rightarrow \pm i$. The eigenvalues for $-u^+$ are $-\lambda_{1,2,3}$. 
We conclude that $\dim(W_{u}(u^{+}))=2$, $\dim(W_{s}(u^{+}))=1$, $\dim(W_{u}(- u^{+}))=1$, $\dim(W_{s}(-u^{+}))=2$. Therefore, there is a neighbourhood of the point $\bigr(-u^+,u^+\bigl)$ in the $(u_\mathrm{a},u_\mathrm{b})$-plane in which the kink solution exists only for $u_\mathrm{a}=-u^+$ and $u_\mathrm{b}=u^+$, and this kink solution is $u_0^+(x)$, given by (\ref{Eq.7}). Note that there may exist other isolates points in the $(u_\mathrm{a},u_\mathrm{b})$-plane which correspond to kink solutions, and some of these solutions were computed by Zaks et al.\ \cite{Ref.51}. Regarding anti-kink solutions, we conclude that there exists a one-parameter family of such solutions corresponding to a curve in some neighbourhood of the point $\bigr(u^+,-u^+\bigl)$ in the $(u_\mathrm{a},u_\mathrm{b})$-plane for each $D\geq 0$. We also note that although kink solutions exist for $D\in[0,\bar D)$
and anti-kink solutions exist for any $D\geq 0$, the flat parts of such solutions become linearly unstable (in the sense of temporal linear stability analysis) on a sufficiently long spatial domain when $D>\check  D=2\sqrt{2}/3$. 

Double-interface (and, in fact, many-interface) solutions can be analysed, for instance, by using the Shilnikov-type approach, see, e.g., Glendinning and Sparrow~\cite{Ref.54}, Guckenheimer and Holmes \cite{Ref.53}, Knobloch and Wagenknecht \cite{Ref.55}, Kuznetsov \cite{Ref.44}, Tseluiko et al.\ \cite{Ref.57}. Indeed, let us consider, for example, the case $D\in (\widehat D,\bar D)$. For the points $(-u^+,0,0)$ and $(u^+,0,0)$, there exists a heteroclinic orbit connecting the first point to the second (corresponding to the kink solution) and heteroclinic orbits connecting the second point to the first (corresponding to the anti-kink solutions). Then, we expect that there exists an infinite but countable number of the values of $u^+_{k}$, $k\in\mathbb{N}$, in the neighbourhood of $u^+$ for which there exist homoclinic orbits for the fixed points $(-u^+_{k}, 0, 0)$ that pass near $(u^+, 0, 0)$. Such orbits then correspond to drop solutions, and such drop solutions differ by their lengths. Then, since $\mathrm{Re}\,\lambda_{2,3}<-\lambda_1$ {(note that, given that the dynamical system (\ref{eq:dynsyt1})--(\ref{eq:dynsyt3}) preserves phase space volume, this inequality is automatically satisfied)}, Shilnikov's theory implies the existence of an infinite but countable number of subsidiary homoclinic orbits in the vicinity of the primary orbit that pass near $(u^+, 0, 0)$ several times before achieving homoclinicity. Such subsidiary homoclinic orbits correspond to multi-drop solutions. In addition, Shilnikov's theory implies the existence of an infinite number of {periodic orbits} in the vicinity of the primary homoclinic orbits. Such {periodic orbits} correspond to periodic arrays of drops. In a similar way, we can analyze hole solutions and can obtain finite or periodic arrays of hole solutions (of course, periodic arrays of hole solutions are equivalent to periodic arrays of drop solutions). We note, however, that for $D>\bar D$, kink solutions do not exist, and, therefore, the double-interface or multi-interface solutions that are typical of the standard CH equation do not exist for such values of $D$. Nevertheless, there may still exist homoclinic orbits corresponding to pulse or anti-pulse solutions (also referred to as hump or hollow solutions, respectively). Shilnikov's theory then implies the existence of bound states or (a)periodic arrays of such pulses or anti-pulses.  These solutions may still be characterized as localized drops or holes, but the nature of these solutions is different from that for the standard CH equation.

\section{Numerical computation of periodic one-drop solutions}
\label{sect:double_numerics}

To obtain solutions of equation (\ref{Eq.14a_steady2}) numerically, we use the continuation and bifurcation software Auto07p \cite{Ref.40}, see, e.g., Refs.~\cite{Tseluiko_etal_2018,EGUW2019springer,Tseluiko_etal_2016b,Lin_etal_2018,TVNB2001pre}, for more details on numerical implementation of such equations. For hands-on tutorials see \cite{cenosTutorial}.

\subsection{One-drop solutions for the standard CH equation}
\label{Sect:single_droplets_CH}

In this section, we review the structure of one-period solutions for the standard CH equation, when $D=0$, for different values of $\bar u$, and, in particular, we compute solutions for $\bar u=0.4$, $0.55$ and $0.6$. Note that much more exhaustive results are available in the literature for the standard CH equation, e.g., \cite{MaMW2007ijbc,MMMW2008rmc,NovickCohen_Segel1984,Novi1985jsp,ThMF2007pf}.
We characterize the solutions by their norms $\| \delta u_0 \|=\sqrt{(1/L)\int_0^L(u_0-\bar u)^2\,dx}$ and their free energies $F[u_0]$ defined by (\ref{Eq.B65}).  
Note that for $|\bar u| <1/ \sqrt 3$, the flat solution $u_0= \bar u$ becomes unstable when $L > L_c={2 \pi}/{k_c}$, where $k_c = \sqrt{1-3 \bar u^2}$. We find that $L_{c} =8.7$ and $20.66$  for $\bar u =0.4$ and $0.55$, respectively. Whereas for $\bar u =0.6$ the flat solution is linearly stable for any domain size.

\begin{figure}
 \centering
 \includegraphics [height=4cm]{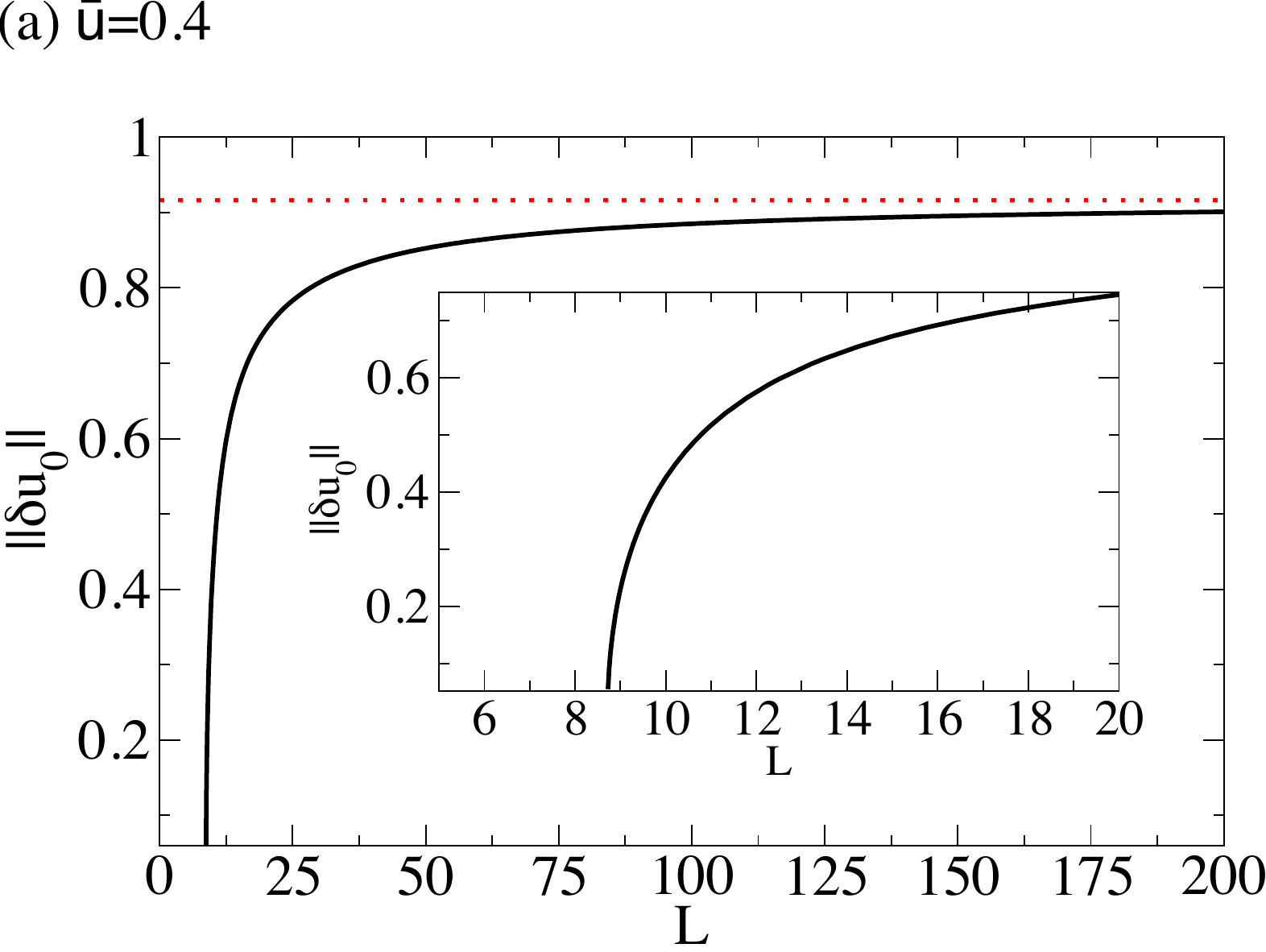}\hspace{-0.1cm}
 \includegraphics [height=4cm]{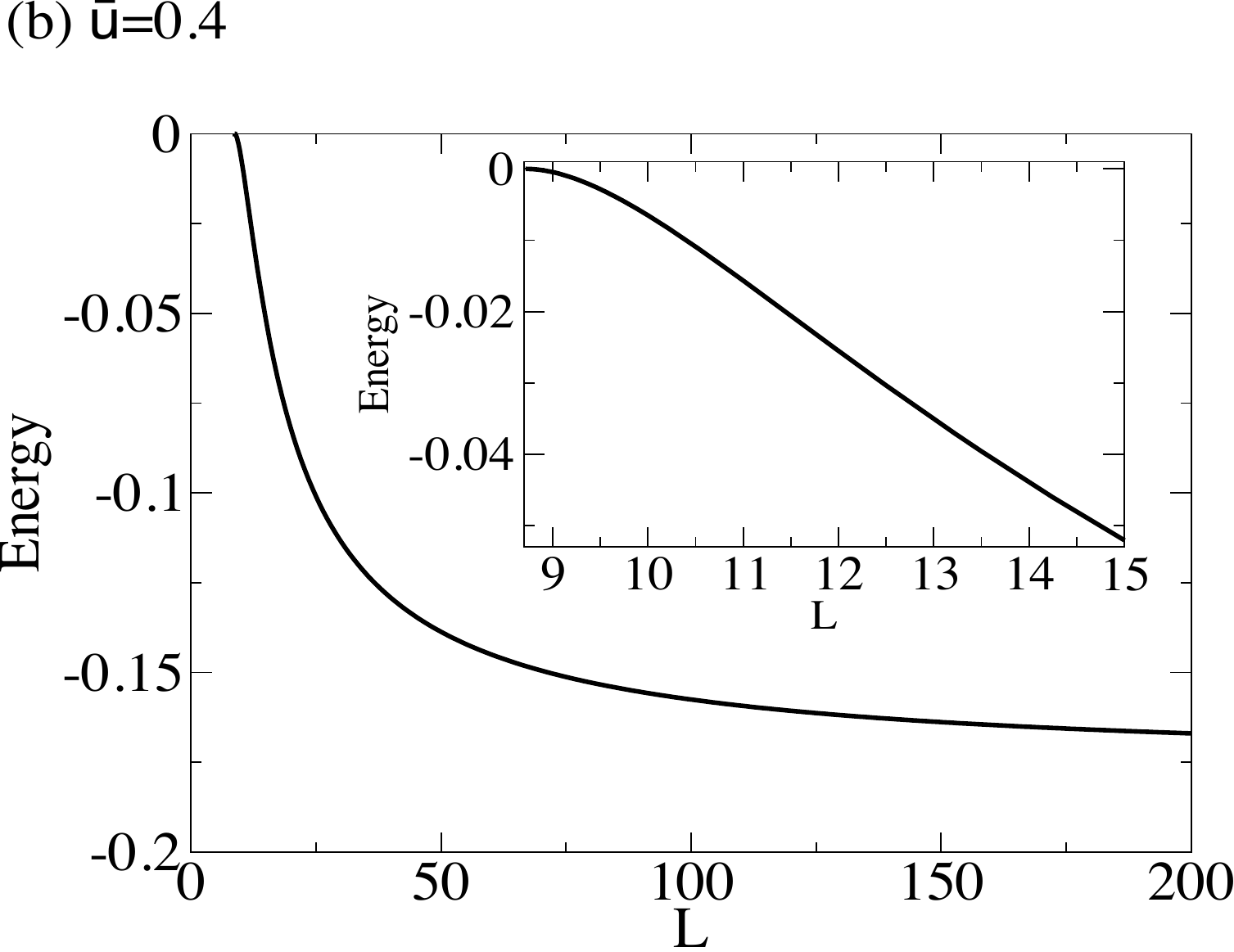}\hspace{-0.1cm}
 \includegraphics [height=4cm]{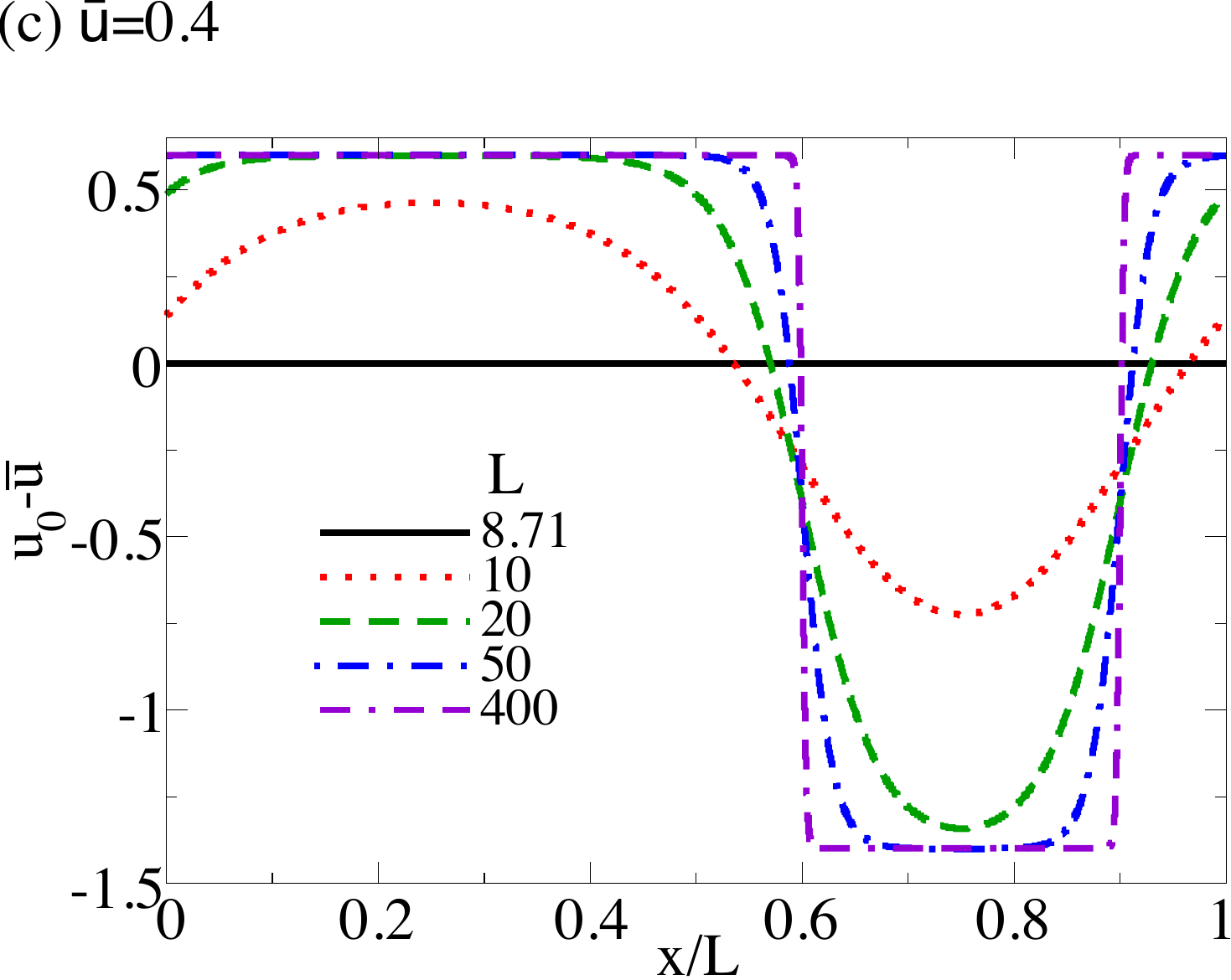}
 \includegraphics [height=4cm]{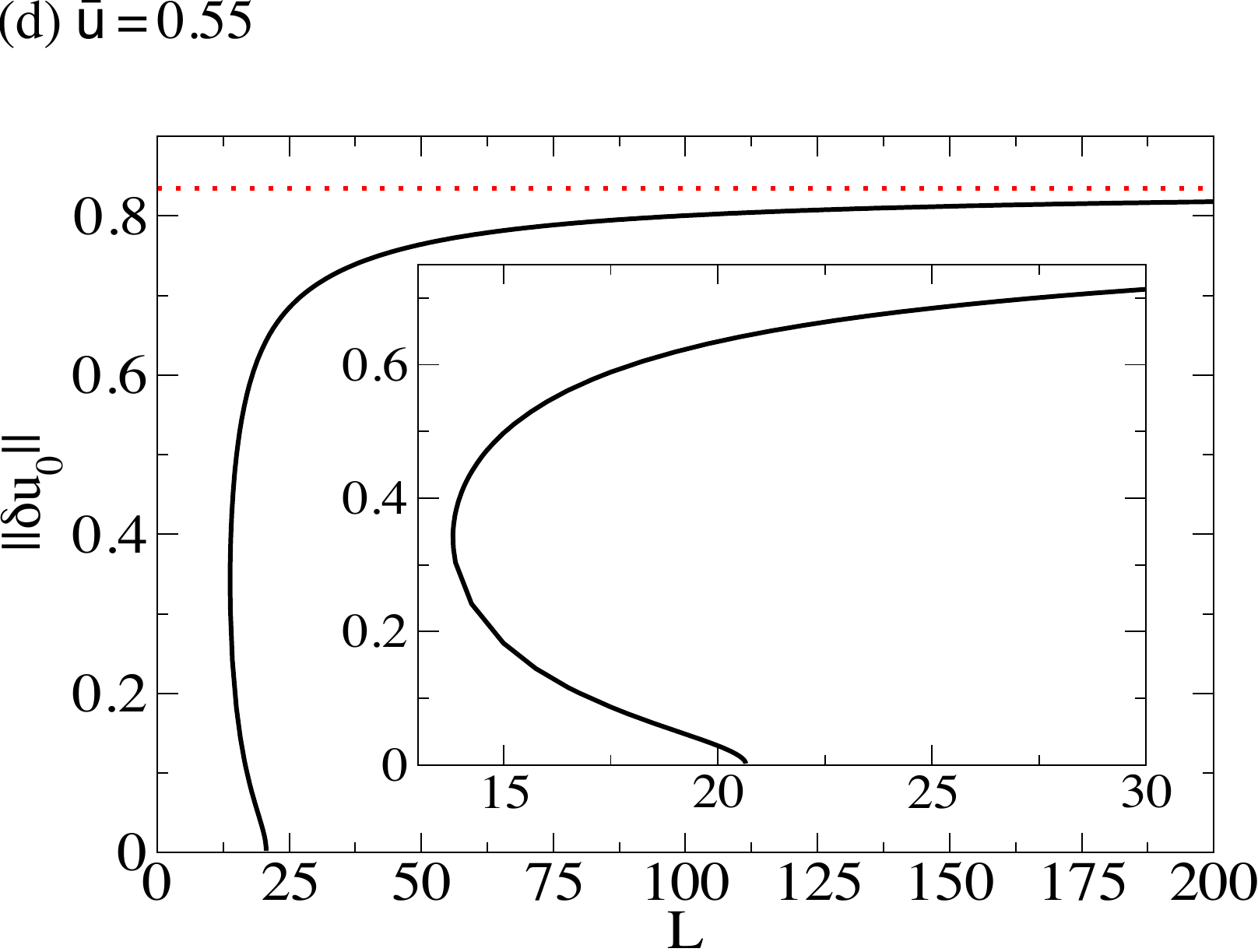}\hspace{-0.1cm}
 \includegraphics [height=4cm]{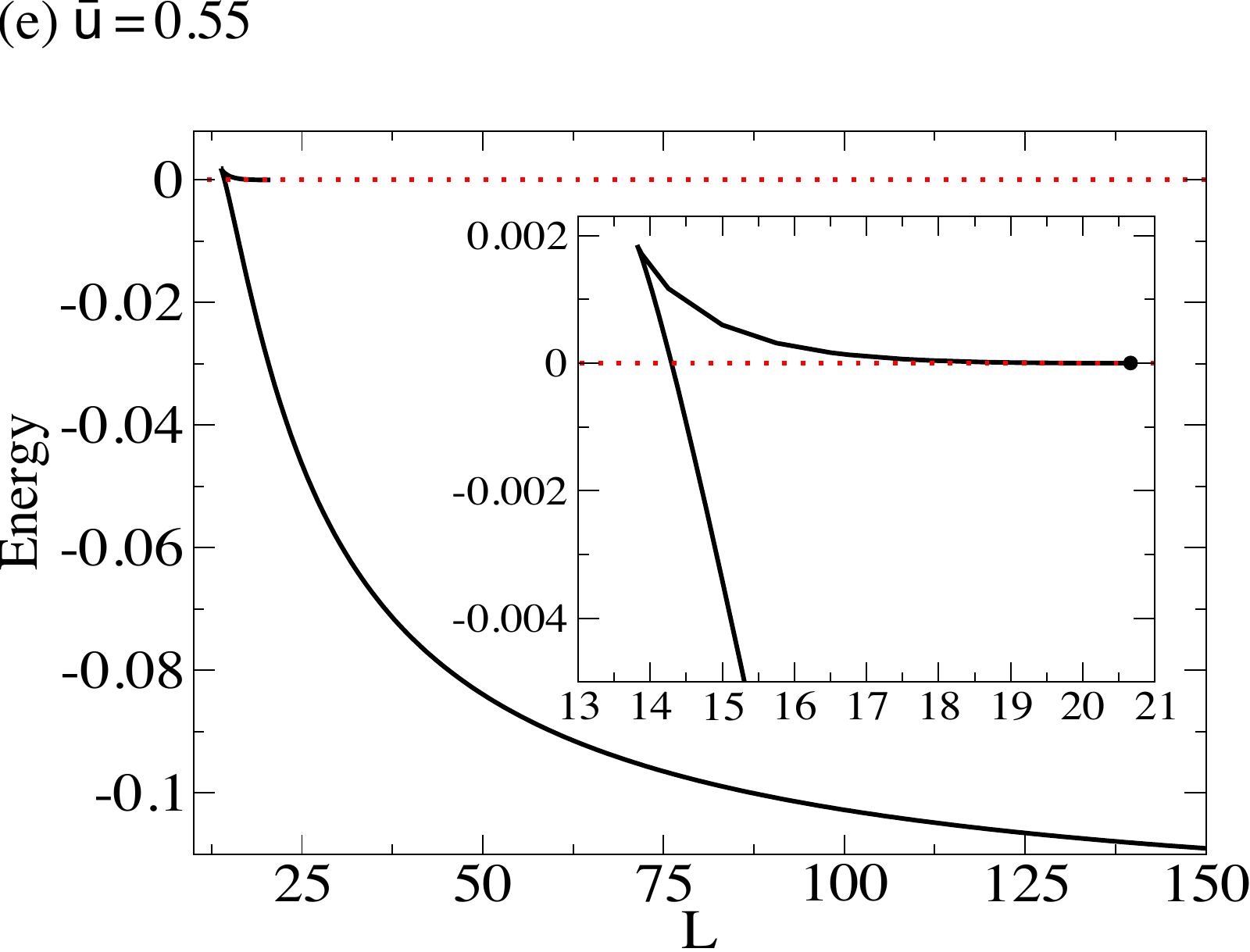}\hspace{-0.1cm}
 \includegraphics [height=4cm]{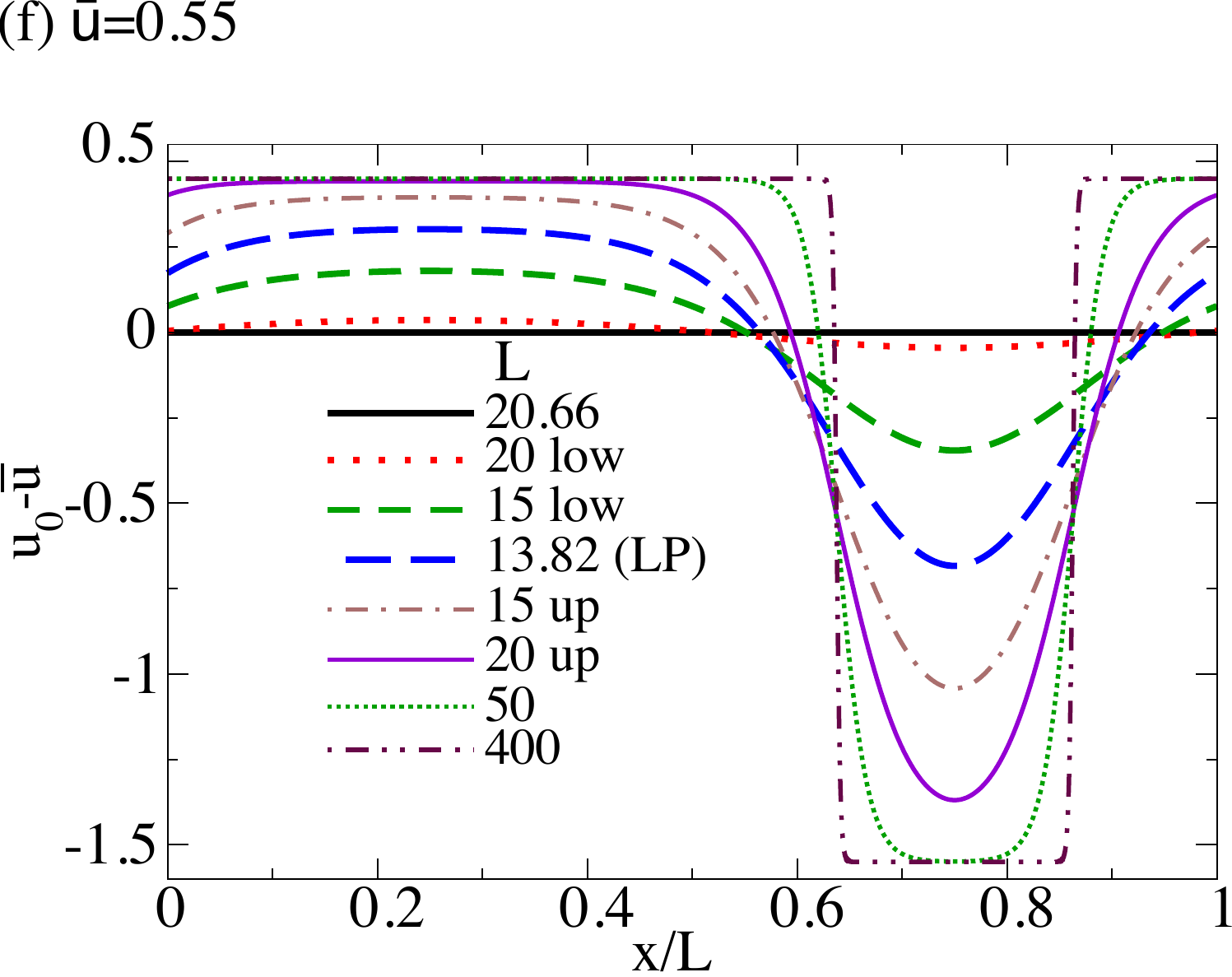}
 \includegraphics [height=4cm]{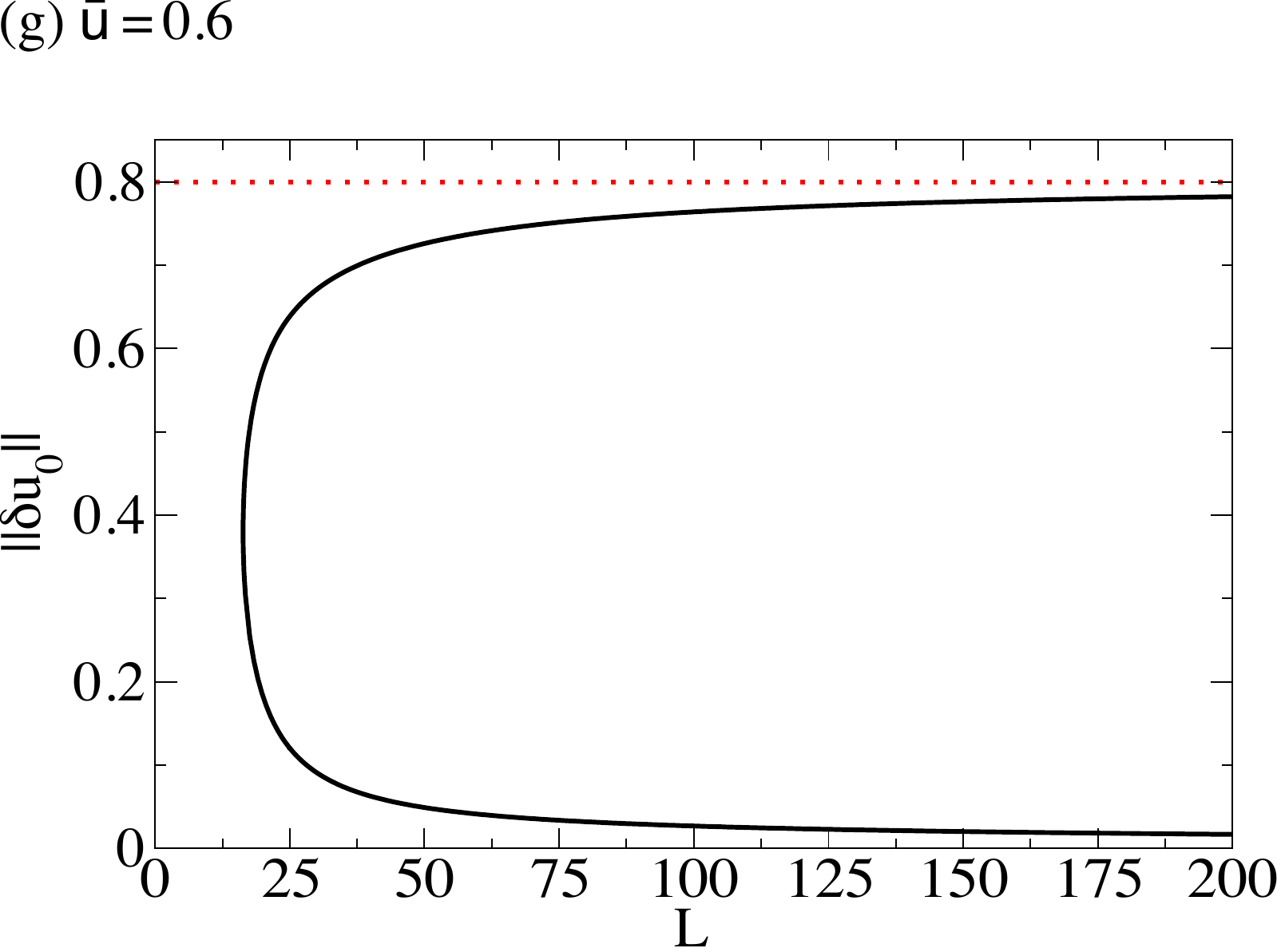}\hspace{-0.1cm}
 \includegraphics [height=4cm]{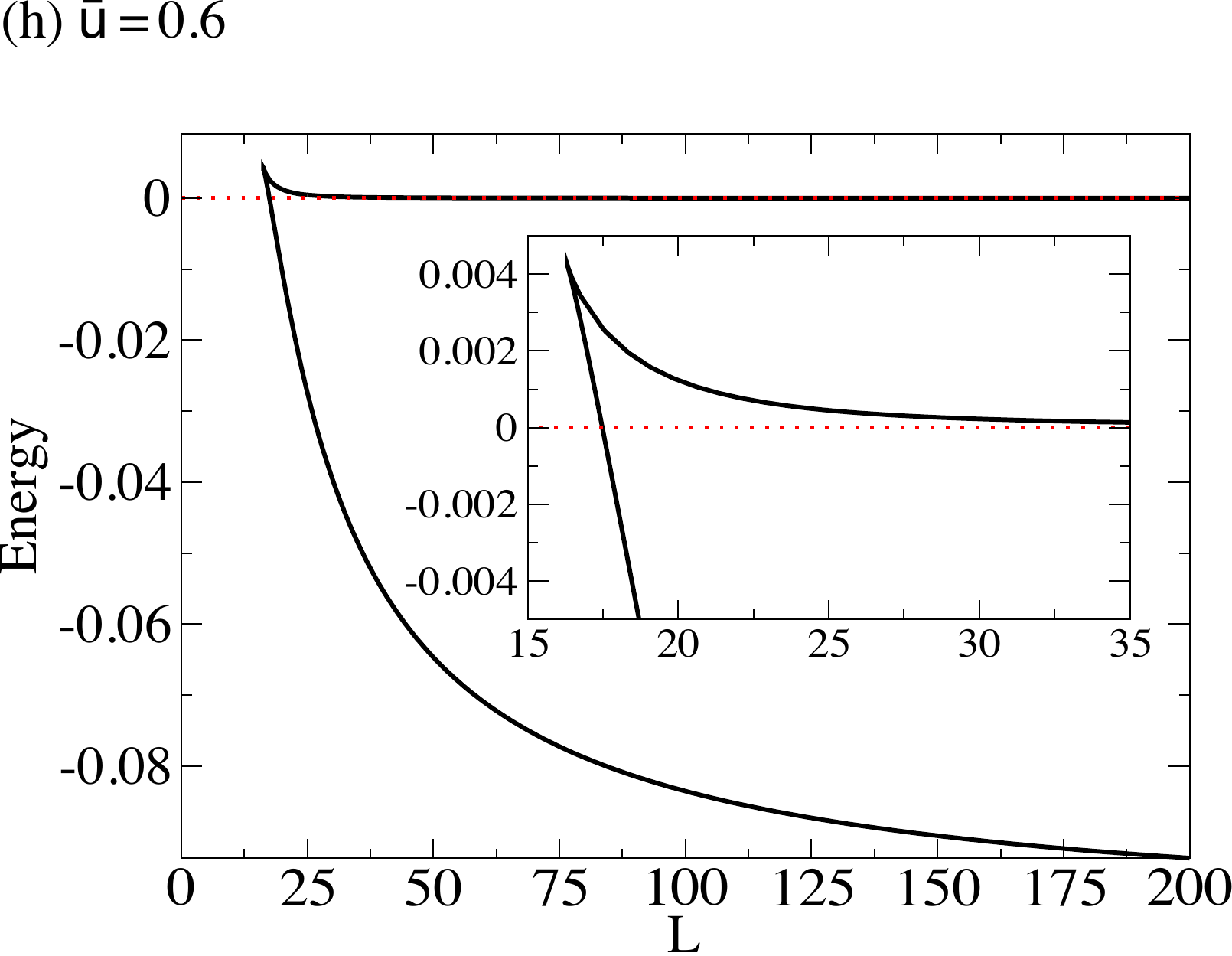}\hspace{-0.12cm}
 \includegraphics [height=4cm]{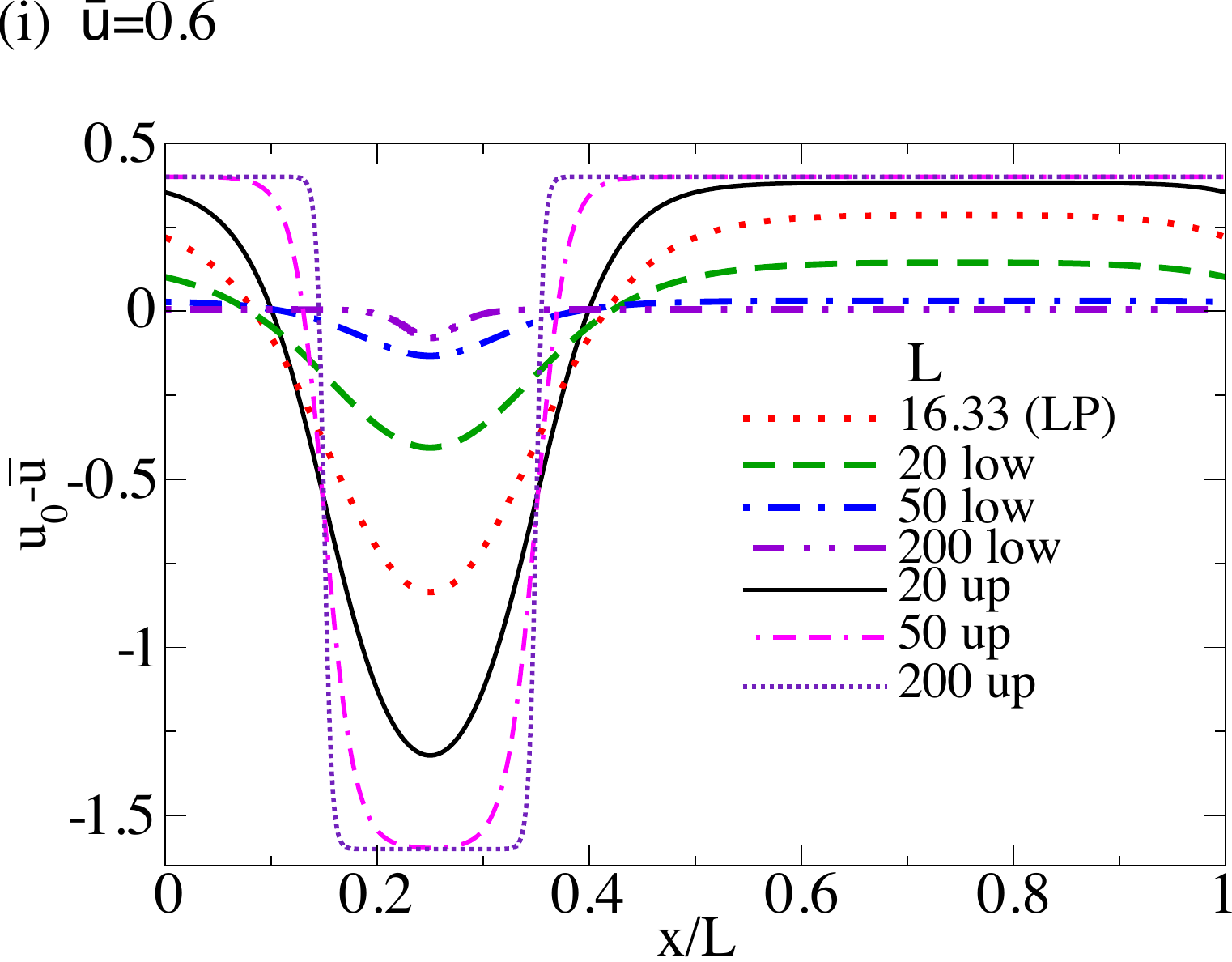}\\
 \vspace{-0.3cm}
 \caption{Shown are (a) the dependence of $\| \delta u_0 \|$ on $L$, (b) the dependence of $F[u_0]$ on $L$, and (c) profiles at different values of $L$, as indicated in the legend, of steady one-drop solutions $u_0$ of the standard CH equation (\ref{Eq.2}), when $D=0$, for the case when $\bar{u}=0.4$. The respective results for $\bar{u}=0.55$ are given in panels (d), (e) and (f), and for $\bar{u}=0.6$ -- in panels (g), (h) and (i).
}
 \label{fig:1.1}
 \vspace{-0.3cm}
\end{figure}

The results for $\bar u=0.4$ showing the dependence of the norm $\| \delta u_0 \|$ on $L$, the dependence of the energy $F[u_0]$ on $L$ and solutions for several values of $L$ are shown in Figs.~\ref{fig:1.1}(a), (b) and (c), respectively. The respective results for $\bar u=0.55$ are shown in  Figs.~\ref{fig:1.1}(d), (e) and (f), and for $\bar u=0.6$ -- in  Figs.~\ref{fig:1.1}(g), (h) and (i). The dotted lines in panels (a), (d) and (g) correspond to $\sqrt{1-\bar u^2}$, and we can see that in each case the norm approaches this value as $L$ increases.

For $\bar u=0.4$, we can see that the primary bifurcation at $L_{c} =8.7$ is supercritical, and the energy monotonically decreases as $L$ increases. The solution profiles have the form of a single wide drop, or, equivalently, a single narrow hole. As can be seen in Fig.~\ref{fig:1.1}(c), the width of the drop approaches a constant value in the rescaled variable $x/L$, i.e., in the original variable $x$ the width grows linearly with $L$. In fact, the width of the drop grows as $0.5(1+\bar u)L=0.7\, L$ and the width of the hole grows as $0.5(1-\bar u)L=0.3\, L$ as $L$ increases, so that the mean value remains equal to $\bar u=0.4$.

For $\bar u=0.55$, the primary bifurcation at $L_c=20.66$ is subcritical. The branch of nonuniform solutions initially follows to decreasing values of the domain size $L$ and is unstable up to the saddle-node bifurcation at $L=L_s\approx 13.818$.  After this point, the branch turns back and becomes stable. The exact value of $\bar u$ at which the bifurcation switches from supercritical to subcritical can be obtained by the weakly nonlinear analysis given in the Appendix, see equation (\ref{eq:Sub_Sup_cond}). It turns out that this value is $\bar u^*=1/\sqrt{5}\approx 0.45$. Note that for $\bar u=0.55$ the energy of the nonuniform solution first increases monotonically, up to the saddle-node bifurcation, and then decreases monotonically. It remains positive up to a certain value of the domain size, $L_m\approx 14.30$ between $L_s$ and $L_c$, and then becomes negative. The point $L=L_m$ is the so-called Maxwell point. At this point, both linearly stable solutions, i.e., the uniform solution and the nonuniform solution with the larger value of the norm, have the same value of the energy. For $L\in (L_s,\,L_m)$, the uniform solution has lower free energy, whereas for $L>L_m$, the nonuniform solution has lower free energy. In Fig.~\ref{fig:1.1}(f), we can see that, as $L$ increases, the solution profiles for $\bar u=0.55$ behave as in the case of $\bar u =0.4$, except that now the width of the drop grows as $0.775\,L$ and the width of the hole grows to $0.225\,L$ as $L$ increases, so that the mean value remains equal to $\bar u=0.55$.

As mentioned above, since $\bar u= 0.6 > 1/\sqrt 3$, the flat solution $\bar u=0.6$ is linearly stable for any $L$, i.e., there is no primary bifurcation on the uniform solution. To produce the branch of nonuniform solutions, we can first compute the branch of nonuniform solutions for, e.g., $\bar u=0$, and then select a solution on this branch at a sufficiently large value of $L$ (e.g., $L=100$). We then keep $L$ fixed and perform a continuation in $\bar u$, until we reach the value $\bar u=0.6$. This produces the nonuniform solution for $\bar u=0.6$ at $L=100$. After that, we again keep $\bar u$ fixed and perform a continuation in $L$, going in both directions, which produces the whole branch of nonuniform solutions. We can observe that the branch of nonuniform solutions has a turning point at $L=L'_s \approx 16.327$. For each $L>L'_s $, there are two nonuniform solutions, one is unstable and is of smaller norm while the other one is stable and is of larger norm. The energy of the linearly unstable nonuniform solution monotonically decreases from some positive value to zero as $L$ increases from $L'_s$. Whereas the energy of the linearly stable nonuniform solution decreases monotonically from a positive value to negative values crossing zero at the Maxwell point, $L_m'\approx 17.466$.  In Fig.~\ref{fig:1.1}(i), we can see that, as $L$ increases, the solution profiles of the upper branch of nonuniform solutions for $\bar u=0.6$ behave as in the previous cases, except that now the width of the drop grows as $0.8\,L$ and the width of the hole grows to $0.2\,L$ as $L$ increases, so that the mean value remains equal to $\bar u=0.6$. The behaviour of the solutions of the lower branch of nonuniform solutions is, however, different. As $L$ increases, their amplitude decreases approaching a constant value, and the width in the rescaled variable $x/L$ also decreases approaching a constant value in the original variable $x$, so that the solution tends to an anti-pulse shape.
Note that a recent study investigates how the Maxwell construction at phase coexistence emerges from bifurcation diagrams like the ones in Fig.~\ref{fig:1.1} for finite-size systems when approaching the thermodynamic limit \cite{TFEK2019njp}.

\subsection{One-drop solutions for the cCH equation}

\label{sect:convCH_waves_0p4_0p55_0p6}

\begin{figure}
 \centering
 \includegraphics [width=8cm]{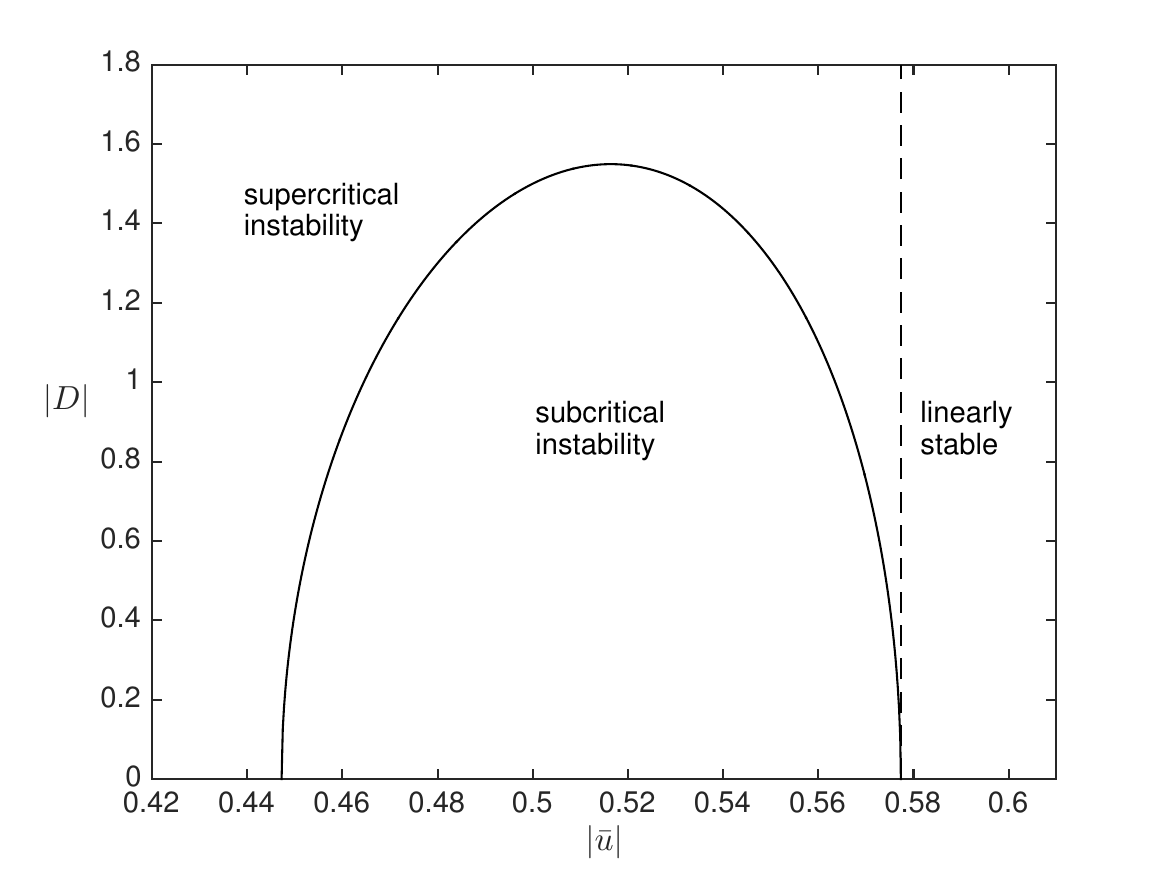}
 \vspace{-0.3cm}
 \caption{The solid line represents the boundary in the $(|\bar u |,  |D|)$-plane separating the regions where the primary bifurcation for the cCH equation (\ref{Eq.2}) is supercritical or subcritical. The region to the right of the vertical dashed line is the region where the homogeneous solution is linearly stable.}
 \label{fig:Dc_u}
 \vspace{-0.3cm}
\end{figure}

We now consider how the driving force affects the one-period steady traveling-wave solutions of the CH equation. We use both the driving force, $D$, and the domain size, $L$, as the control parameters and consider three cases, $\bar{u}=0.4$, $0.55$ and $0.6$, as we did for the standard CH equation. {Note that bifurcations of periodic one-drop  solutions of the cCH equation were previously analysed in detail by Zaks et al.\ \cite{Ref.51} but only for $\bar{u}=0$.} We first notice that the changeover from supercritical to subcritical primary bifurcation that we have discussed in the previous section, is affected by $D$. Using the weakly nonlinear analysis presented in the Appendix (see equation (\ref{eq:Sub_Sup_cond})) we can show that the line separating the regions in the $(\bar u, D)$-plane where the primary bifurcation is supercritical or subcritical is given by the equation
\begin{equation}
540\bar u^4-288\bar u^2+36+D^2=0,
\end{equation}
see Fig.~\ref{fig:Dc_u}. To be more precise, for a fixed $D>0$, the primary bifurcation (if it exists) is subcritical if $\bar u \in(\bar u^*,\bar u^{**})$ and supercritical otherwise, where
\begin{equation}
\bar u^*=\frac{\sqrt{240-10\sqrt{36-15 D^2}}}{30}, \quad \bar u^{**}=\frac{\sqrt{240+10\sqrt{36-15 D^2}}}{30},
\end{equation} 
We remind here that we consider only nonnegative values of $\bar u$ and $D$. Equivalently, for a fixed $\bar u$, the bifurcation is subcritical if $D<D_c$ and supercritical otherwise, where
\begin{equation}
D_c=\sqrt{-540\bar u^4+288 \bar u^2-36}.
\label{eq:Dc}
\end{equation}
Note that the expression under the square root in (\ref{eq:Dc}) is positive only when $1/\sqrt{5}<\bar u <1/\sqrt{3}$ (considering nonnegative values of $\bar u$), i.e., the driving force can switch the type of the bifurcation only when $1/\sqrt{5}<\bar u<1/\sqrt{3}$. If $0\leq \bar u<1/\sqrt{5}$, the primary bifurcation is supercritical for any value of the driving force. We also remind that if $\bar u>1/\sqrt{3}$, there is no primary bifurcation and the uniform solution is linearly stable for any value of the driving force. It can also be easily concluded that if $D\geq D_c^{\mathrm{max}}=2\sqrt{3/5}\approx 1.55$, the primary bifurcation can only be supercritical.

\begin{figure}
 \centering
  \includegraphics [width=5.1cm]{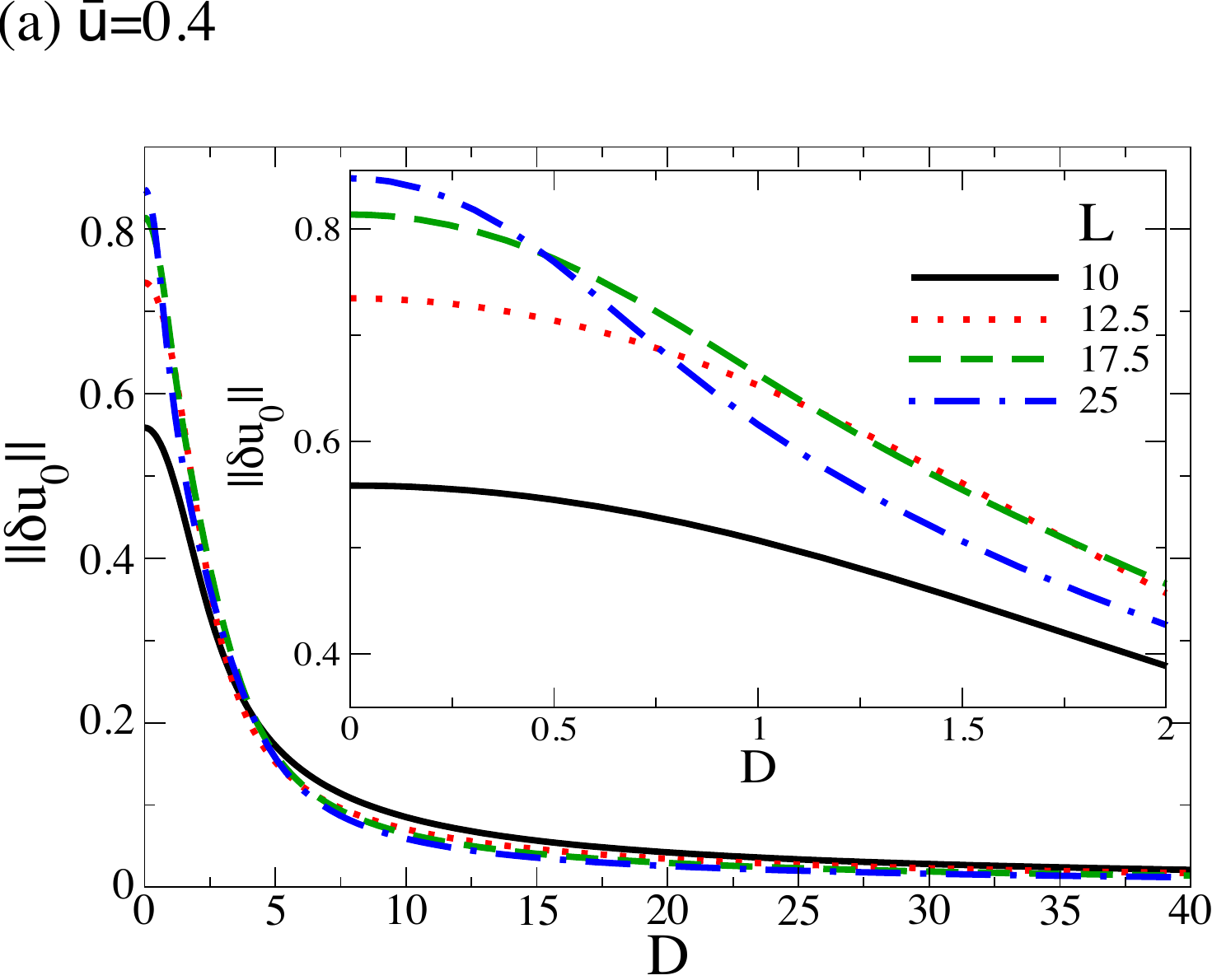}\hspace{-0.1cm}
 \includegraphics [width=5.1cm]{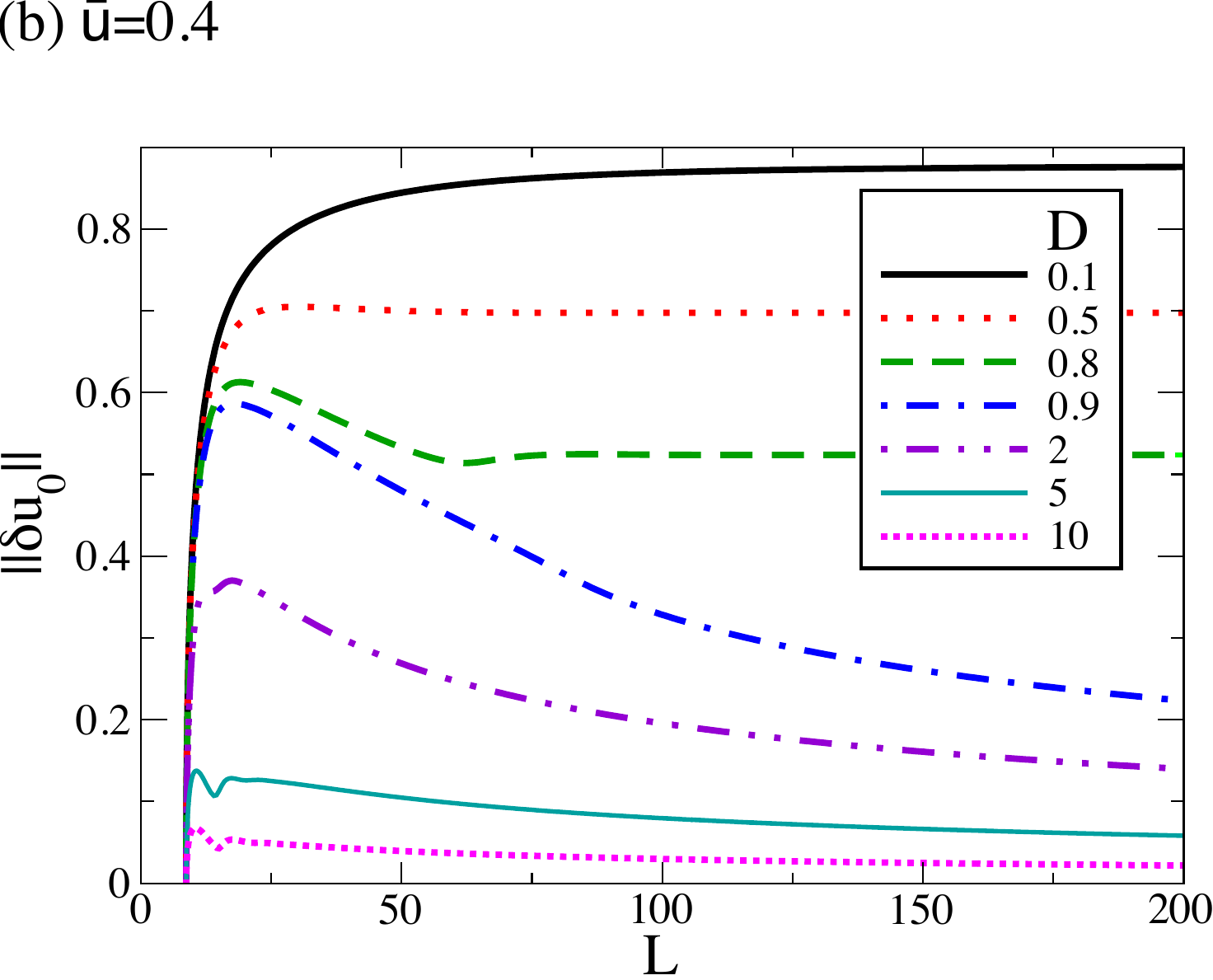}\hspace{-0.1cm}
 \includegraphics [width=5.1cm]{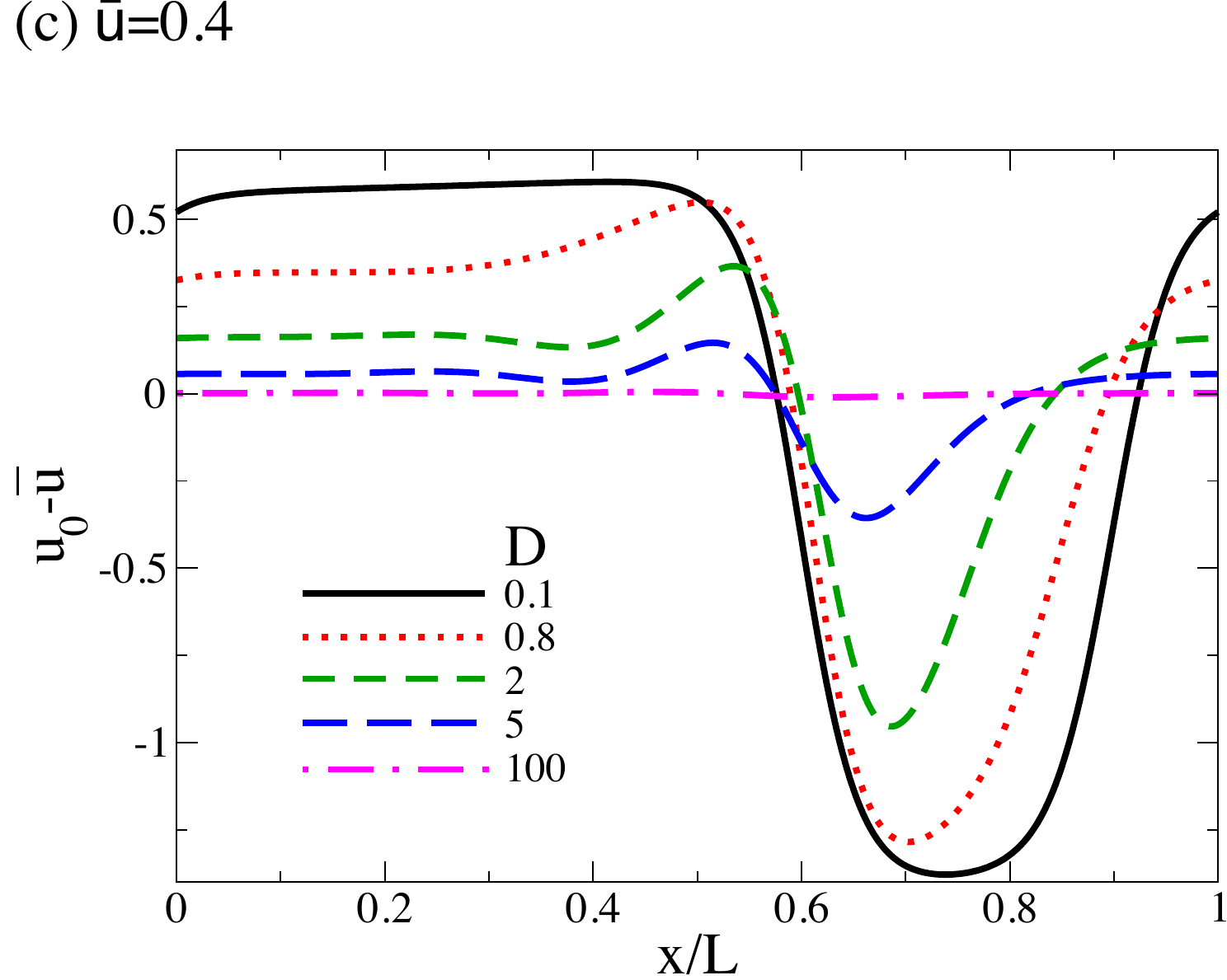} 
 \includegraphics [width=5.1cm]{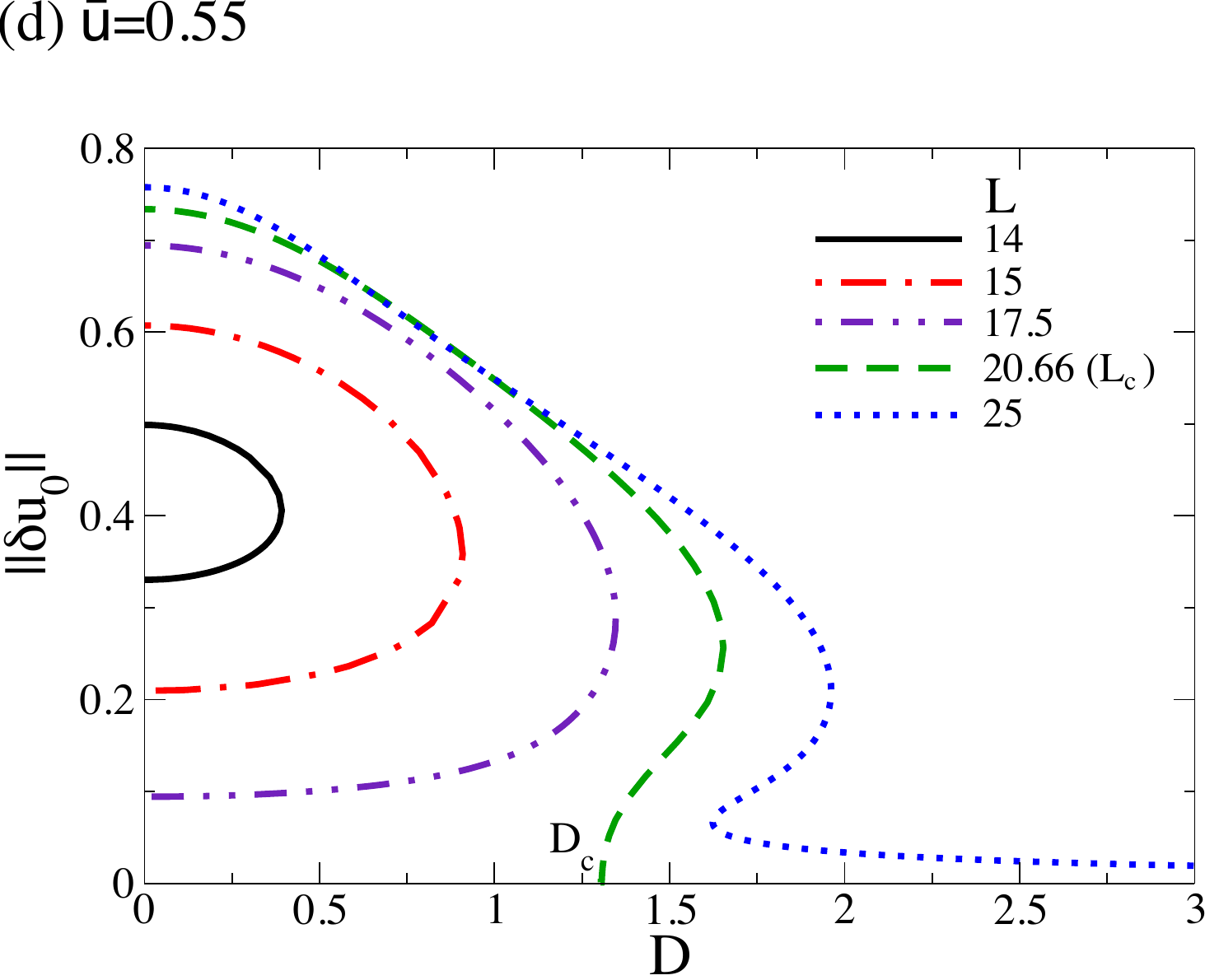}\hspace{-0.1cm}
 \includegraphics [width=5.1cm]{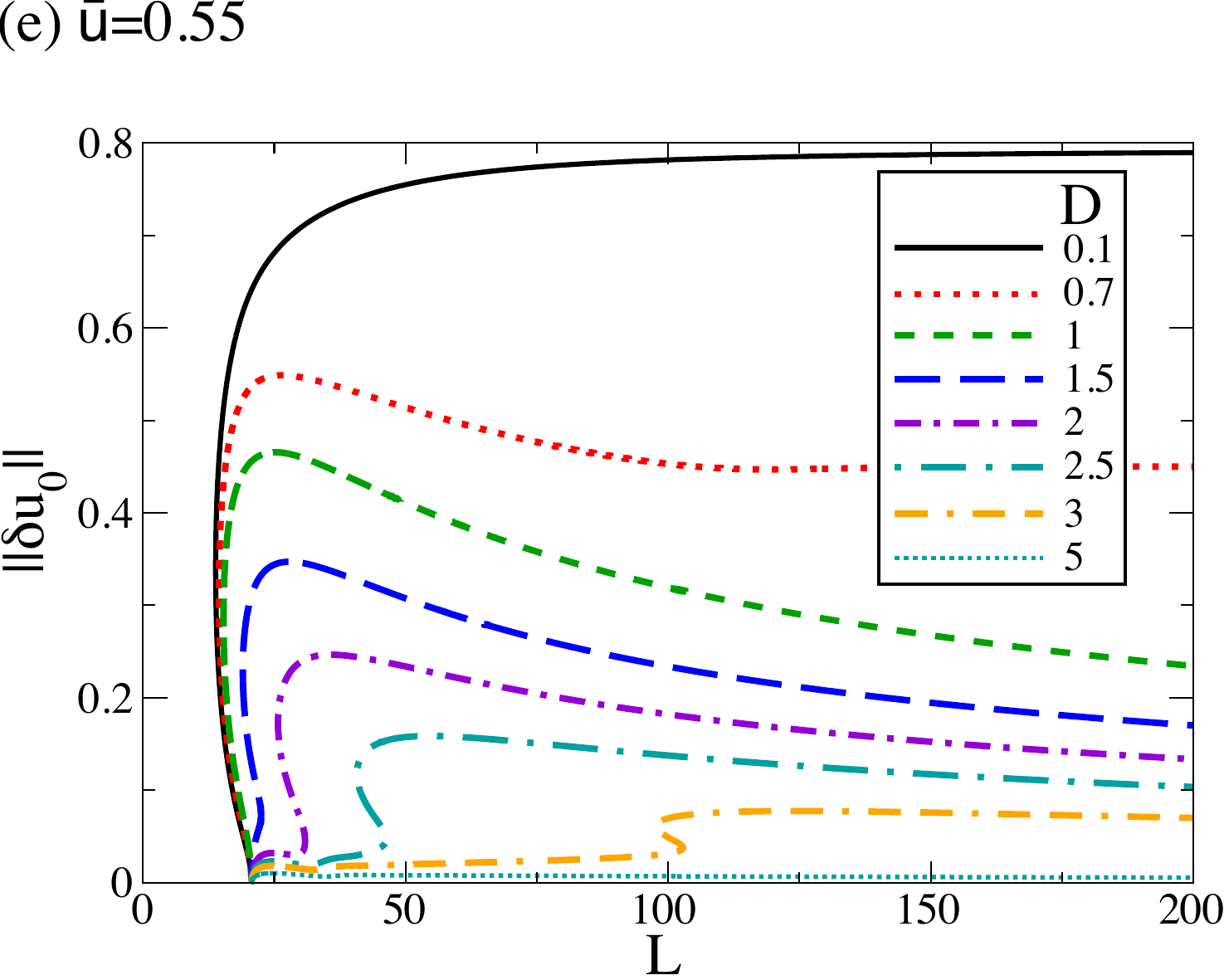}\hspace{-0.1cm}
 \includegraphics [width=5.1cm]{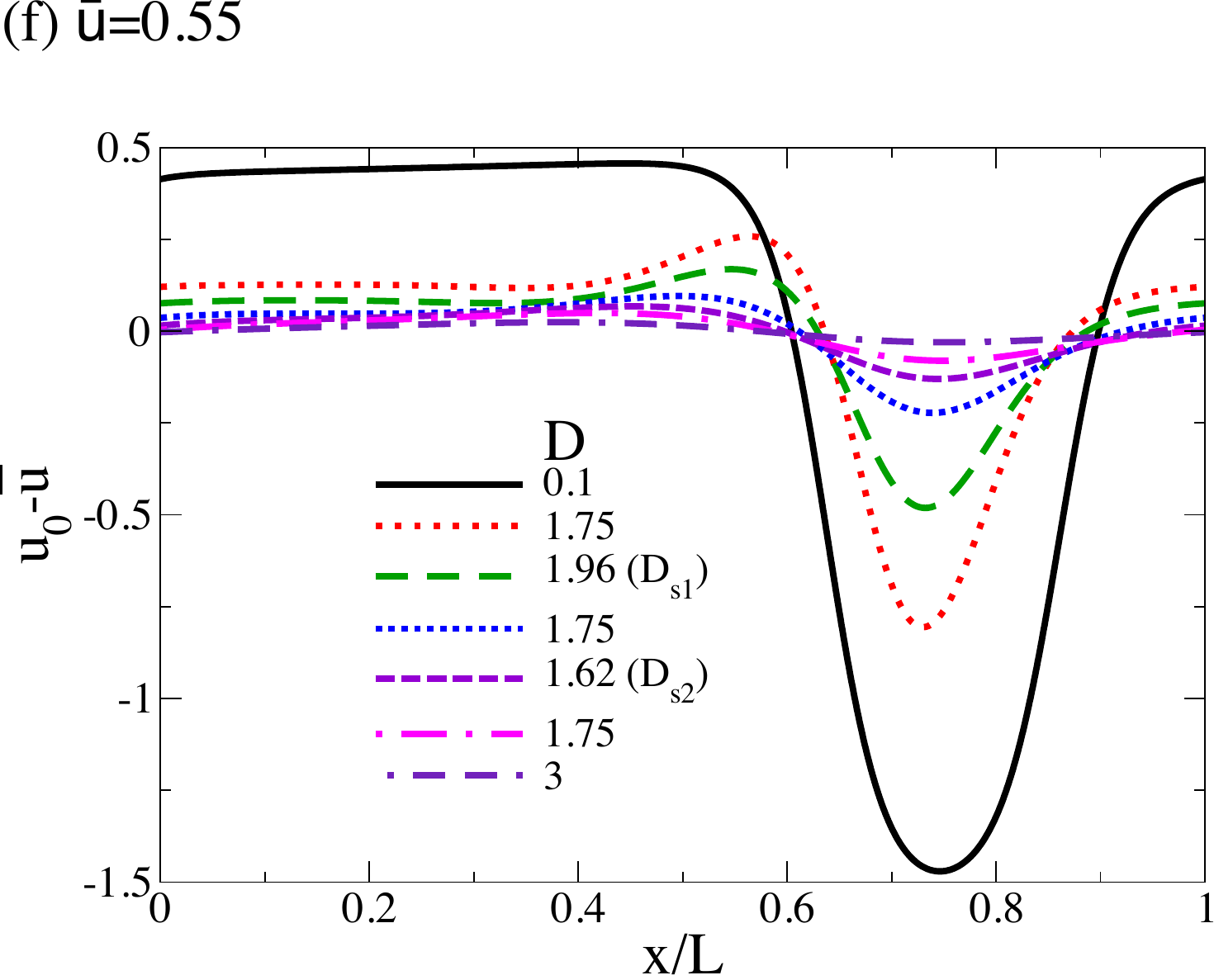} 
 \includegraphics [width=5.1cm]{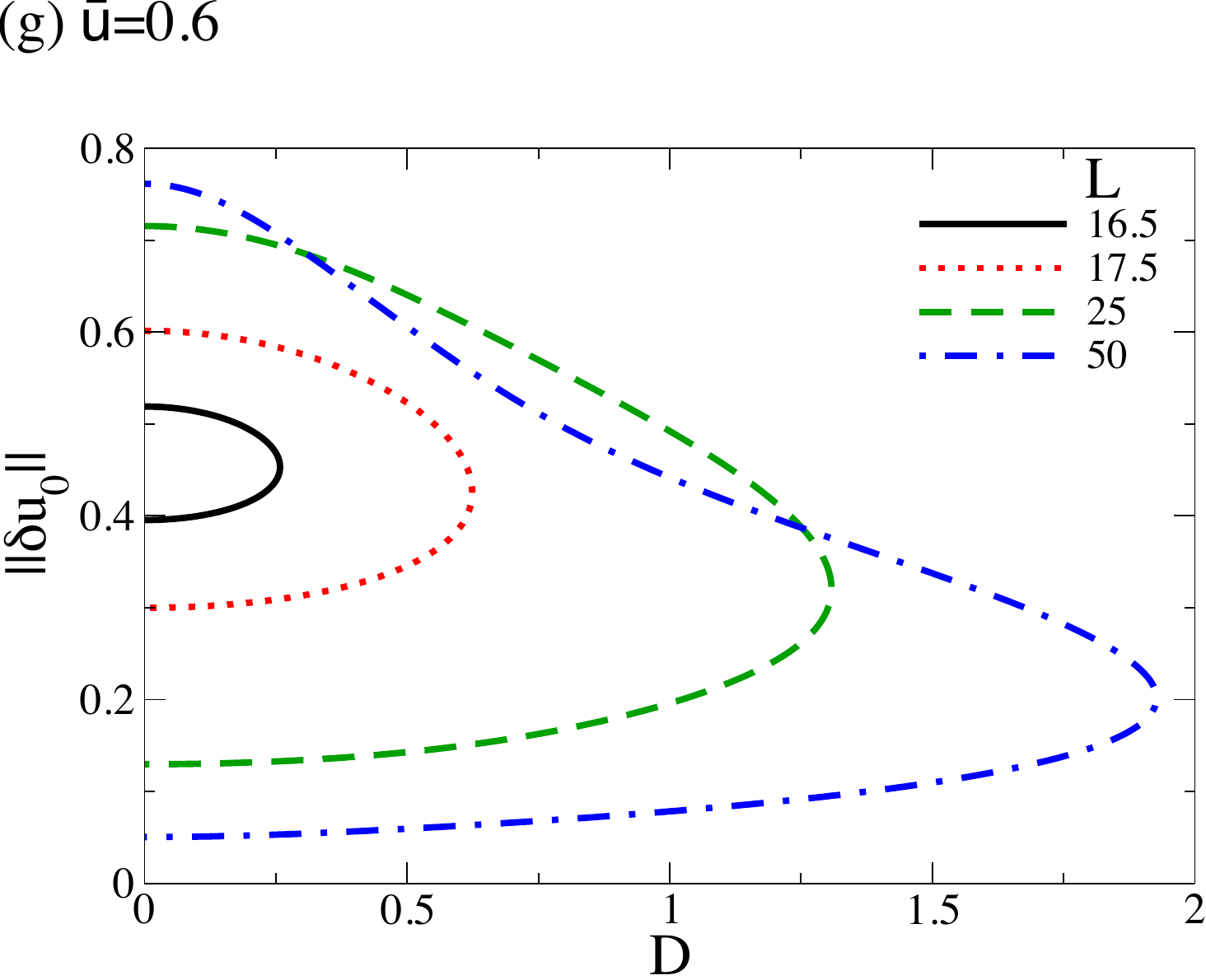}\hspace{-0.1cm}
 \includegraphics [width=5.1cm]{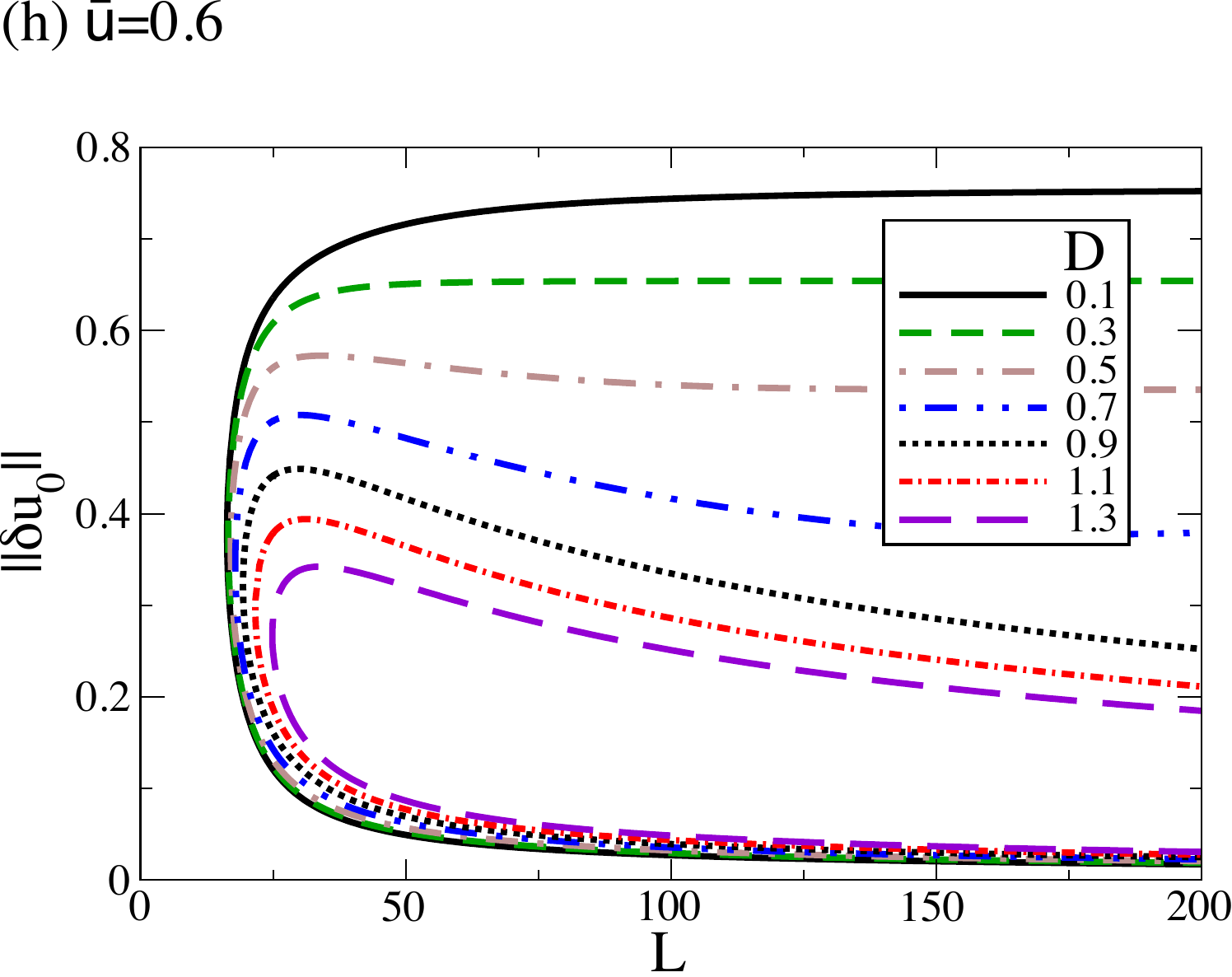}\hspace{-0.1cm}
 \includegraphics [width=5.1cm]{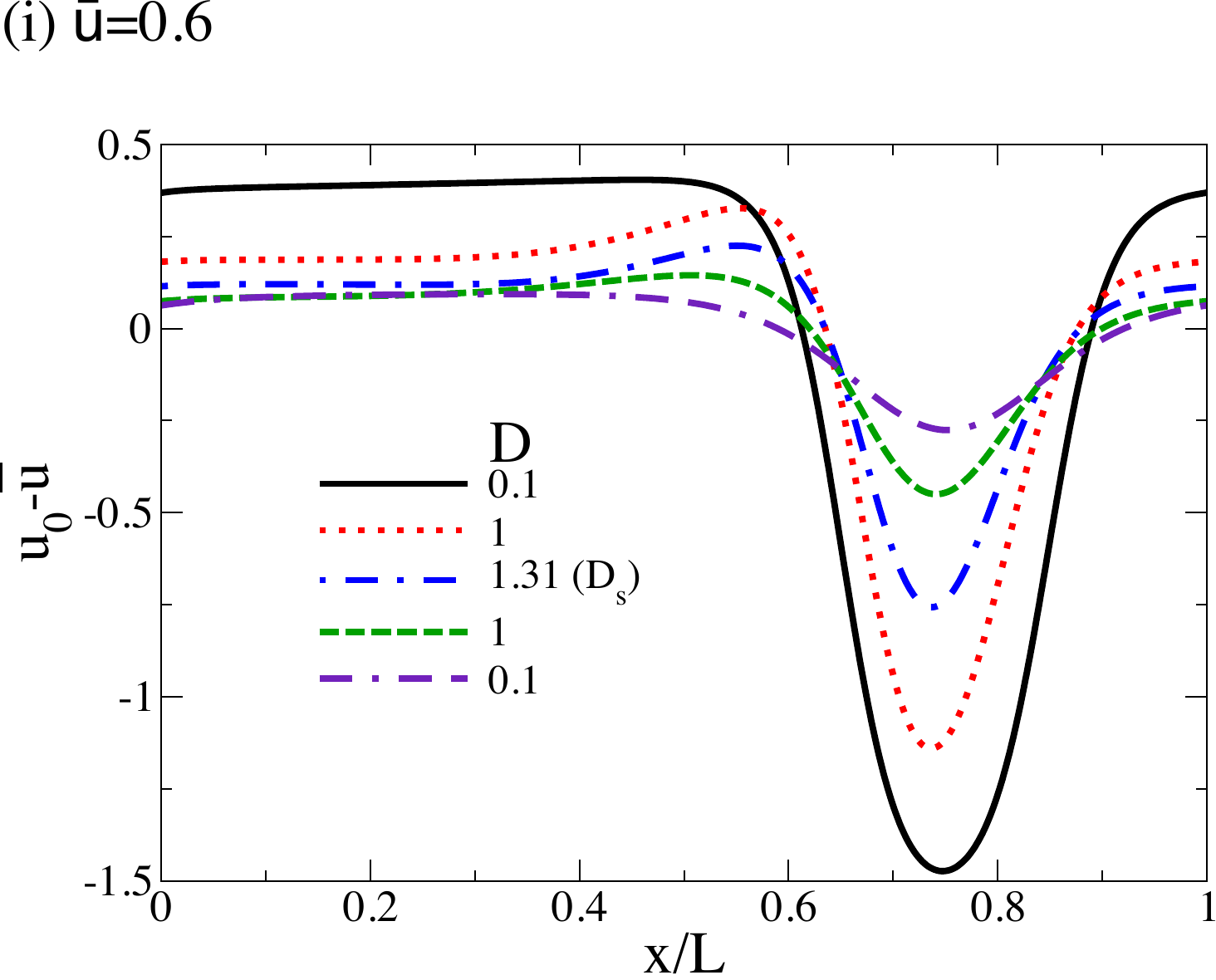} \\
 \vspace{-0.3cm}
\caption{Shown are (a) the dependence of $\| \delta u_0 \|$ on $D$ for several values of $L$, (b) the dependence of $\| \delta u_0 \|$ on $L$ for several values of $D$, and (c) profiles at $L=25$ for several values of $D$ of one-period steady traveling-wave  solutions $u_0$ of the cCH equation (\ref{Eq.2}), when $\bar{u}=0.4$.  The respective results for $\bar{u}=0.55$ are given in panels (d), (e) and (f), and for $\bar{u}=0.6$ -- in panels (g), (h) and (i).
}
\label{fig:1.4}
\vspace{-0.3cm}
\end{figure}

The results for $\bar u=0.4$ showing the dependence of the norm $\| \delta u_0 \|$ on $D$ for several values of $L$, the dependence of the norm $\| \delta u_0 \|$ on $L$ for several values of $D$, and profiles at $L=25$ for several values of $D$ of one-period steady traveling-wave  solutions $u_0$ of the cCH equation (\ref{Eq.2}) are shown in Figs.~\ref{fig:1.4}(a), (b) and (c), respectively. The respective results for $\bar u=0.55$ are shown in  Figs.~\ref{fig:1.4}(d), (e) and (f), and for $\bar u=0.6$ -- in Figs.~\ref{fig:1.4}(g), (h) and (i). 

For $\bar u=0.4$, Fig.~\ref{fig:1.4}(a) shows that for all the considered values of $L$, the norm $\| \delta u_0 \|$ is a monotonically decreasing function of $D$. In Fig.~\ref{fig:1.4}(b), we can observe that for $\bar u=0.4$ all the branches of spatially nonuniform solutions (when $L$ is used as the control parameter) bifurcate supercritically from the homogeneous branch at $L=L_{c}$, consistent with the weakly nonlinear analysis discussed above. We can also observe that for small values of $D$, the norm increases monotonically and tends to a constant as $L$ increases. As $D$ increases, the norm becomes a nonmonotonic function of $L$ but still tends to a constant as $L$ increases (see, for example, the line for $D=0.8$). For even larger values of $D$ this behaviour changes -- the norm first monotonically increases, then it may undergo a few oscillations before monotonically decreasing. This is consistent with the fact that the one-drop or multi-drop solutions that are typical of the standard CH equation do not exist for $D>\bar D\equiv\sqrt{2}$, as discussed at the end of Sect.~\ref{sec:single_double}. Instead, we obtain solutions of a different nature, namely, localized traveling-wave solutions, whose width remains almost unaffected by the increasing domain size, and whose norm, therefore, tends to zero according to the law $1/\sqrt{L}$ as $L$ increases (this has been verified numerically). In Fig.~\ref{fig:1.4}(c), we can see that for smaller values of $D$, the solution profile has a drop shape. As $D$ increases, the solution becomes flatter and the drop is deformed, namely, a ridge develops at the right-hand side of the drop. For larger values of $D$, the ridge first becomes more pronounced and then decreases in amplitude. Further, there appear additional visible oscillations in the profile that decay upstream.  The appearance of such oscillations can be understood through the spatial linear stability analysis. Also, it can be observed that for any value of $D$, the width of the drop in the rescaled coordinate $x/L$ increases as $D$ increases and the cavity narrows down. In fact, as discussed above, proper drop solutions exist only for $D<\sqrt{2}$, and the solution profiles for $D>\sqrt{2}$ should rather be classified as localized anti-pulse or hollow (or as pulse or hump solutions for negative values of $\bar u$) than as drop solutions.

For $\bar u= 0.55$, using $D$ as the control parameter, we can see in Fig.~\ref{fig:1.4}(d) that for $L<L_c$ the branches start at $D=0$, then have saddle-node bifurcations at some positive values of $D$, and then return to $D=0$. As $L$ increases, the saddle-node bifurcation shifts to the left. For $L= L_{c}$, the branch starts at $D=0$, then has one saddle-node bifurcation at a positive value of $D$. However, it does not go back to $D=0$. Instead, the branch terminates at the horizontal axis, where $\| \delta u_0 \|=0$, at some positive value of the driving force, $D=D_c \approx 1.3064$. For $L> L_{c}$, the branches start at $D=0$, but are characterized by two saddle-node bifurcations. After the second saddle-node bifurcation, the branch continues to infinity. For sufficiently large $L$, both saddle-node bifurcations annihilate each other, as is below discussed in more detail. In fact, the value $D_c$ is precisely the value at which the primary bifurcation changes from subcritical to supercritical when the domain size $L$ is used as the control parameter, as given by equation (\ref{eq:Dc}). In Fig.~\ref{fig:1.4}(e), when $L$ is used as the control parameter, we can observe that for $\bar u=0.55$ the primary bifurcation is indeed subcritical for $D < D_c$ while it is supercritical otherwise, in agreement with the weakly nonlinear analysis. When $D < D_c$, there is only one saddle-node bifurcation. On the other hand, when $D > D_c$, there are two saddle-node bifurcation -- the branch bifurcates supercritically from the uniform solution, then turns back at the first saddle-node bifurcation, and then turns again at the second saddle-node bifurcation and goes off to infinity. This is consistent with the results presented in Fig.~\ref{fig:1.4}(d), which show that for moderately large values of $L>L_c$ there exist three different solutions for a certain range of the driving force $D$. In Fig.~\ref{fig:1.4}(f), we can see that for $L=25$ and $\bar u=0.55$ there exist three different solutions at the same values of $D$ between the two saddle-node bifurcations that occur at $D_{s1}\approx 1.96$ and $D_{s2}\approx 1.62$.  For $D=1.75$ the solutions with larger and smaller amplitudes belong to the respective upper and the lower parts of the branch shown in Fig.~\ref{fig:1.4}(e) and are stable, whereas the solution with the intermediate value of the amplitude belongs to the middle part of the branch and is unstable. As $D$ increases further, we can see that the solution becomes flatter, and the ridge that was pronounced for smaller values of $D$ decreases in amplitude. We also remind here that the solution profiles that we observe for $D>\sqrt{2}$ and sufficiently large $L$ should be classified rather as anti-pulse or hollow solutions than drop solutions.

For $\bar u= 0.6$, using $D$ as the control parameter, we can see in Fig.~\ref{fig:1.4}(g) that for all the considered values of $L$, the branches start at $D=0$ then have one saddle-node bifurcation at some positive values of $D$ and return to $D=0$.  In Fig.~\ref{fig:1.4}(h), when $L$ is used as the control parameter, we can observe that for $\bar u=0.6$ there are no primary bifurcations for all the values of $D$, and we always find a saddle-node bifurcation. For smaller values of $D$, the upper parts of the branches monotonically increase as $L$ increases, whereas for larger value of $D$, the upper parts of the branches first monotonically increase and then monotonically decrease. In Fig.~\ref{fig:1.4}(i), we can see that when $\bar u=0.6$ and $L=25$ there are two different solutions for $D<D_s\approx 1.31$.  In particular, for $D= 0.1$ and  $1$ the solutions with larger amplitudes belong to the upper part of the branch for $L=25$ shown in Fig.~\ref{fig:1.4}(g) (these solutions are stable), whereas solutions with smaller amplitudes belong to the lower part of this branch (these solutions are unstable).

\begin{figure}
 \centering
  \includegraphics [width=7cm]{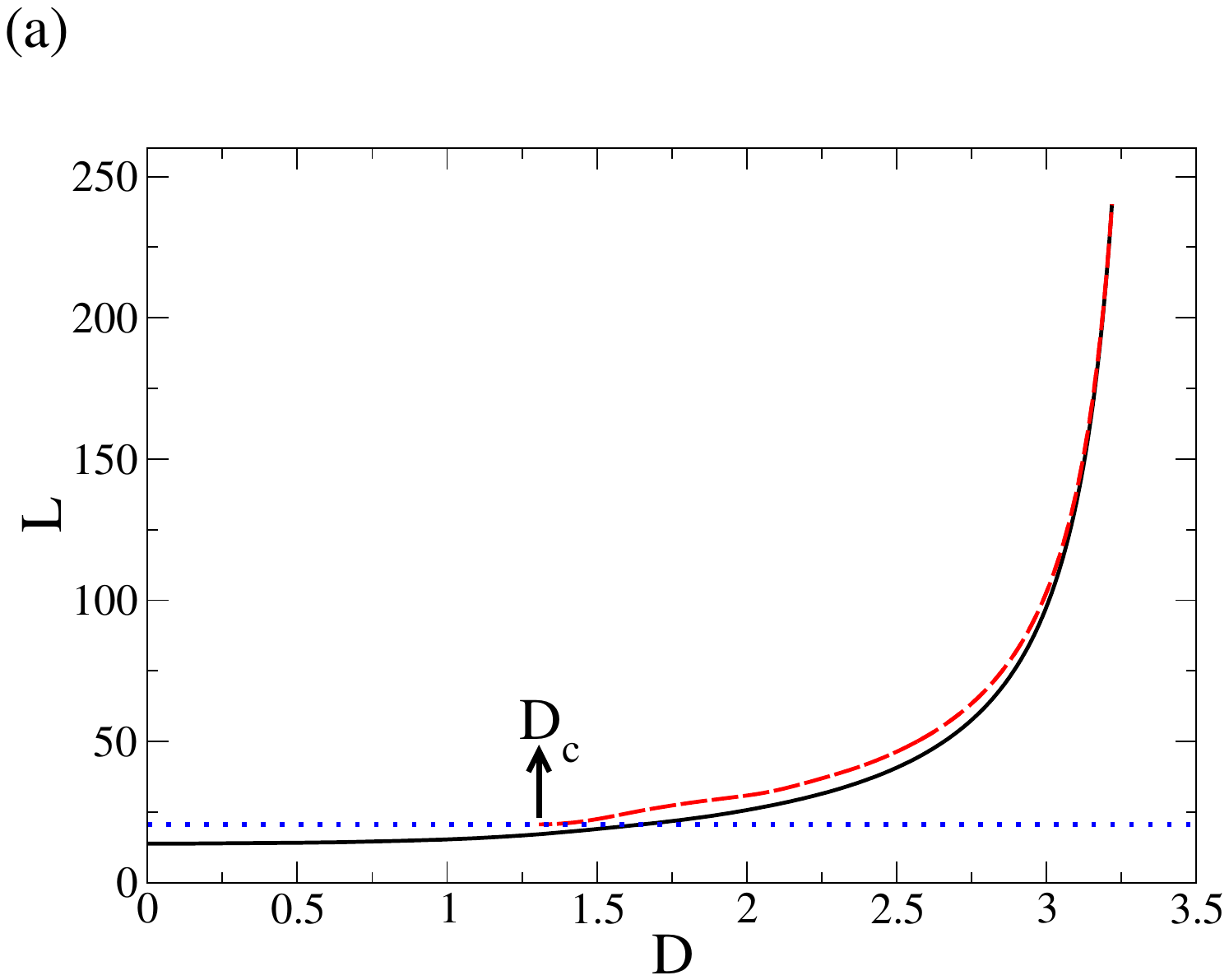}
  \includegraphics [width=7cm]{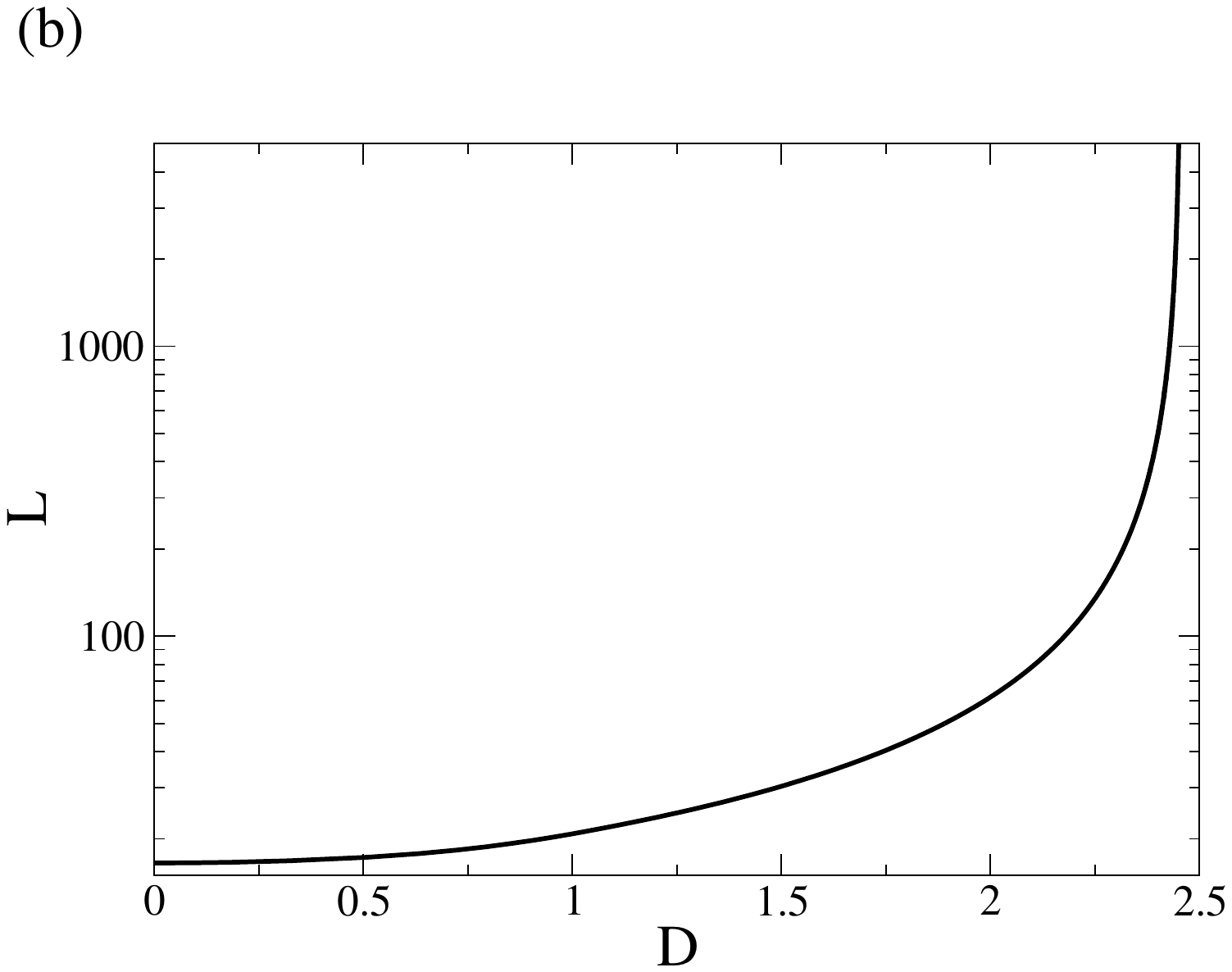}
  \vspace{-0.3cm}
   \caption[The loci of saddle-node bifurcations on the one-drop solution branches in the $(D, L)$-plane for $\bar u=0.55$ and $0.6$]{The loci of saddle-node bifurcations on the one-drop ($n = 1$) solution branches in the $(D, L)$-plane at (a) $\bar u=0.55$ and (b) $\bar u=0.6$. The horizontal dotted line in panel (a) indicates the cutoff period $L_{c} = 2 \pi /k_{c}$ for the linear stability of the uniform solution $\bar u=0.55$.
}
 \label{fig:1.20}
 \vspace{-0.3cm}
\end{figure}

From Fig.~\ref{fig:1.4}(d), it is difficult to infer where exactly the saddle-nodes appear. To understand this process better, we follow in Fig.~\ref{fig:1.20}(a) the loci of saddle-node bifurcations for $\bar u=0.55$ in the $(D,L)$-plane. The horizontal dotted line indicates the cutoff period $L_{c} = 2 \pi /k_{c}$ for the linear stability of the uniform solution $\bar u=0.55$. We see that for $L<L_c$ there is only one saddle-node bifurcation. 
On the other hand,  for $L>L_c$, there are two saddle-node bifurcations. 
For sufficiently large $L$, the two saddle-node bifurcations annihilate each other. Figure~\ref{fig:1.20}(b) shows the loci of the saddle-node bifurcations for $\bar u=0.6$ in the $(D, L)$ plane. We see that for all the values of $L \geq L_{sn}$, where $L_{sn}$ is the locus of the saddle-node bifurcation at $D=0$ (cf. Fig.~\ref{fig:1.1}(g)--(i)), there is one saddle-node bifurcation. 

\section{Linear stability, coarsening and time-periodic behaviour of two-drop solutions}
\label{Sect5}

{In this section, we construct detailed bifurcation diagrams of one- and two-drop solutions of the standard CH and cCH equations and study in detail linear stability properties and coarsening behavior of such solutions. We note that formerly coarsening dynamics of the cCH equation was analysed by Watson et al.\ \cite{Ref.3} (for $D\ll 1$) and Podolny et al.\ \cite{PODOLNY2005291} (for any $D< \sqrt{2}/3$) who derived a nearest-neighbour interaction theory for phase boundaries (kinks and anti-kinks) and revealed an important role of kink triplets in the coarsening process. Namely, they showed that due to mass conservation binary coalescence of phase boundaries is not possible. However, when an anti-kink is located between two kinks, it attracts them leading to simultaneous annihilation of the triplet and formation of a single kink. Note that Watson et al.\ \cite{Ref.3} and Podolny et al.\ \cite{PODOLNY2005291} considered the cCH equation in the form where the sign in front of the convective term is flipped. Thus, due to the symmetry $(D,u)\rightarrow (-D,-u)$, this implies for our case annihilation of a triplet where a kink is located between two anti-kinks resulting in a single anti-kink. In our study, we consider a periodic systems, i.e., in the simplest coarsening process a two-period or symmetric two-drop solution transforms into a one-period or one-drop solution. 
We take a computational approach with the aim to construct detailed stability diagrams in the parameter planes and to analyse transitions in the behaviour of the solutions not only for $D<\sqrt{2}/3$ but also for larger values of $D$. 
}

Assuming that $u_0$ is a steady solution of (\ref{Eq.2mov}) (i.e., a steady traveling-wave solution of (\ref{Eq.2})) and that $\tilde u$ is a small perturbation, we obtain the following linearized problem for $\tilde u$:
\begin{equation} \label {Eq.30}
\tilde u_{t}= \mathcal{L} [\tilde u],
\end{equation} 
where $\mathcal{L}$ is the following linear differential operator with nonconstant coefficients: 
\begin{equation}
\mathcal{L} [f] = [(v-Du_0)f-([1-3u_0^2]f+f_{xx})_x]_x.
\label {Eq.31}
\end{equation} 
The stability of $u_0$ then depends on the spectrum of $\mathcal{L}$, which typically consists of isolated eigenvalues of finite multiplicity, if $\mathcal{L}$ is defined on a finite periodic domain. Numerically, the eigenvalues can be computed directly using, e.g., a Fourier spectral method, or via numerical continuation, e.g., utilizing the continuation and bifurcation software Auto07p \cite{Ref.40}.  In addition to analysing steady traveling-wave solutions, we also construct branches of solutions that are time-periodic in a moving frame (modulated traveling waves, here referred to as time-periodic branches). Such solutions are also known as relative period orbits, see, e.g., \cite{Duguet_etal_2008}. We construct such branches using the procedure described in \cite{Lin_etal_2018}. This allows us to obtain a more complete understanding of the various transitions in the solutions.

\subsection{The case of the standard CH equation}
\label{LinStab_standardCH}

\begin{figure}
\centering
\includegraphics [width=6.0cm]{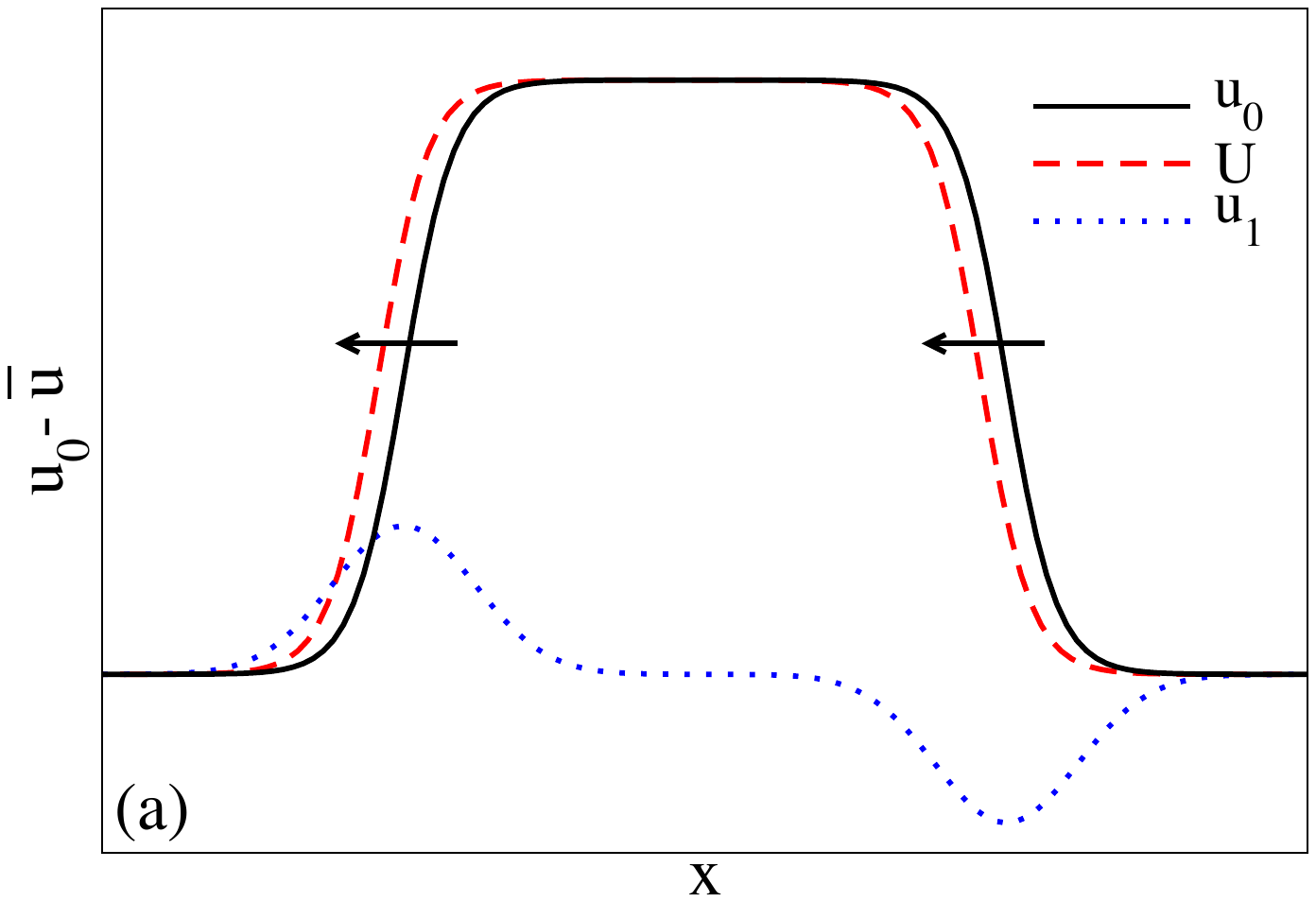} \hspace{-1.2cm}
\includegraphics [width=6.0cm]{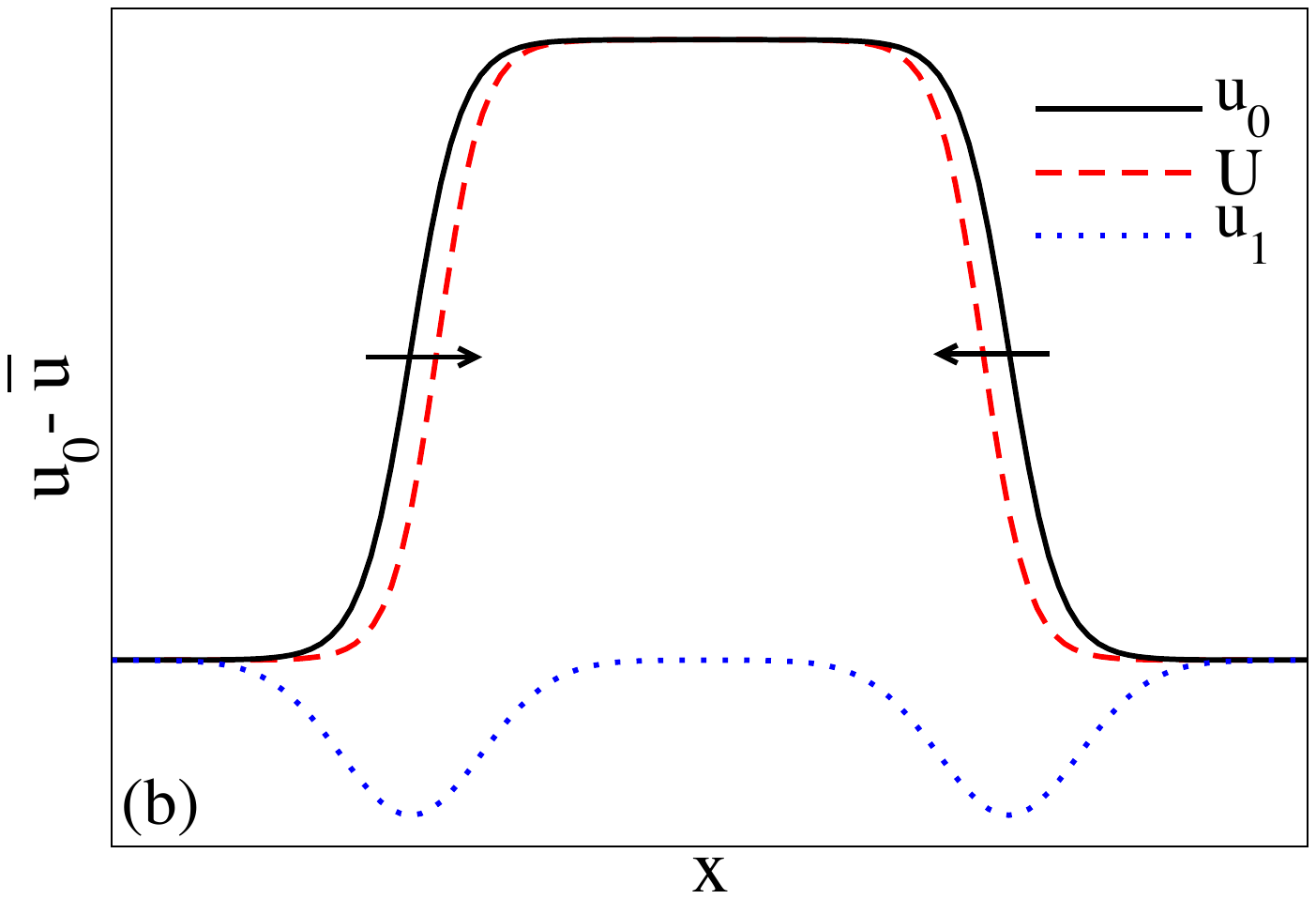}\\[-0.7cm]
\includegraphics [width=6.0cm]{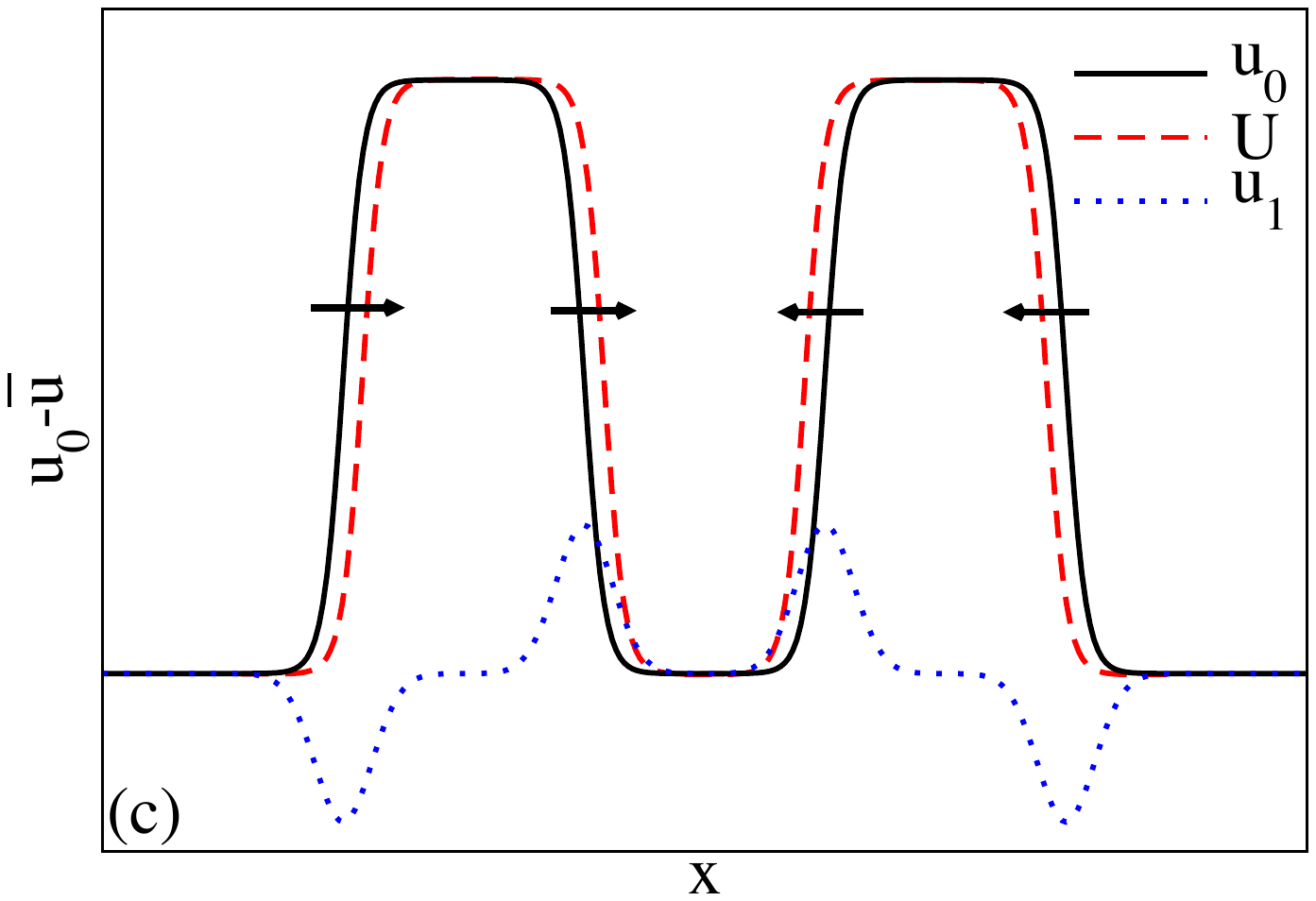} \hspace{-1.2cm}
\includegraphics [width=6.0cm]{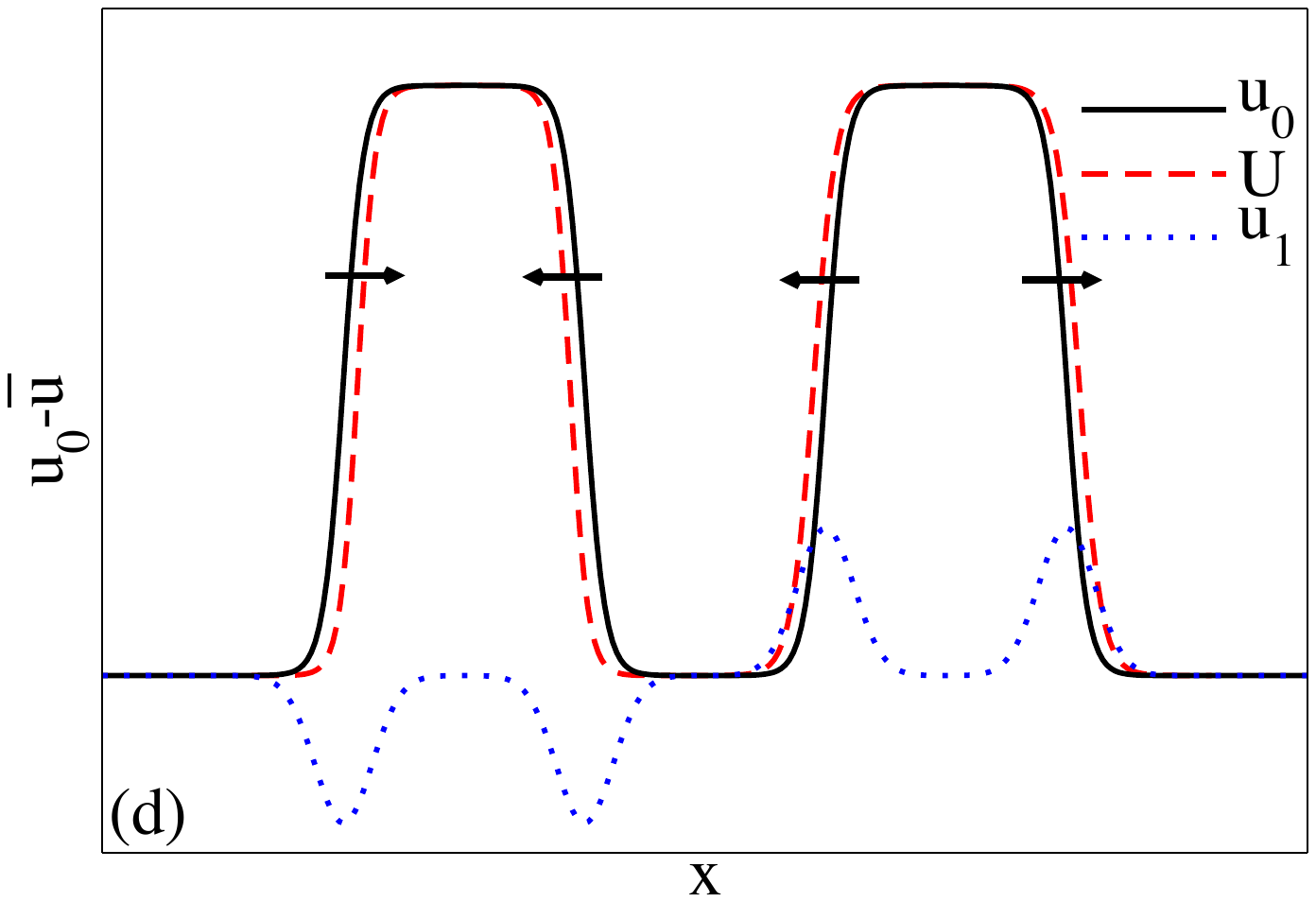}\\[-0.7cm]
   \caption{Shown are schematic representations of symmetry modes for (a,b) one-drop solutions and (c,d) two-drop solutions. Panels (a,c) represent the translation mode and panel (b,d) represent the volume mode. The solid lines correspond to the solutions $u_0(x)$. The dotted lines correspond to the eigenfunctions $u_1(x)$. The dashed lines correspond to the solution $u_0(x)$ superimposed with eigenfunction $u_1(x)$ multiplied by a small coefficient $\epsilon$, i.e., $U(x)=u_0(x)+\epsilon u_1(x)$. The arrows indicate the directions of shift for the various fronts when the eigenfunctions are added.}
 \label{fig:c00}   
 \vspace{-0.3cm}
\end{figure}

First, we note that branches of two-period solutions for the standard CH equation can be obtained  from the branches of one-period solutions discussed in Sect.~\ref{Sect:single_droplets_CH} by considering the identical periodic solution in a domain twice as large as the period. Our calculations show that for the standard CH equation the resulting two-drop branch has no side branches. Therefore, there is actually no need to recompute the primary branches. However, one still needs to individually analyze the linear stability as it may change when going from the one-drop to the two-drop states.  First, we note that zero is always an eigenvalue of the linearized problem with the eigenfunction given by $u_1(x)=u_0'(x)$, and it is associated with the translational invariance of the equation.  The emergence of the various coarsening mechanisms can then be explained by the following consideration (see Thiele et al.\ \cite{Ref.50,  Ref.66}). Each of the two-drop solutions can be considered as a superposition of four fronts (two kink and two anti-kink solutions). Each of these solutions, when considered individually, has a zero eigenvalue with the eigenfunction given by the derivative of the solution. 
When the fronts are superimposed, the corresponding eigenfunctions are also superimposed (with small corrections). For one drop, the superimposed eigenfunctions result in two qualitatively different cases: either both fronts are shifted in the same direction, which results in the overall translation of the drop, or the fronts are shifted in the opposite directions, which results in the decrease [increase] of the volume of the drop. For a single drop with imposed volume conservation (i.e., when the drop is considered on a periodic domain), this mode, of course, disappears as otherwise it would violate volume conservation. However, for a two-drop state on a periodic domain due to volume conservation the decrease [increase] of the volume of one drop implies the increase [decrease] of the volume of the other drop. Schematic representations  are shown in Figs.~\ref{fig:c00}(a) and (b).  For a pair of drops on a periodic domain, only the three (up to the positive or negative sign) possible combinations corresponding to the overall mass conservation should be considered.

One of these combinations results in the overall translation of both drops in the same direction, and it must correspond to the zero eigenvalue. The other two correspond to the two coarsening modes, namely, the translation mode and the volume mode, see schematic representations in Figs.~\ref{fig:c00}(c) and (d). The arrows in these figures indicate the directions in which the fronts are shifted when the eigenfunctions are added.

For the translation mode, the drops move towards each other, and for the volume mode the volume of one of the drops decreases while the volume of the other one increases accordingly. The eigenvalues for these modes correspond to the perturbed zero eigenvalue. The larger the separation distances between the fronts are, the closer to zero these eigenvalues become. It is also interesting to note that the translation [volume] mode for a two-drop solution turns out to be the volume [translation] mode for the corresponding two holes. 

\begin{figure}
\includegraphics [width=5.15cm]{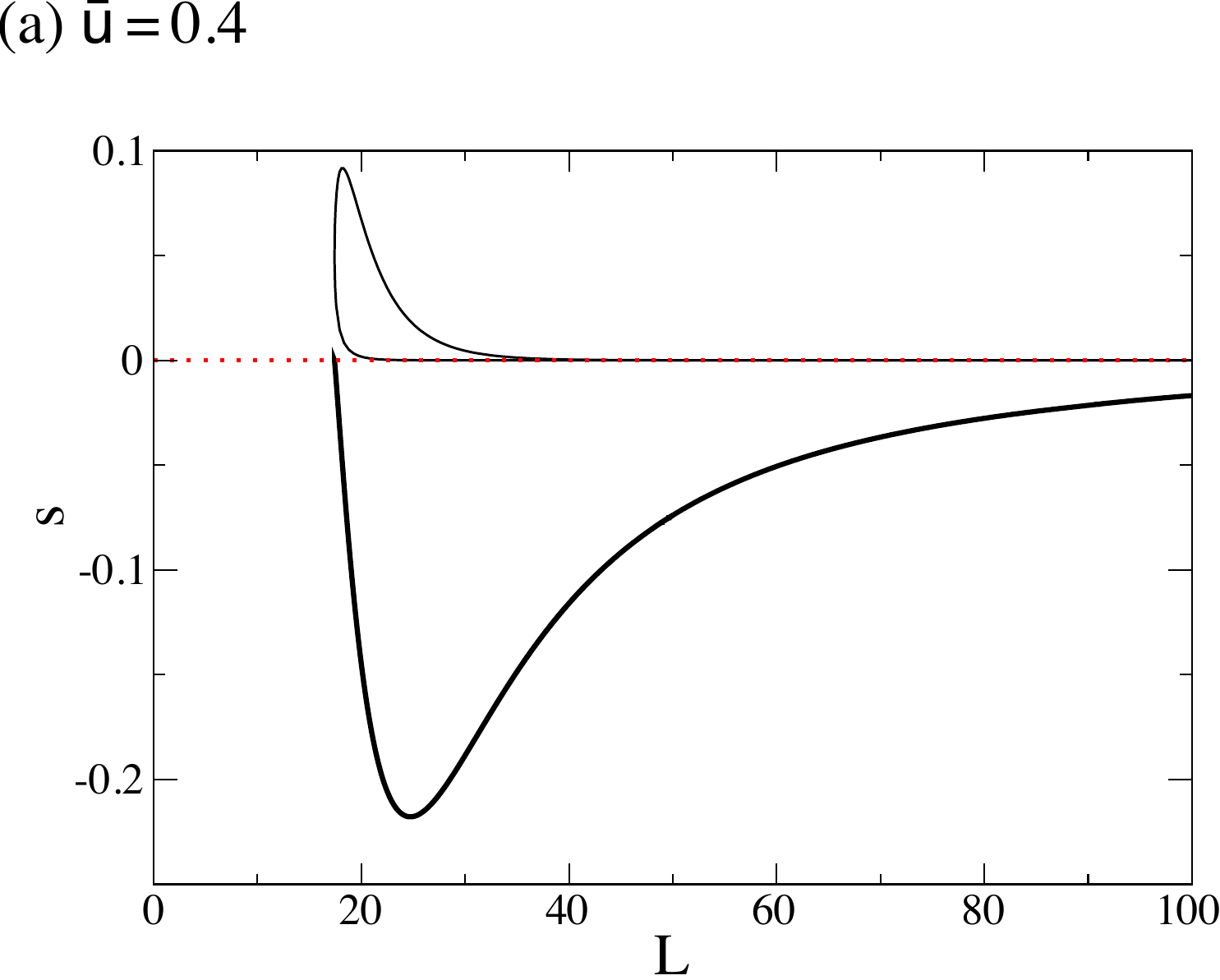}
\includegraphics [width=5.1cm]{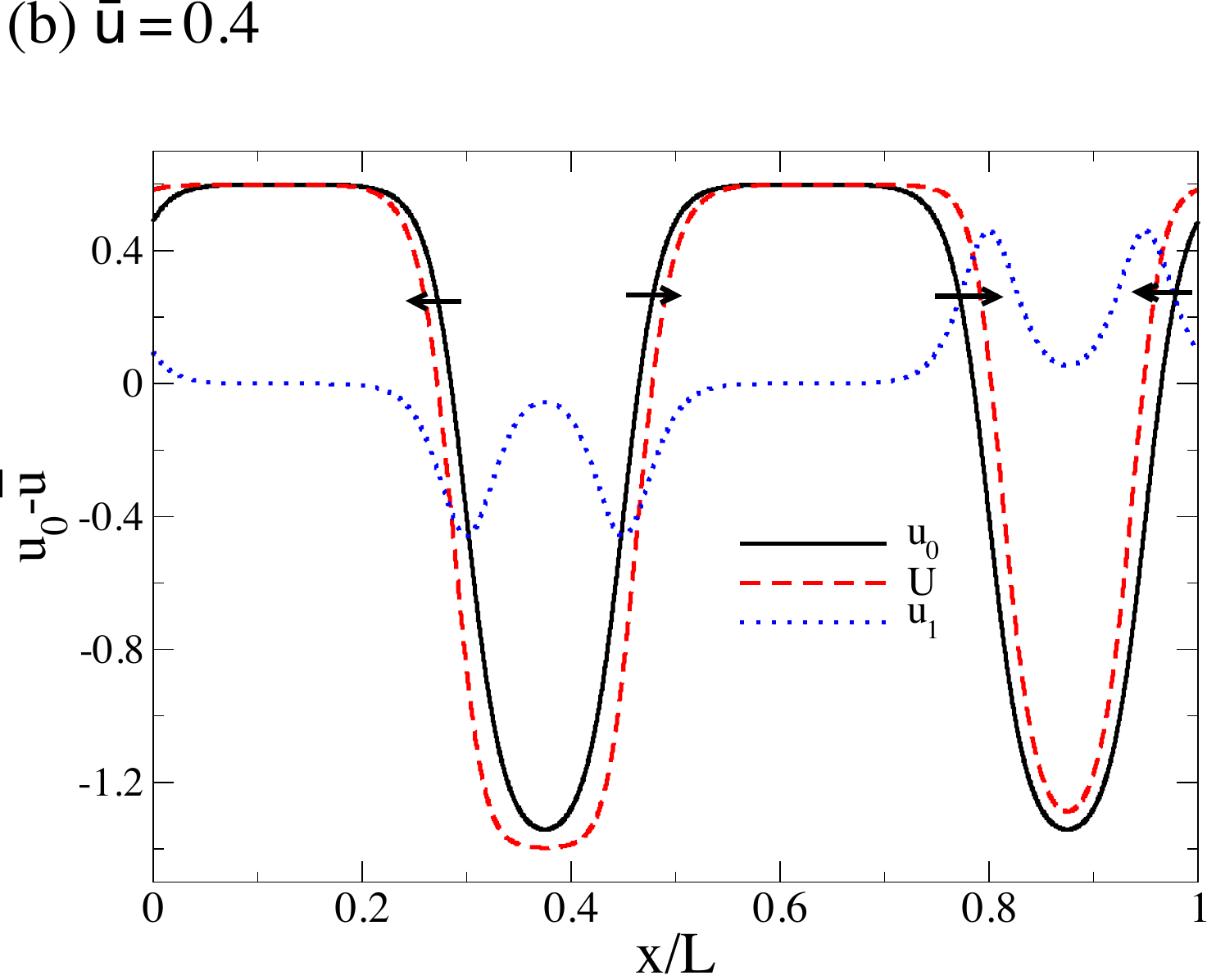}
\includegraphics [width=5.1cm]{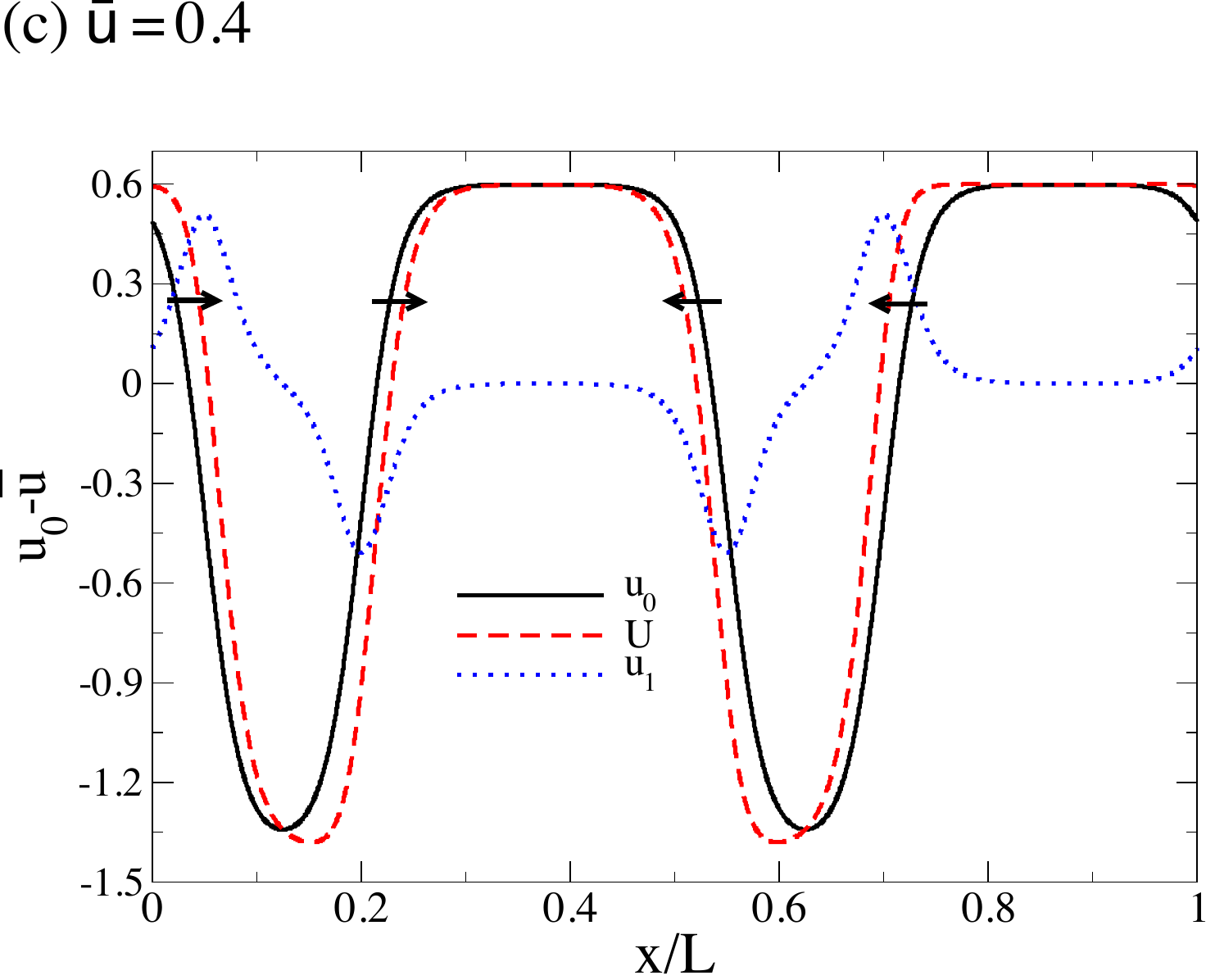}
\includegraphics [width=5.15cm]{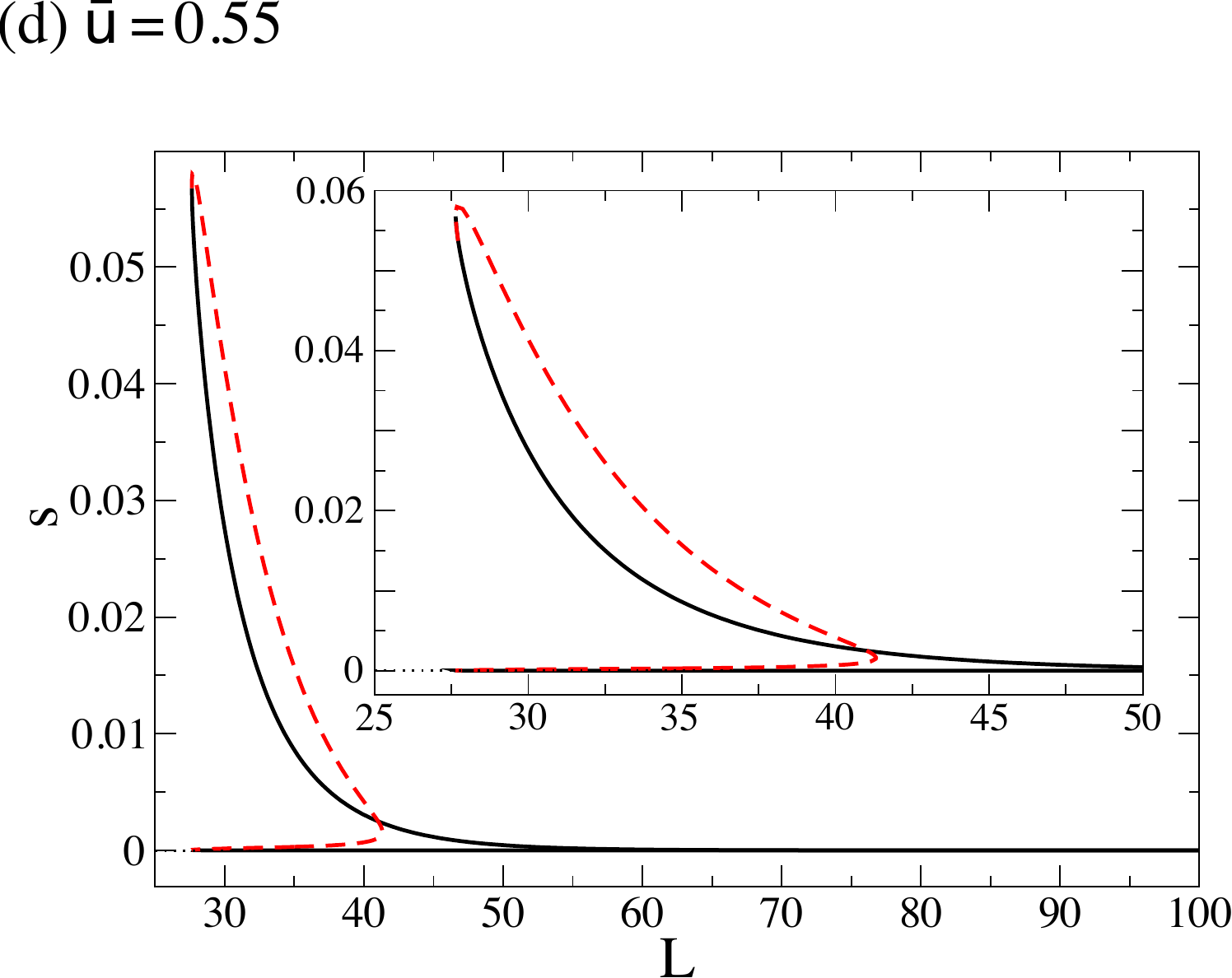}
\includegraphics [width=5.1cm]{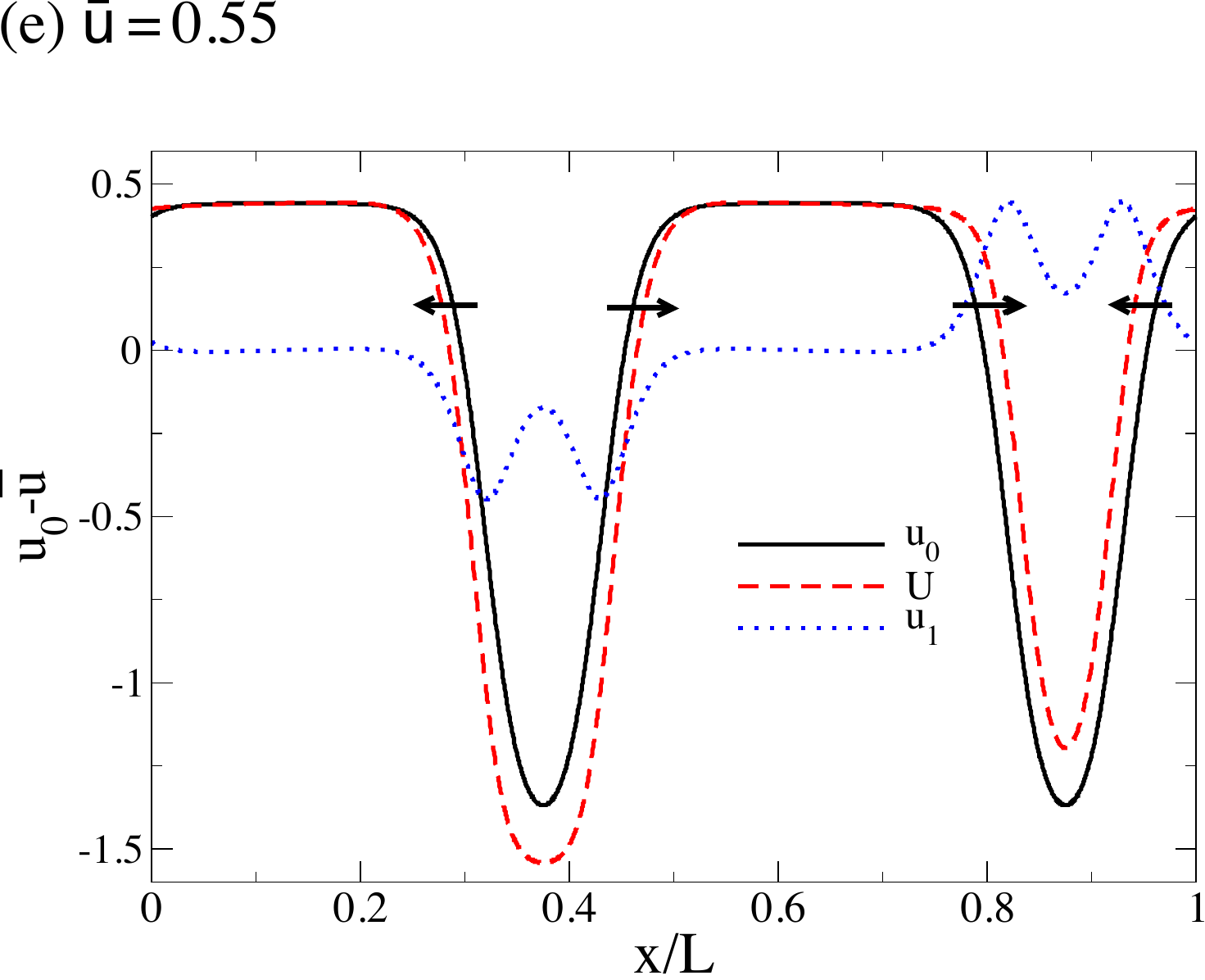}
\includegraphics [width=5.1cm]{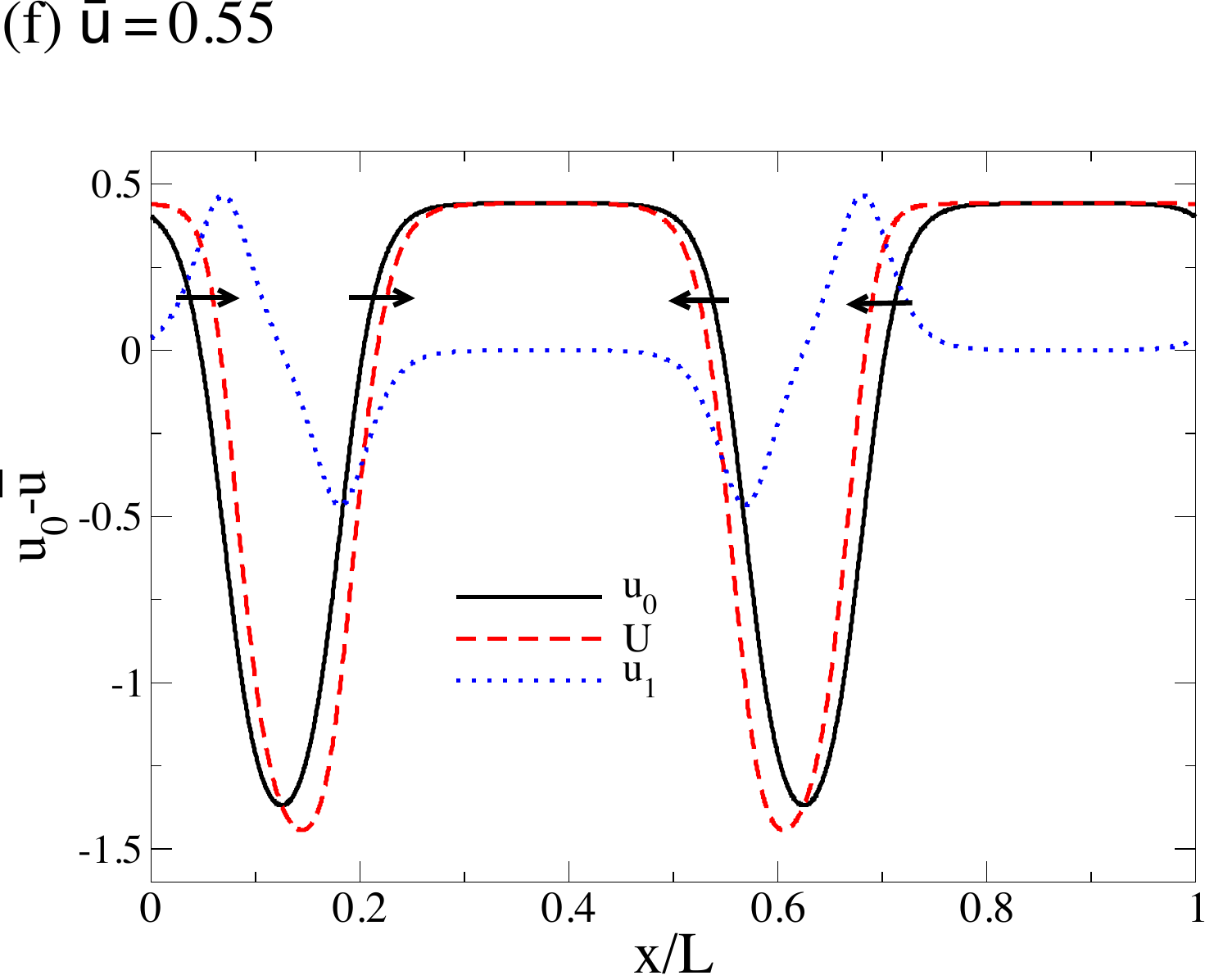}

\vspace{-0.3cm}
 \caption{Shown are (a) he dependence of the dominant eigenvalues $s$ of two-drop solutions of the standard CH equation on the domain size $L$, (b) the dominant translation coarsening mode, and (c) the volume coarsening mode for $\bar u=0.4$. The thin solid lines in panel (a) correspond to two positive eigenvalues, and the thick solid line shows the dominant negative eigenvalue. The results in panels (b) and (c) correspond to $L=40$, and the solid lines correspond to the solutions $u_0(x)$, the dotted lines -- to the eigenfunctions $u_1(x)$, the dashed lines -- to the solution $u_0(x)$ superimposed with eigenfunction $u_1(x)$ multiplied by a small coefficient $\epsilon$, i.e., $U(x)=u_0(x)+\epsilon u_1(x)$. The respective results for $\bar u =0.55$ are shown in panels (d), (e) and (f). Note that in panel (d) the black solid lines correspond to the upper branch of solutions, and the red dashed lines correspond to the lower branch of solutions that bifurcates subcritically from the homogenous solution.
} \label{fig:standardCH_eig}
\vspace{-0.3cm}
\end{figure}

The calculations confirm that for a two-period solution there are additionally two positive eigenvalues close to zero. The dependence of the dominant eigenvalues on $L$ and the dominant coarsening mode and the nondominant coarsening mode for $L=40$ are shown in the first row of Fig.~\ref{fig:standardCH_eig} for $\bar u=0.4$ (panels (a), (b) and (c), respectively) and in the second row for $\bar u=0.55$ (panels (d), (e) and (f), respectively). For $\bar u=0.4$, we can see in Fig.~\ref{fig:standardCH_eig}(a) that the two positive eigenvalues annihilate in a saddle-node bifurcation at the linear stability threshold for the homogeneous solution. For $\bar u=0.55$, we remind that the primary bifurcation is subcritical and there exists a range of $L$ values for which there exist two solutions, see Fig.~\ref{fig:1.1}(d). The one-drop solutions of smaller norm (lower brach) are linearly unstable even on the domain equal to the solution period. The one-drop solutions of larger norm (upper branch) are linearly stable when considered on the domain equal to the solution period, but become unstable to coarsening modes when two-period domains are considered. In Fig.~\ref{fig:standardCH_eig}(d), the black solid [red dashed] lines correspond to the eigenvalues of the solutions of the upper [lower] branch. The black solid lines in Figs.~\ref{fig:standardCH_eig}(b), (c), (e), and (f) show the two-drop solutions, $u_0$, the corresponding coarsening modes (eigenfunctions corresponding to the positive eigenvalues) are shown by the blue dotted lines, the red dashed lines show the two-drop solutions superimposed with the eigenfunctions, $U=u_0+\epsilon u_1$ for sufficiently small $\epsilon$, and the arrows indicate the directions in which the corresponding fronts shift. Panels (b) and (e) show that both for $\bar u=0.4$ and for $\bar u=0.55$, the dominant coarsening modes (corresponding to the largest eigenvalue) are the translation ones, while in panels (c) and (f) we can see that the other nondominant coarsening modes (corresponding to positive but not the largest eigenvalues) are the volume ones. This, in fact, holds for any positive value of $\bar u$.

Finally, let us point out that if $u_0$ is a two-period steady solution of the standard CH equation for a certain value of $\bar u$, then $-u_0$ is again a steady solution of the standard CH equation for the mean value equal to $-\bar u$. More interestingly, the eigenvalues and the eigenfunctions are exactly the same as for the mean value equal to $\bar u$, since it can be shown that the linearized operator does not change. For the mean value $-\bar u$, we, therefore, again obtain two coarsening modes (which are exactly the same as for the mean value $\bar u$). However, when the steady solutions are superimposed with the eigenfunctions, the roles of the coarsening modes are interchanged, namely, the dominant coarsening mode is now the volume one and the other one is now the translation mode. 

\subsection{The case of the cCH equation}

\subsubsection{Symmetry breaking}

First, we employ continuation to compute branches of two-period solutions in dependence of the driving force $D$ for several fixed values of  $L$ and $\bar u$. As for the standard CH equation, branches of two-period solutions can in fact be obtained  from the branches of one-period solutions (that were discussed in Sect.~\ref{sect:convCH_waves_0p4_0p55_0p6}) by considering domain sizes that are twice the solution period. We call the resulting solution branches two-drop primary branches. The symmetric two-drop states on such branches have the discrete internal translation symmetry. Solution branches bifurcating from these primary branches in secondary bifurcations we call secondary branches. Secondary pitchfork bifurcations break the discrete translation symmetry and, therefore, result in solutions with a larger spatial period. Hence, if such solutions are stable, the corresponding secondary bifurcations are associated with coarsening of the pattern. However, we emphasize here that at least for $D<\sqrt{2}$ for a two-drop solution given on a domain of certain length there exists a one-drop solution of the period equal to that domain length, and true coarsening would correspond to evolution towards such a one-drop solution. For completeness of the bifurcation diagrams, we also include the branches of one-drop states.

\begin{figure}
 \centering
\includegraphics [width=6cm]{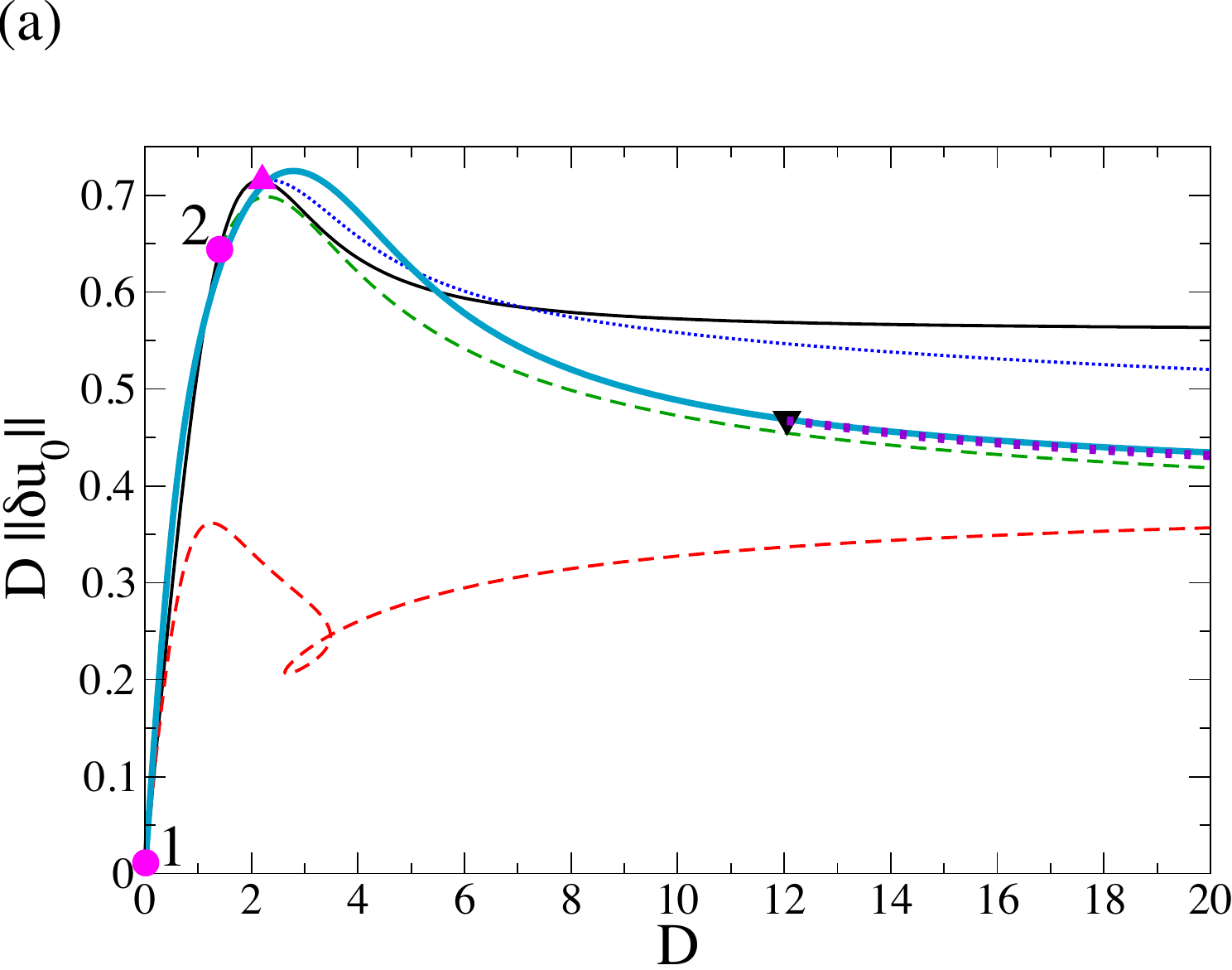}\hspace{0.5cm}
\includegraphics [width=6cm]{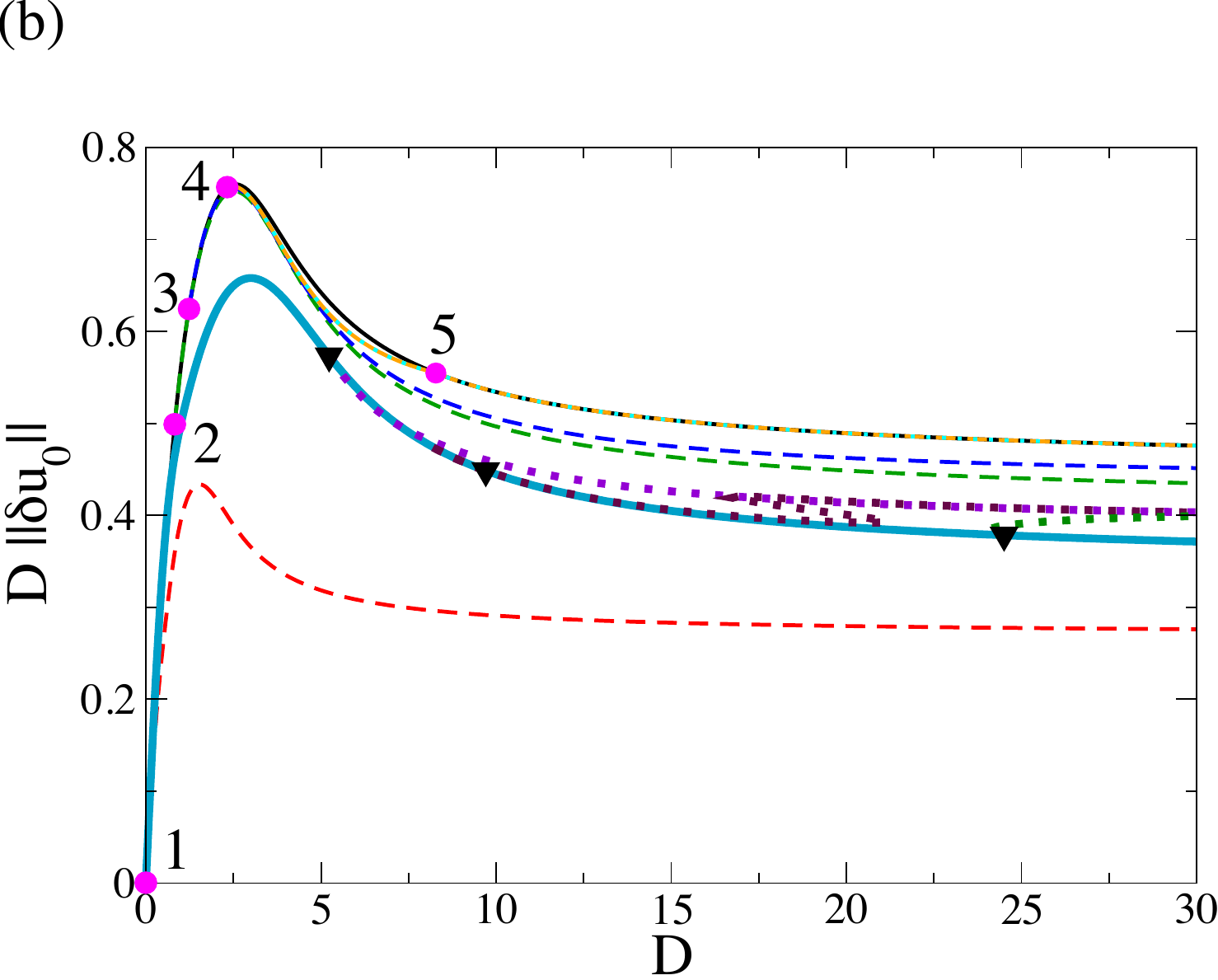}

\vspace{-0.3cm}
 \caption{Shown is the dependence of $D\| \delta u_0 \|$ on  $D$ for one- and two-drop solutions of the cCH equation (\ref{Eq.2}) when $\bar u=0.4$ and (a)~$L=25$ and (b)~$L=35$. The various line styles and markers correspond to the various solution types and bifurcation points, respectively, as explained in the text. 
}
\label{fig:A1.1}
\vspace{-0.3cm}
\end{figure}

\begin{figure}
\centering
\includegraphics [width=6cm]{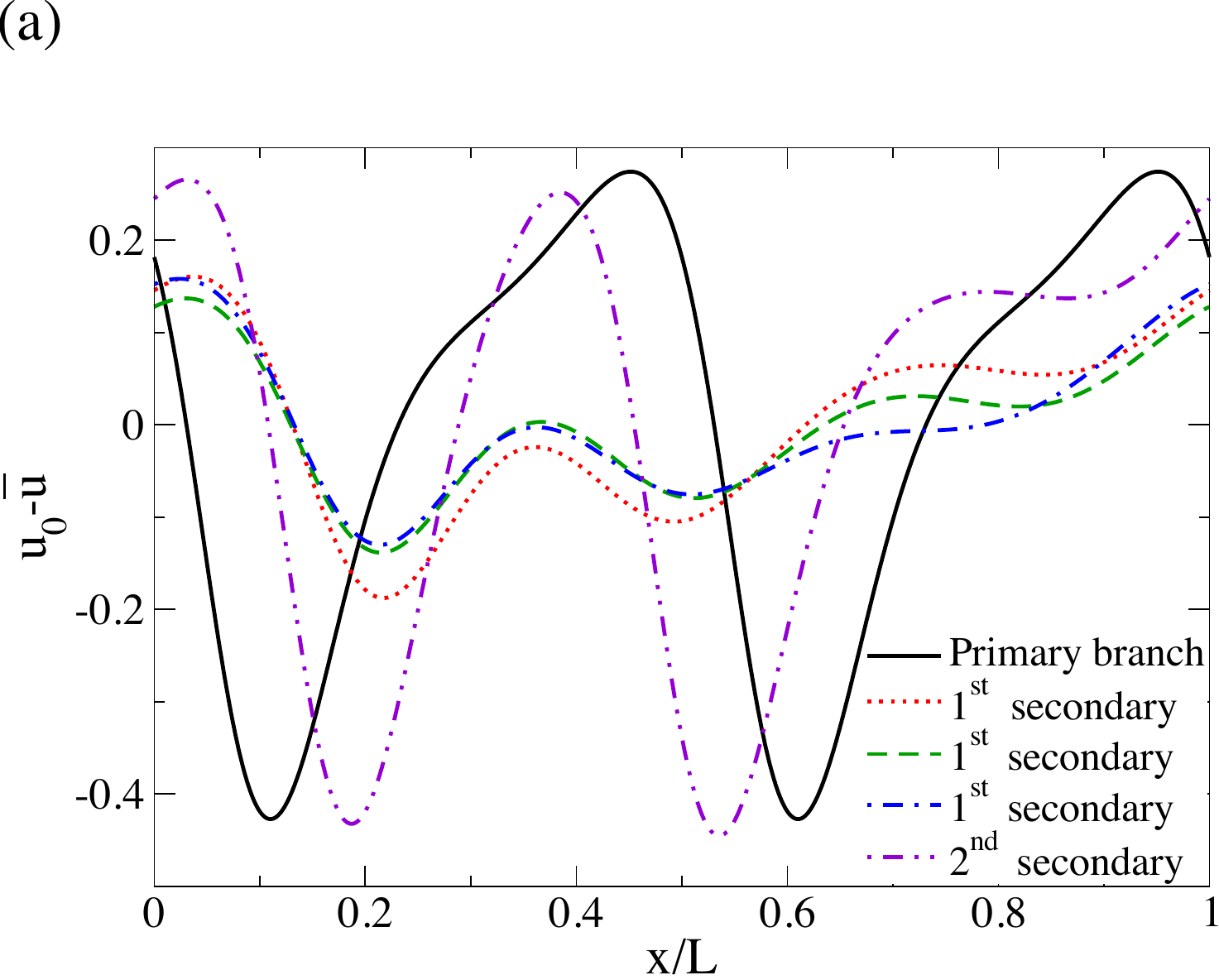}\hspace{0.5cm}
\includegraphics [width=6cm]{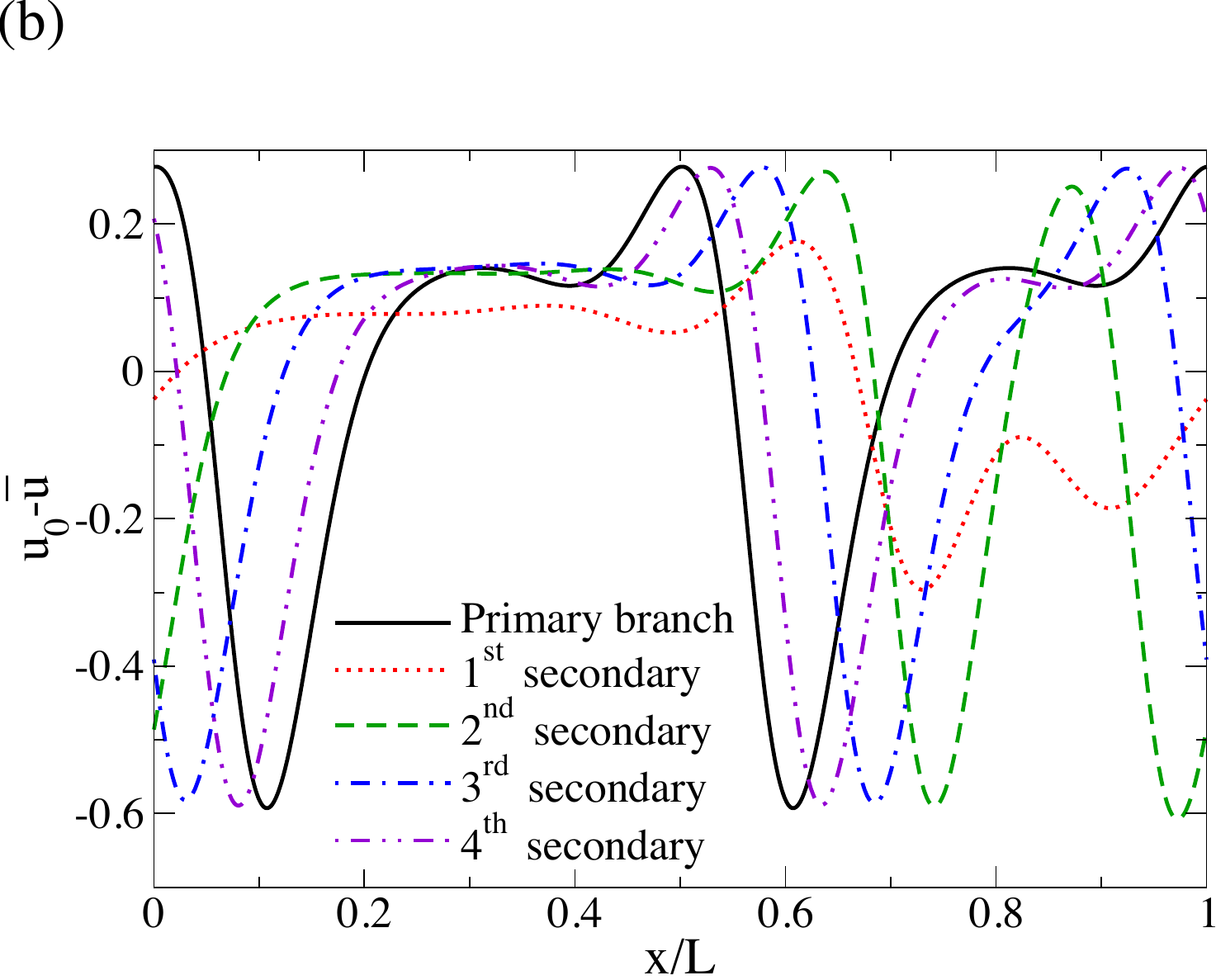} 

\vspace{-0.3cm}
   \caption{Solution profiles from the two-drop primary and secondary branches for $\bar u=0.4$ when $D=3$ and (a)~$L=25$ and (b)~$L=35$.}
 \label{fig:p1}  
%
\centering
\includegraphics [width=8cm]{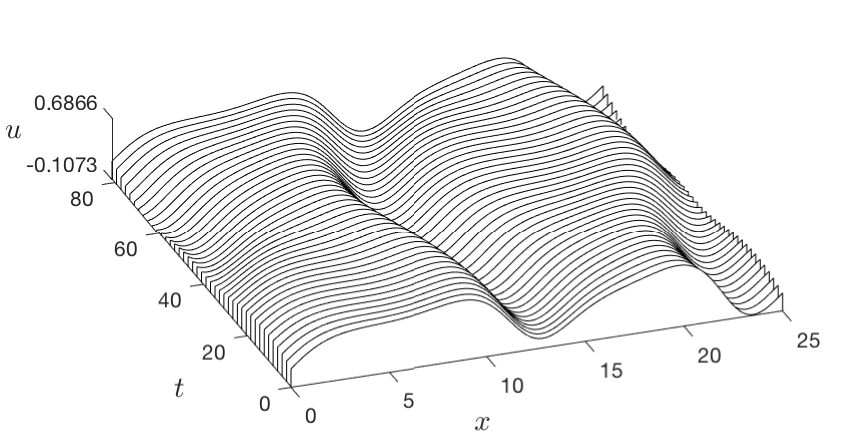} 

\vspace{-0.3cm}
\caption{Time evolution over one period of time of the time-periodic solution for $\bar u=0.4$, $L=25$ when $D=3$ (see Fig.~\ref{fig:A1.1}(b)).}
 \label{fig:tper_bu_0_4_L_25_LAB_14_D3}  
 \vspace{-0.3cm}
\end{figure}

\begin{figure}
\centering
\includegraphics [width=6cm]{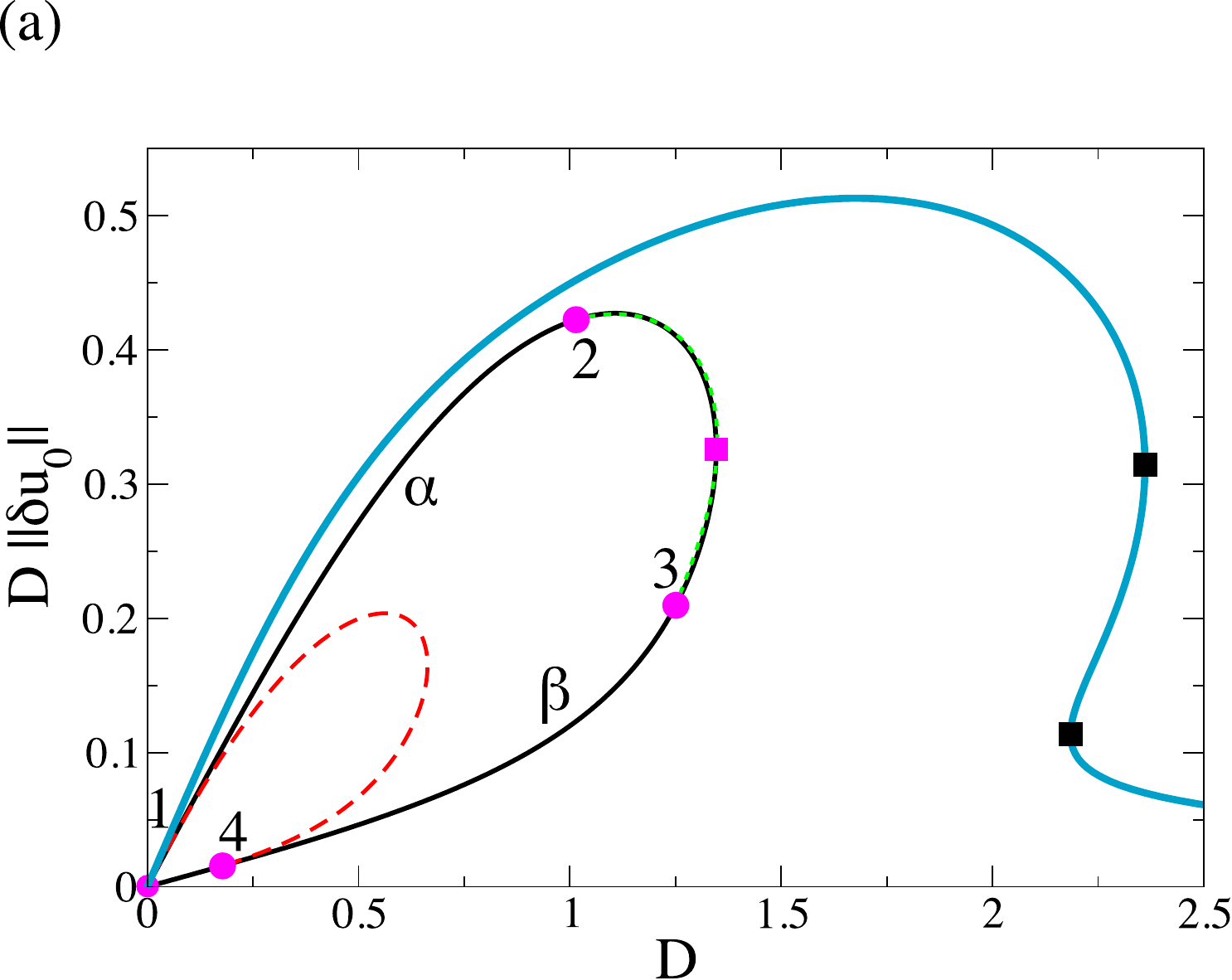}\hspace{0.5cm}
\includegraphics [width=6cm]{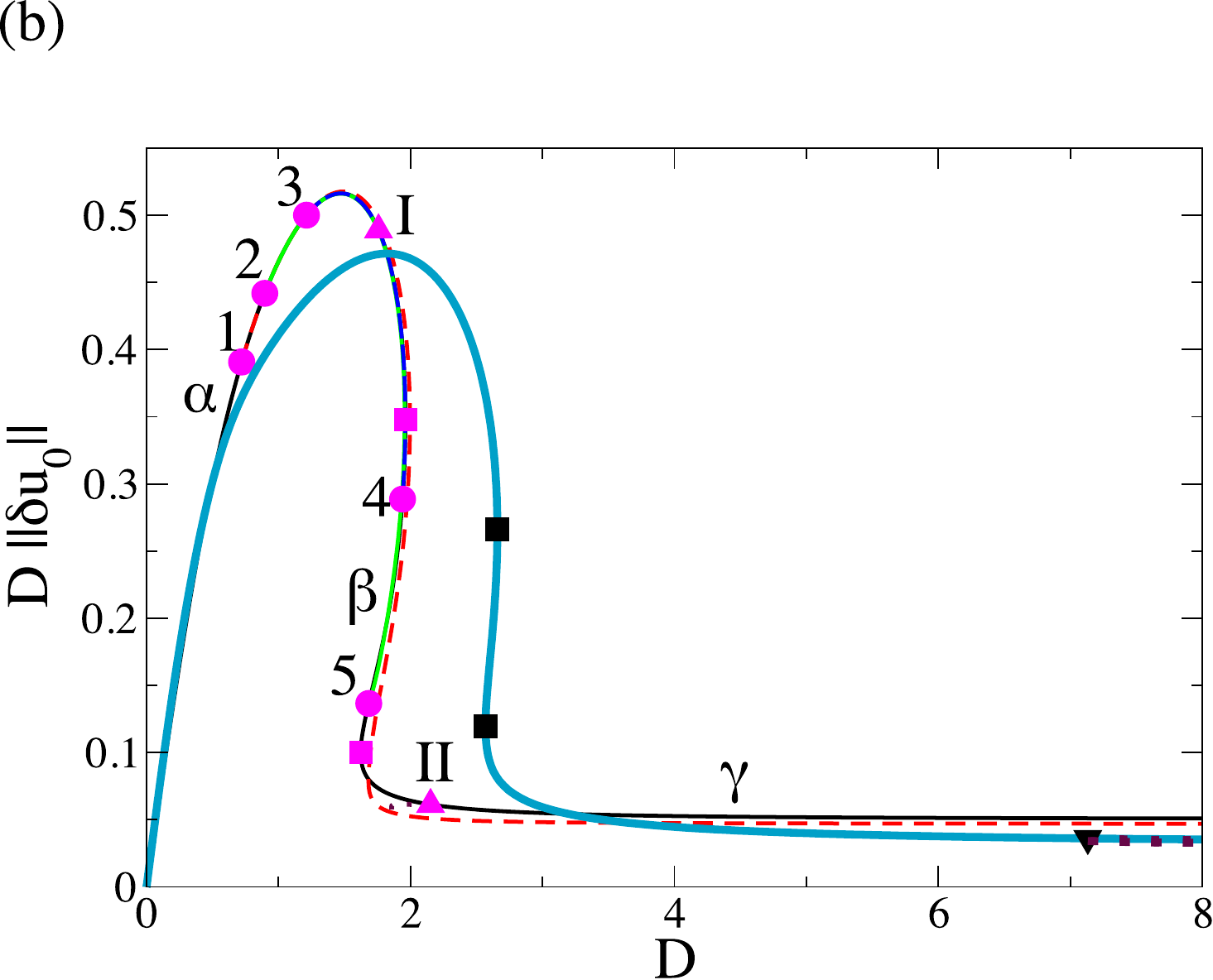}
\vspace{-0.5cm}
 \caption{Shown is the dependence of $D\| \delta u_0 \|$ on $D$ for one- and two-drop solutions of the cCH equation (\ref{Eq.2}) when $\bar u=0.55$ and (a)~$L=35$ and (b)~$L=50$. The various line styles and markers correspond to the various solution types and bifurcation points, respectively, as explained in the text. 
 In panel (a) [panel (b)], the upper and lower [upper, middle and lower] parts of the primary branch are denoted by letters $\alpha$ and $\beta$ [$\alpha$,  $\beta$ and $\gamma$], respectively. }
 \label{fig:A1.2}
 \vspace{0.3cm}
 
\includegraphics [width=6cm]{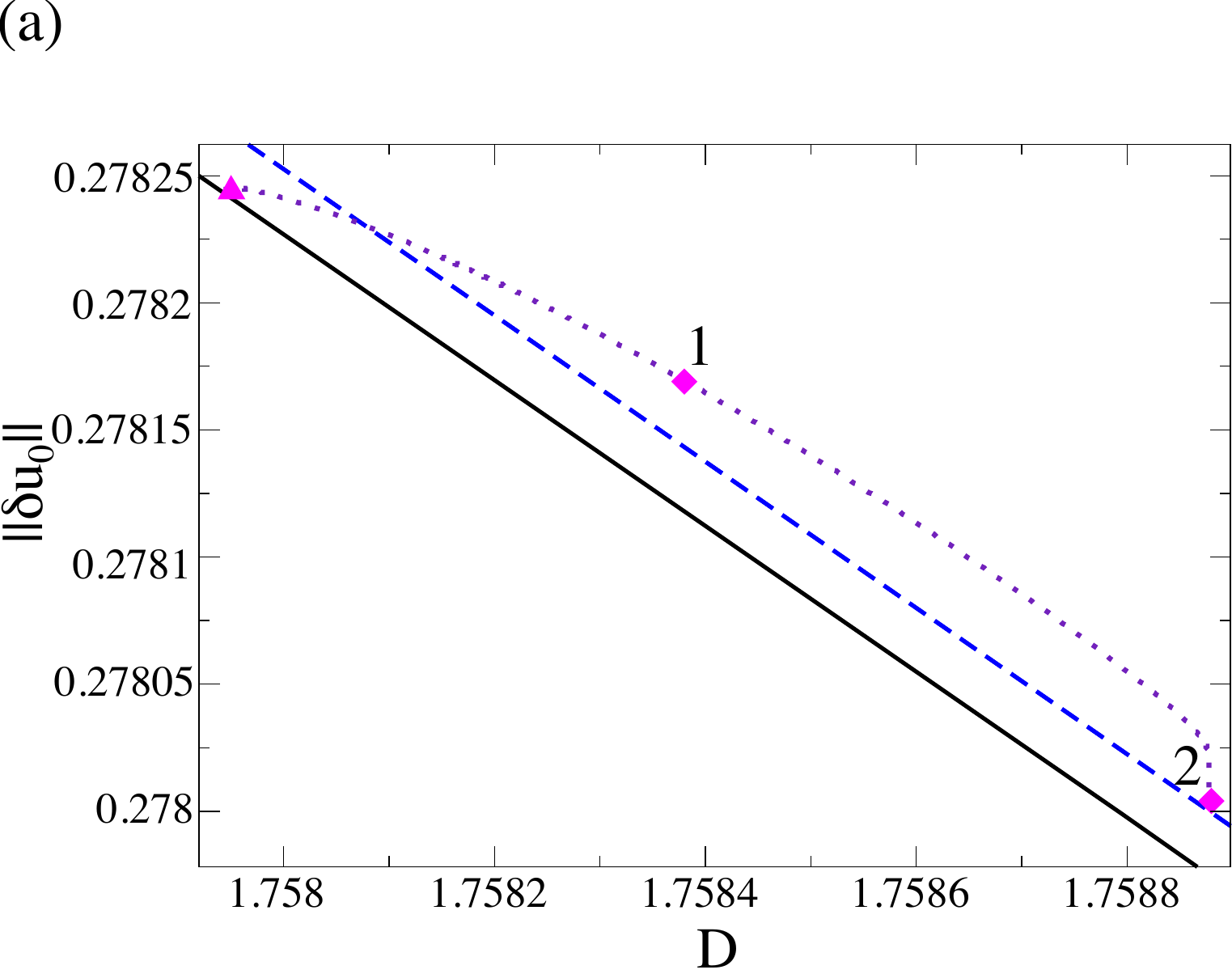}\hspace{0.5cm}
\includegraphics [width=6cm]{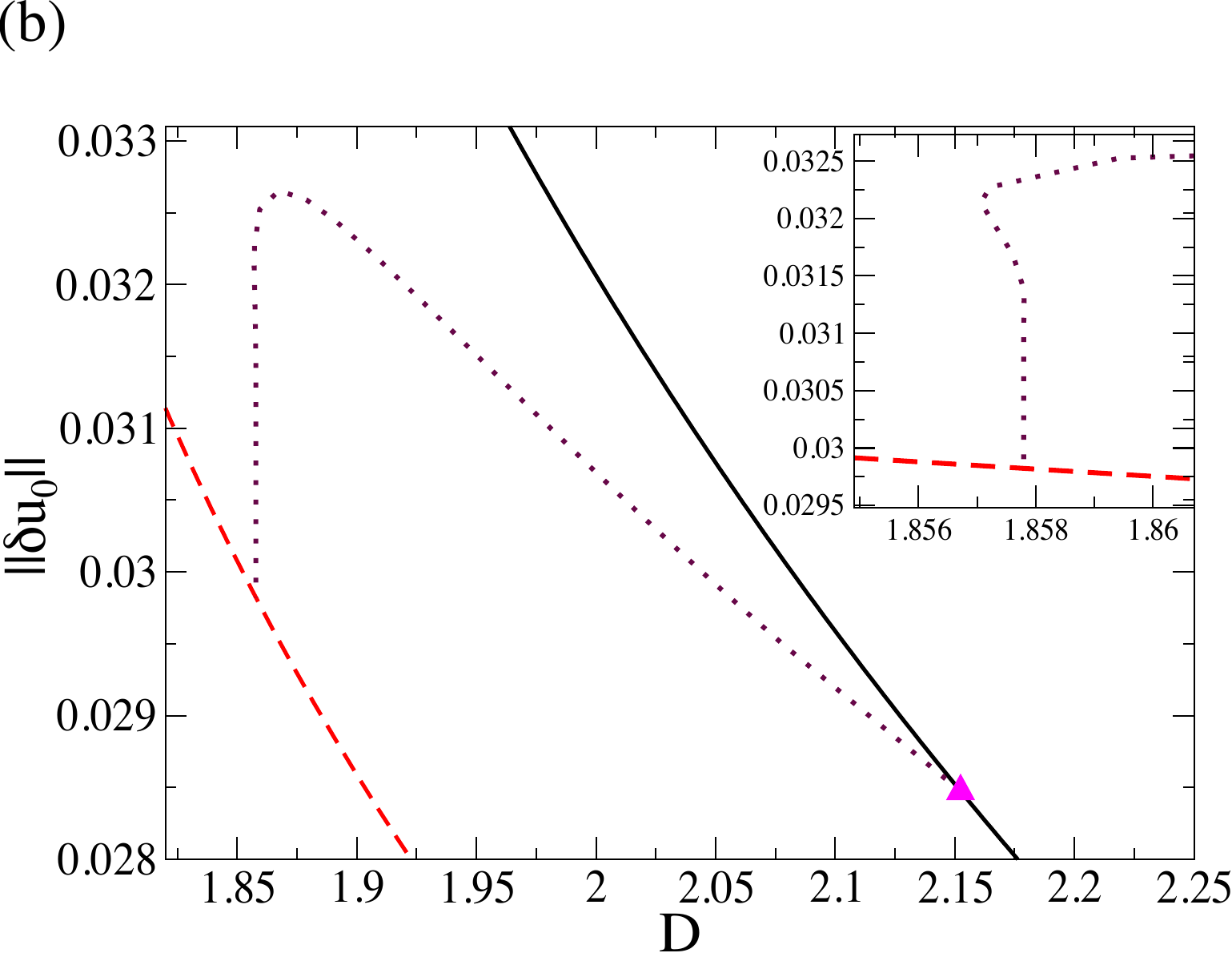}
\vspace{-0.5cm}
 \caption[Zooms of the time-periodic branches shown in Fig.~\ref{fig:A1.2}]{Zooms of the time-periodic branches shown in Fig.~\ref{fig:A1.2}(b) and starting from points I and II (panels (a) and (b), respectively). 
 The red diamonds 1 and 2 in panel (a) correspond to time-periodic solutions shown in Figs.~\ref{fig:tper_bu_0_55_L_50_I_LAB_23}(a) and (b), respectively. }
 \label{fig:zooms}
%

 \vspace{0.3cm}
\centering
\includegraphics [width=7.2cm]{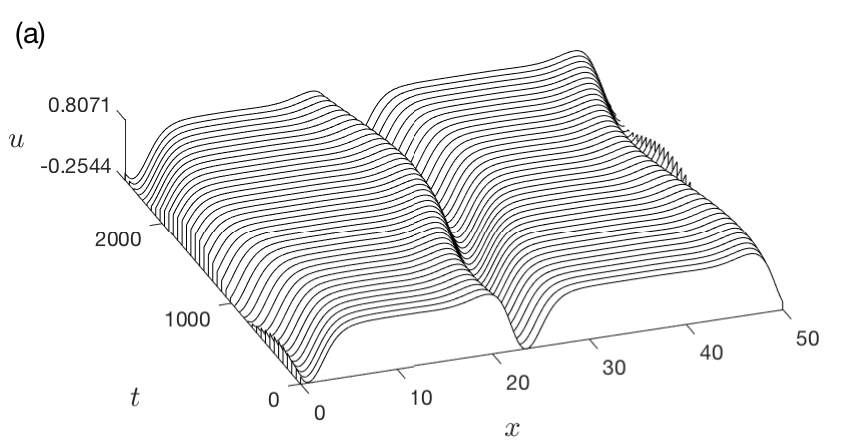} 
\includegraphics [width=7.2cm]{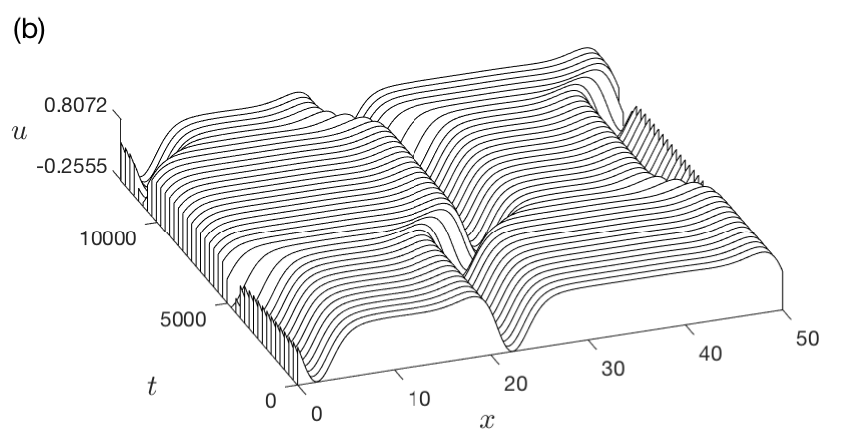} 
\vspace{-0.5cm}
\caption{Time evolution over one period of time of the time-periodic solution for $\bar u=0.55$ and $L=50$ corresponding to (a) point~1 and (b) point~2 shown in Fig.~\ref{fig:zooms}(a).
}
 \label{fig:tper_bu_0_55_L_50_I_LAB_23}  
 \vspace{-0.3cm}
\end{figure}

Figures~\ref{fig:A1.1}--\ref{fig:tper_bu_0_55_L_50_I_LAB_23} show the results of the calculations (bifurcation diagrams and solution profiles) for several values of $L$ and for $\bar u=0.4$ and $0.55$. In the bifurcation diagrams, we use black thin solid lines to show the two-drop primary branches. The secondary branches are shown by dashed lines, and the dotted lines show branches of time-periodic solutions. The bifurcation points to secondary branches of steady states are indicated by red circles, the red solid squares indicate saddle-node bifurcations, and the red solid triangles indicate Hopf bifurcations to branches of time-periodic solutions. In addition, blue thick solid lines show the branches of one-drop solutions of the period equal to the domain length $L$.  The black solid squares indicate saddle-node bifurcations on these one-drop branches and the black solid triangles indicate Hopf bifurcations. The thick dotted lines show the branches of time-periodic states that bifurcate from such points.

Figures~\ref{fig:A1.1}(a) and (b) show the bifurcation diagrams for $\bar u=0.4$ and $L=25$ and $35$, respectively. For presentational purposes, we show the dependence of $D\|\delta u_0\|$ (instead of $\|\delta u_0\|$) on $D$. 
We observe in Fig.~\ref{fig:A1.1}(a) that for $L=25$ there are two bifurcation points on the two-drop primary branch, and the secondary branches that start at these bifurcation points continue towards large values of $D$. There is also one Hopf bifurcation on the two-drop primary branch, and the time-periodic branch starting at this point also extends to large values of $D$. Figure~\ref{fig:A1.1}(b) shows that for $L=35$ there are five bifurcation points on the two-drop primary branch. Some of the secondary branches that start at these points reach large values of $D$ and may continue to infinity, whereas secondary branches starting at other bifurcation points reconnect to the same primary branch. In particular, the secondary branches starting at bifurcation points $1$, $2$ and $3$ continue to infinity, while  bifurcation points $4$ and $5$ are connected to each other by a secondary branch.

Regarding the one-drop branches, 
we find that for $L=25$, there exists one Hopf bifurcation, and
the time-periodic branch emanating at this Hopf bifurcation extends to large values of $D$. 
For $L=35$ there are three Hopf bifurcation points on the one-drop branch, and the time-periodic branches starting at these bifurcation points all extend to large values of $D$.

Figs.~\ref{fig:p1}(a) and (b) show selected solution profiles for $\bar u=0.4$ when $D=3$ for $L=25$ and $35$, respectively. We exclude the solution profiles for the one-drop branches. Note that at $L=25$, there are three different solutions on the first secondary branch that correspond to $D=3$. These solutions are shown by the dotted, dashed and dot-dashed lines and are ordered in the decreasing norm $\|\delta u_0\|$, i.e., the dotted line corresponds to the solution with the largest norm and the dot-dashed line corresponds to the solution with the smallest norm. In general, we can observe that the solutions of the secondary branches that are located closer to the primary branch have profiles that are similar to the profiles of the solutions of the primary branch.

An example of a time evolution over one period of time of a solution from a time-periodic branch is shown in Fig.~\ref{fig:tper_bu_0_4_L_25_LAB_14_D3}. In particular, this figure shows the time-periodic solution corresponding to Fig.~\ref{fig:A1.1}(b) for $\bar u=0.4$, $L=25$ when $D=3$. We can see that the solution looks like a superposition of two drops (a smaller one and a bigger one) periodically exchanging mass.

Figures~\ref{fig:A1.2}(a) and (b) show bifurcation diagrams for $\bar u=0.55$ and $L=35$ and $50$, respectively. 
For $L=35$, we observe that the one-drop branch has two saddle nodes, where as for $L=50$ the one-drop branch has two saddle-node bifurcations and one Hopf bifurcation. The branch of time-periodic solutions starting at this bifurcation point  extends to larger values of $D$.
 
Regarding the two-drop branches, we observe in Fig.~\ref{fig:A1.2}(a) that for $L=35$ there are four bifurcation points and one saddle-node bifurcations on the two-drop primary branch. The secondary branches that start at these bifurcation points reconnect to the two-drop primary branch. Also, we denote the upper and the lower parts of the primary branch by letters $\alpha$ and $\beta$, respectively. 
We can observe that points~$1$ and~$2$ on the upper part are connected to points~$4$ and~$3$, respectively on the lower part. On the one-drop branch
we find two saddle-nodes, but there are no other bifurcation points. Figure~\ref{fig:A1.2}(b) shows that for  $L=50$ there are five bifurcation points and two saddle-node bifurcations on the two-drop primary branch. Some of the secondary branches that start at these points, reach large values of $D$ and may continue to infinity, whereas secondary branches starting at other bifurcation points reconnect to the primary branch. We call the upper part of the primary branch (up to the first saddle node) part~$\alpha$, the part connecting the two saddle nodes part~$\beta$, and the lower part (starting from the second saddle node) part~$\gamma$. We find that the secondary branch starting at bifurcation point~$1$ on part~$\alpha$ continues to infinity, while  bifurcation point~$2$ on part~$\alpha$ is connected to point~$5$ on part~$\beta$, and bifurcation point~$3$ on part~$\alpha$ is connected to point~$4$ on part~$\beta$. For $L=50$, we additionally find that there are two Hopf bifurcations on the two-drop primary branch, denoted by symbols I and II. It is interesting to note that these bifurcation points are not connected to each other by a time-periodic branch, and the time-periodic branches that emerge from these points do not extend to large values of $D$. Instead, these time-periodic branches are connected to side branches (the dashed blue and red branches, respectively). This is confirmed in Figs.~\ref{fig:zooms}(a) and (b) for the time-periodic branches starting at points~I and~II, respectively.  Moreover, the inset in Fig.~\ref{fig:zooms}(b) indicates a possible exponential snaking behaviour of the time-periodic branch -- one saddle-node is clearly visible, and one more can be obtained by another zoom.
We conjecture that the time-periodic branch starting at point~I results from a Takens-Bogdanov-type codimension-2 bifurcation at the pitchfork bifurcation point~3 (we note that for the usual Takens-Bogdanov bifurcation the time-periodic branch emerges from a saddle-node bifurcation, not from a pitchfork bifurcation, see, for example, Kuznetsov~\cite{Ref.44}). Similarly, the time-periodic branch starting at point~II results from such a codimension-2 bifurcation, but at a pitchfork bifurcation that has, at the shown value of $L$, moved to larger values of $D$ (or to infinity). 
The time evolutions over one period of solutions corresponding to points~1 and~2 shown by red diamonds in Fig.~\ref{fig:zooms}(a) are shown in Figs.~\ref{fig:tper_bu_0_55_L_50_I_LAB_23}(a) and(b), respectively. In both cases, the solution behaves as a superposition of two drops  periodically exchanging mass. Panel (b) confirms that as the homoclinic bifurcation is approached, the temporal period increases, and now the mass-exchange events happen burst-like over relatively short time intervals while for most of the time the solution is a quasi-steady superposition of two drops of different sizes.

\begin{figure}
 \centering
\includegraphics [width=6cm]{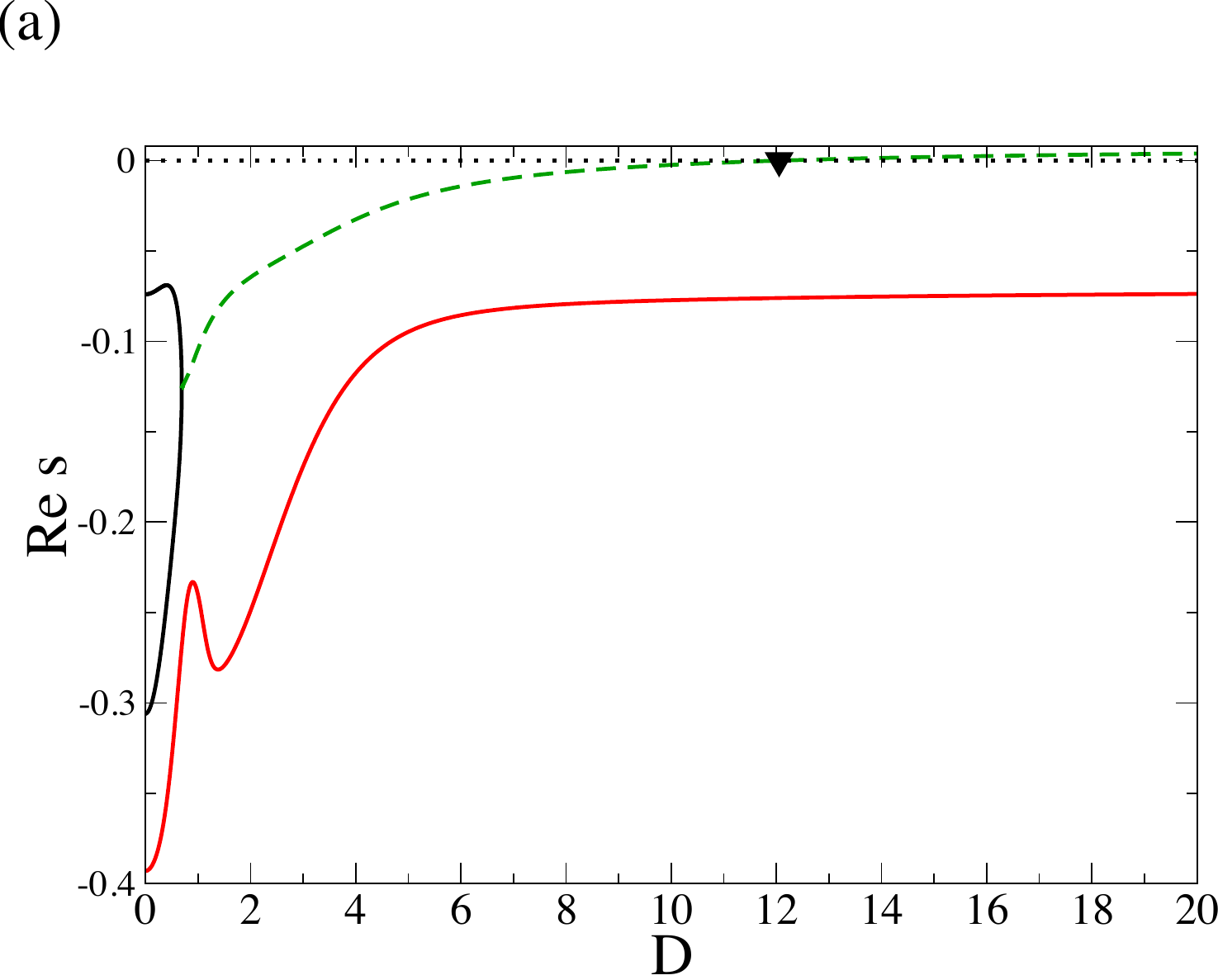}\hspace{0.5cm}
\includegraphics [width=6cm]{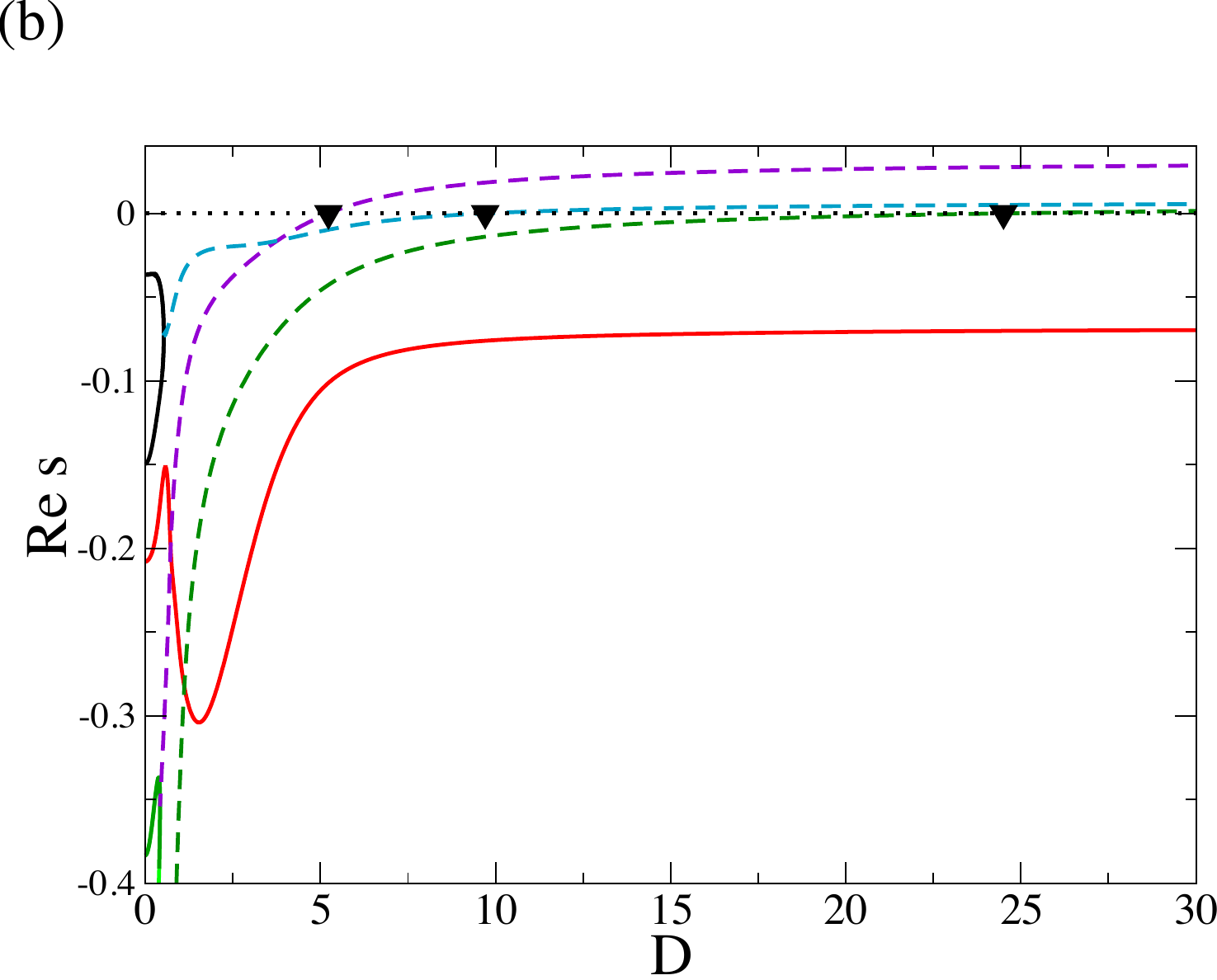}

\vspace{-0.3cm}
 \caption{The dependence of the real parts of the dominant eigenvalues $s$ on  $D$ along the one-drop primary branch when $\bar u=0.4$ and (a) $L=25$ and (b) $L=35$ (cf. Fig.~\ref{fig:A1.1}). 
 }
 \label{fig:AA1.4}
 \vspace{0.3cm}
 \centering
\includegraphics [width=6cm]{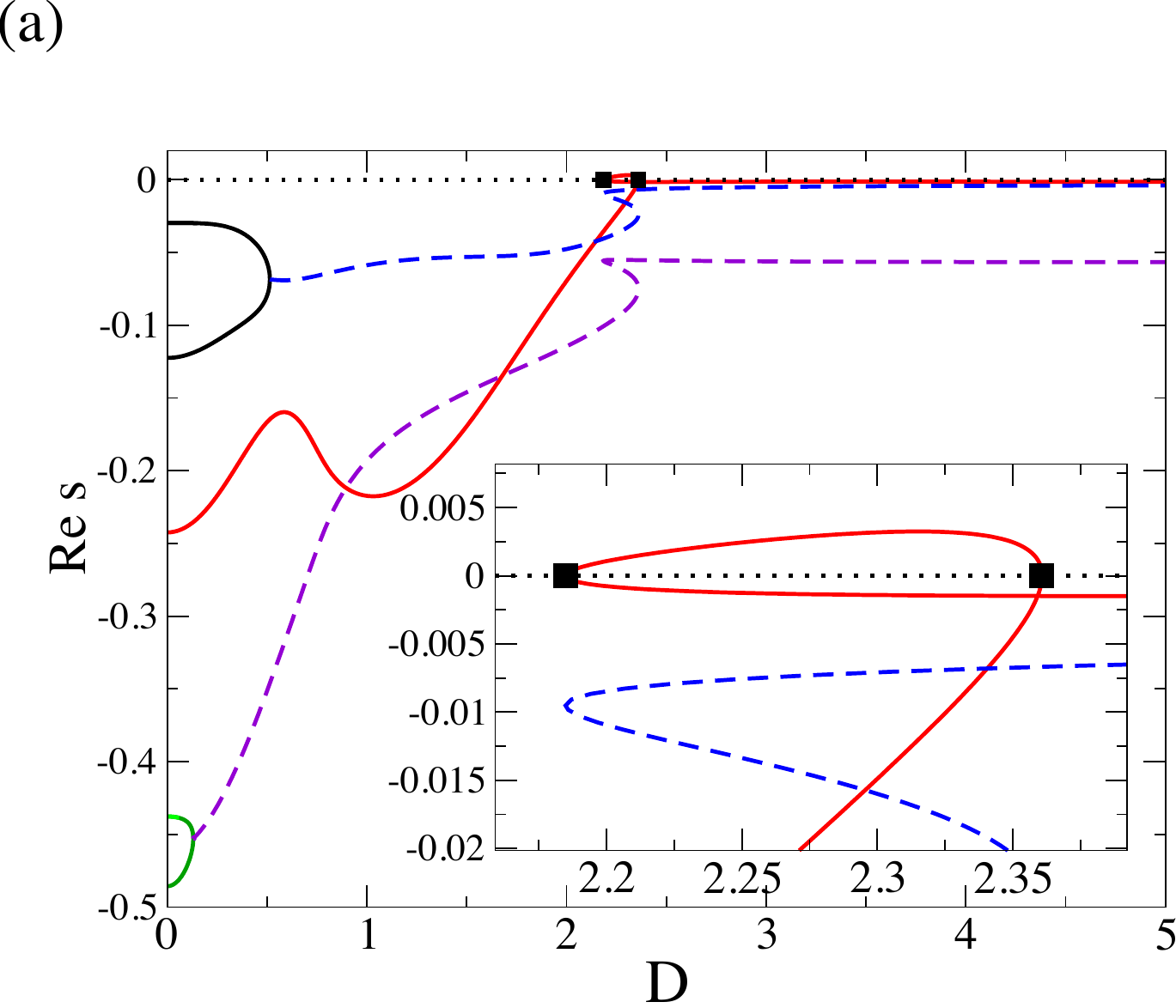} \hspace{0.5cm}
\includegraphics [width=6cm]{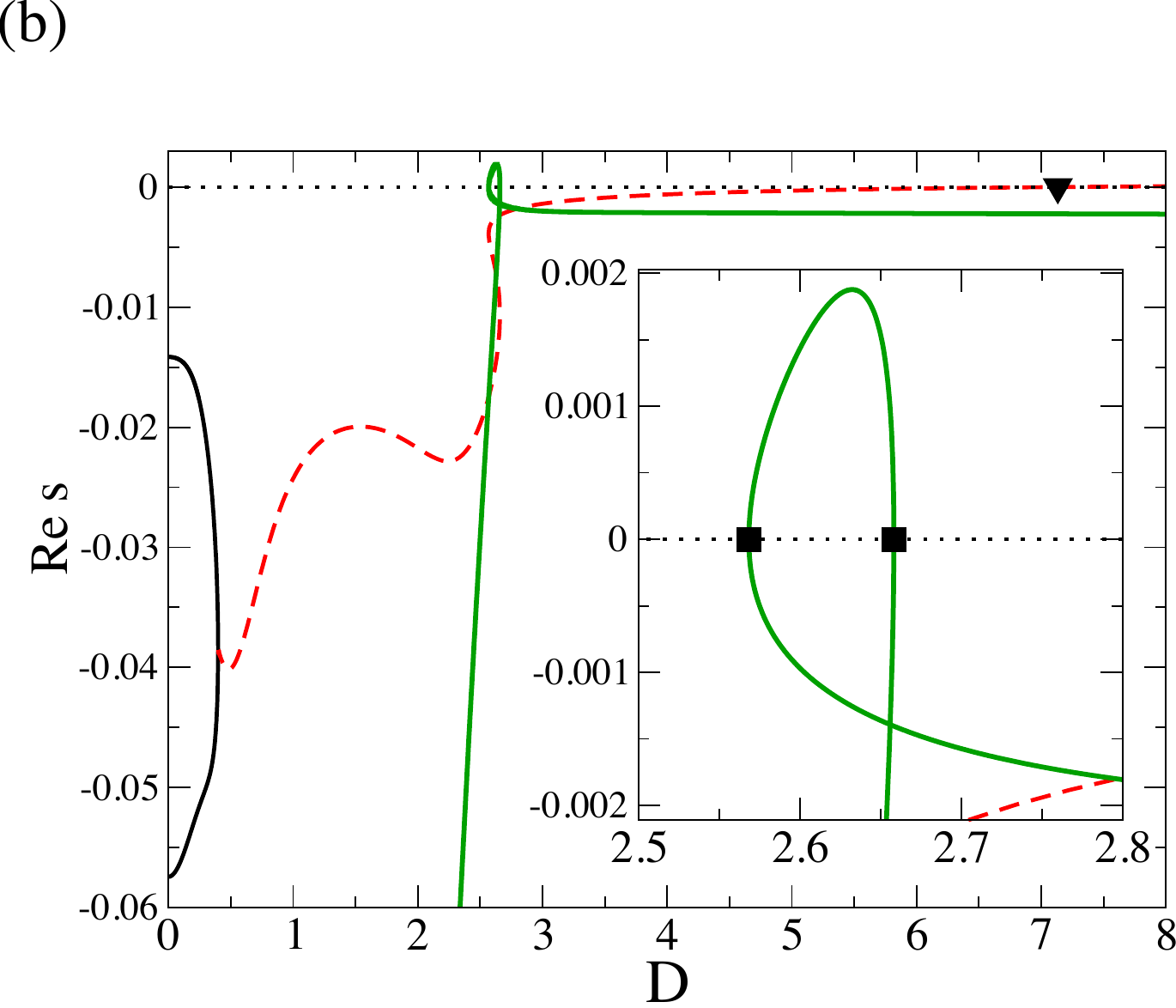} 

\vspace{-0.3cm}
 \caption{The dependence of the real parts of the dominant eigenvalues $s$ on $D$ along the one-drop primary branch when $\bar u=0.55$ and (a) $L=35$ and (b) $L=50$ (cf. Fig.~\ref{fig:A1.2}). 
 }
 \label{fig:AA1.5}
 \vspace{-0.3cm}
\end{figure}


\subsubsection{Linear stability of one-drop branches}

Figures~\ref{fig:AA1.4} and \ref{fig:AA1.5} show the real parts of the dominant eigenvalues along the one-period primary branches presented in Figs.~\ref{fig:A1.1} and \ref{fig:A1.2}, respectively. The solid lines correspond to the real eigenvalues. The dashed lines correspond to the eigenvalues with nonzero imaginary parts. 

Figures~\ref{fig:AA1.4}(a) and (b) correspond to $L=25$ and $35$, respectively, at $\bar u=0.4$. 
In agreement with the results presented in Fig.~\ref{fig:A1.1}, we see that for $L=25$ there is one Hopf bifurcation. We can conclude that the Hopf bifurcation is supercritical and there is a stable interval for one-drop solutions for $D\lesssim 12.05$. For $L=35$, there are three Hopf bifurcations, related to three different pairs of eigenvalues and there is a stable interval for one-drop solutions for $D\lesssim 5.23$. 

Figures~\ref{fig:AA1.5}(a) and (b) correspond to $L=35$ and $50$, respectively, at $\bar u=0.55$ (see Figs.~\ref{fig:A1.2} (a) and (b), respectively). In agreement with the results presented in Fig.~\ref{fig:A1.2}, we see that for $L=35$ there are two saddle-node bifurcations and there are no Hopf bifurcations. The part of the branch connecting the two saddle nodes (for $D$ between $2.18$ and $2.36$) is unstable, but there are stable solutions for all the values of $D$. For $L=50$, there are two saddle-node bifurcations (at $D\approx 2.57$ and $D\approx 2.67$) and there is one supercritical Hopf bifurcation at $D\approx 7.13$, so that there are stable one-drop solutions for $D\lesssim 7.13$. 

We generally observe that sufficiently strong driving may destabilize one-drop solutions if the domain size is sufficiently large. 

\subsubsection{Linear stability of two-drop primary branches and coarsening}
\label{Ch6:Linear stability of primary branches}

\begin{figure}
 \centering
\includegraphics [width=6cm]{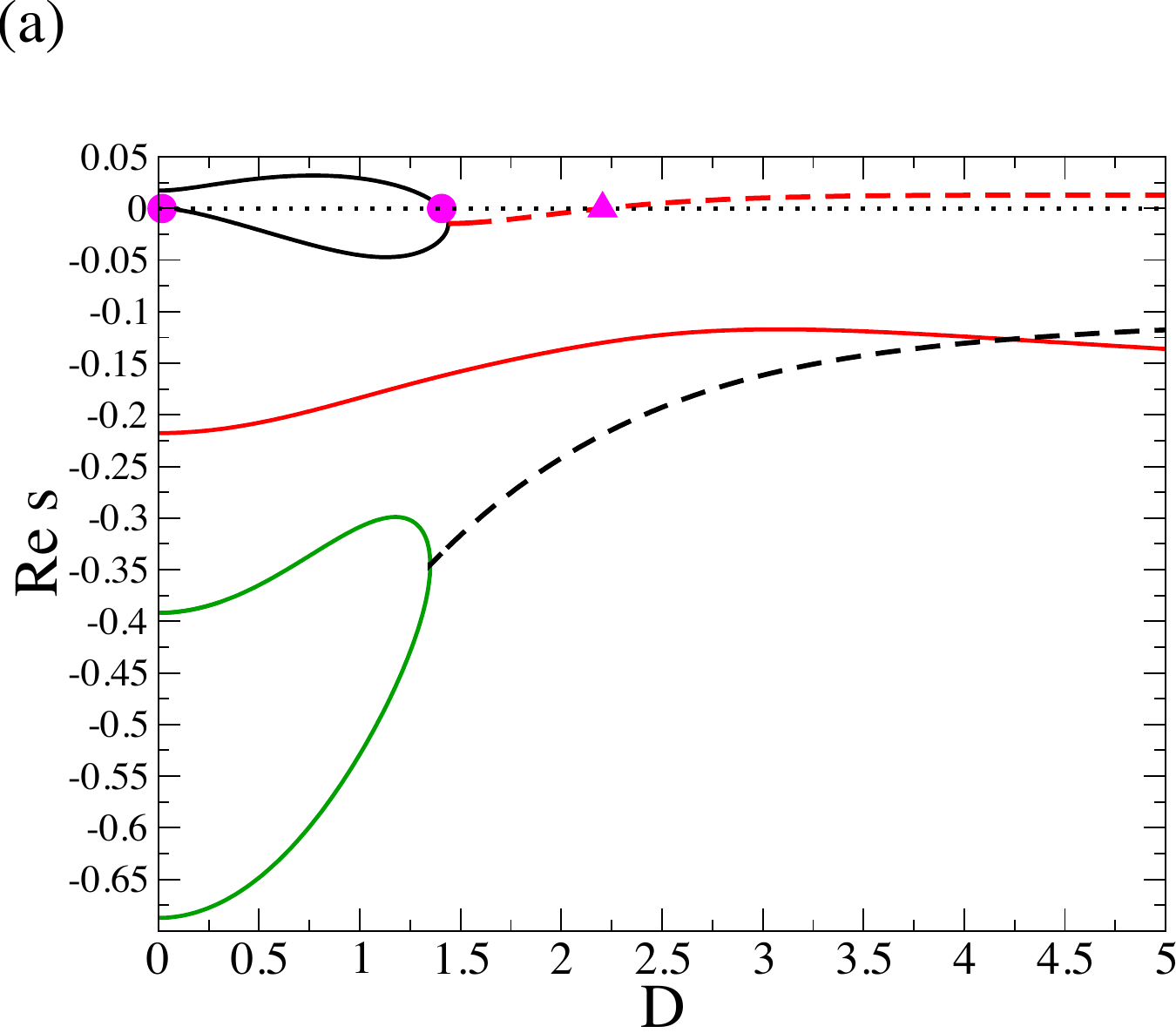} \hspace{0.5cm}
\includegraphics [width=6cm]{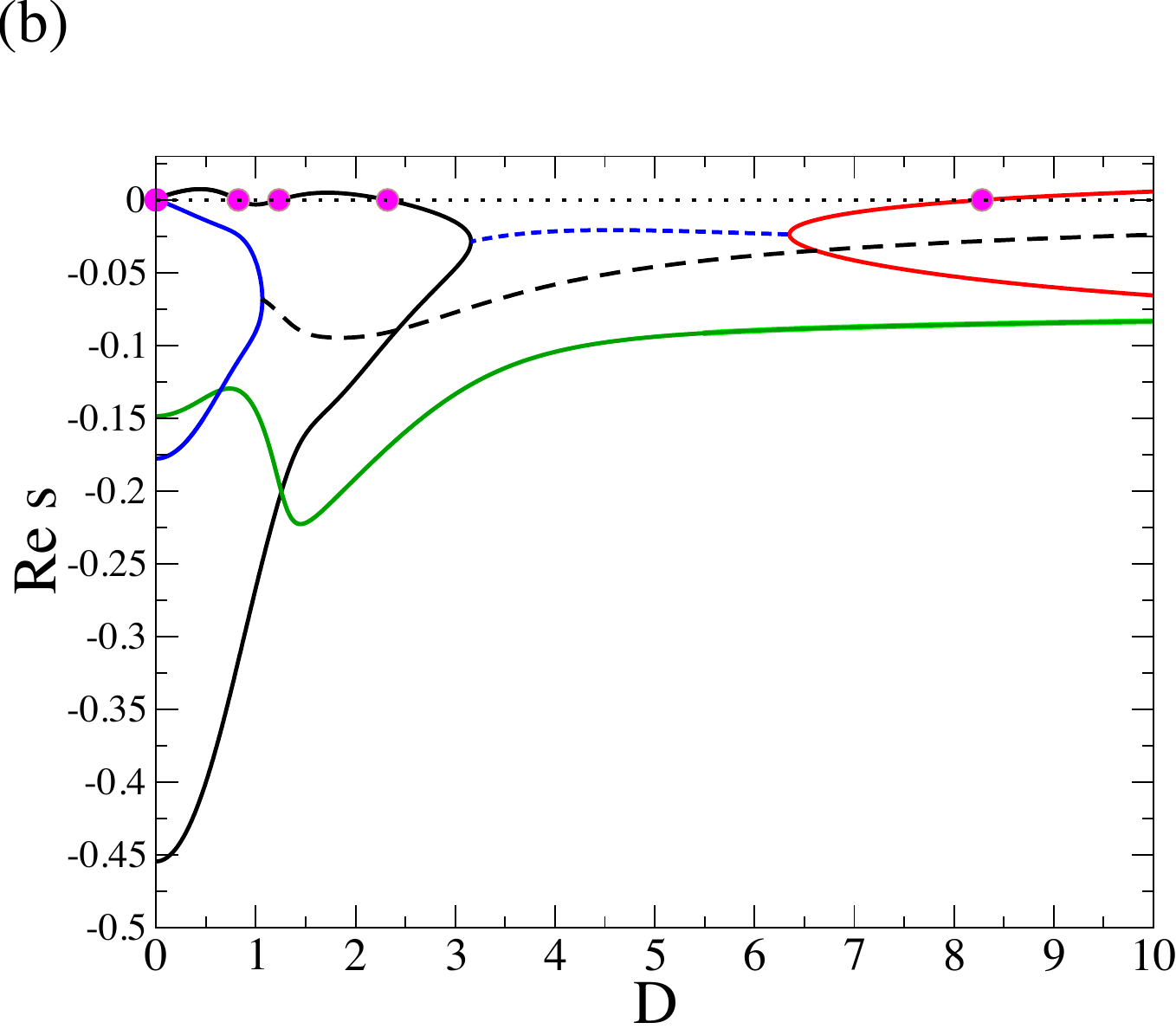} 

\vspace{-0.3cm}
 \caption{The dependence of the real parts of the dominant eigenvalues $s$ on $D$ along the two-drop primary branch when $\bar u=0.4$ and (a) $L=25$ and (b) $L=35$ (cf. Fig.~\ref{fig:A1.1}). 
 }
 \label{fig:A1.4}
 \vspace{-0.3cm}
\end{figure}

\begin{figure}
 \centering
 \includegraphics [width=5.1cm]{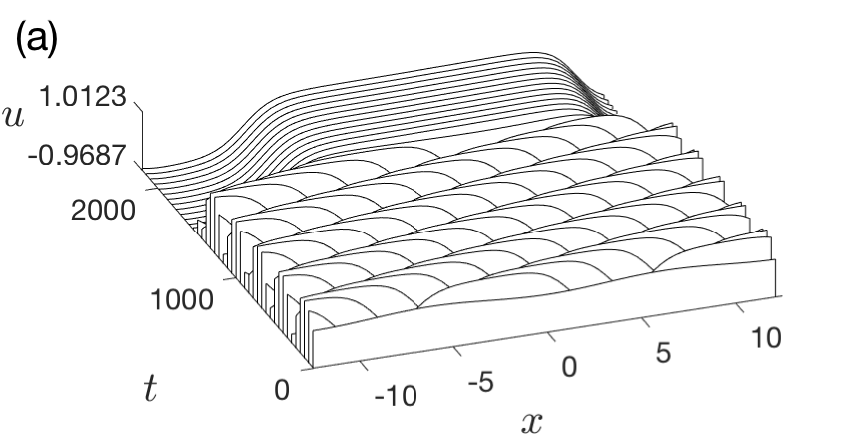}
 \includegraphics [width=5.1cm]{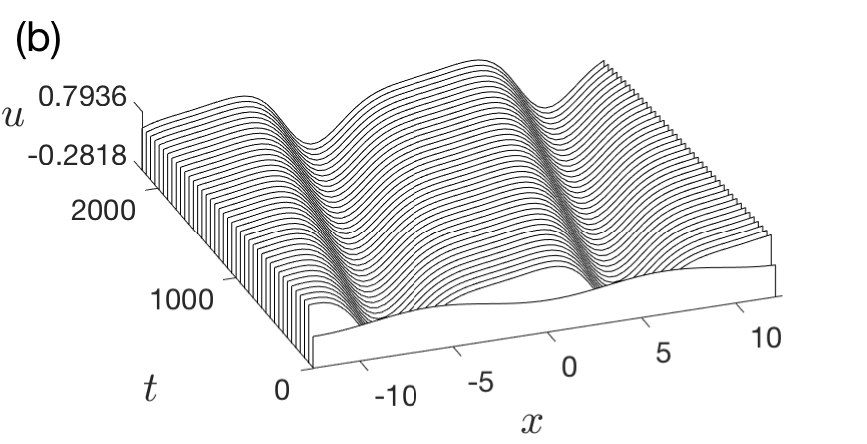}
 \includegraphics [width=5.1cm]{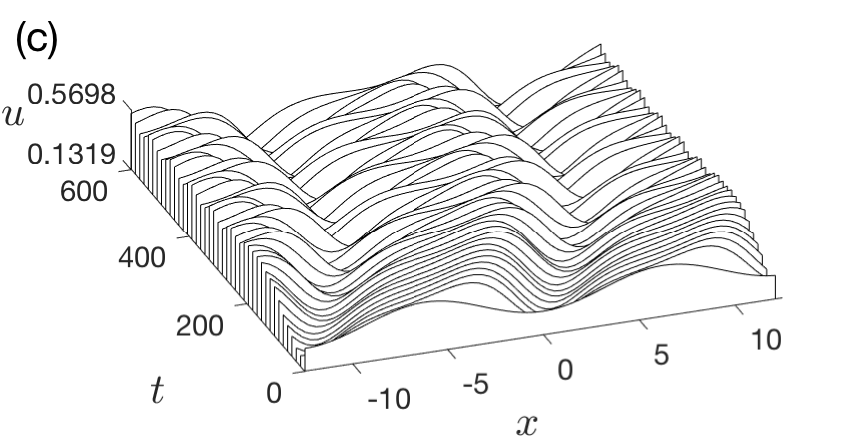}
 \includegraphics [width=5.1cm]{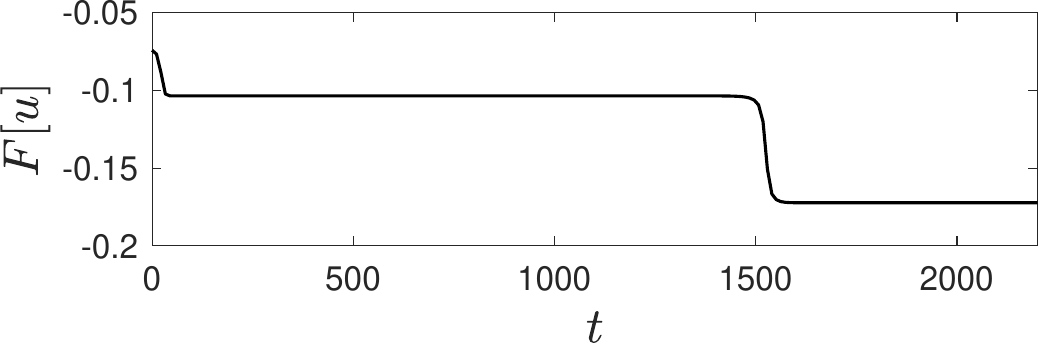}
 \includegraphics [width=5.1cm]{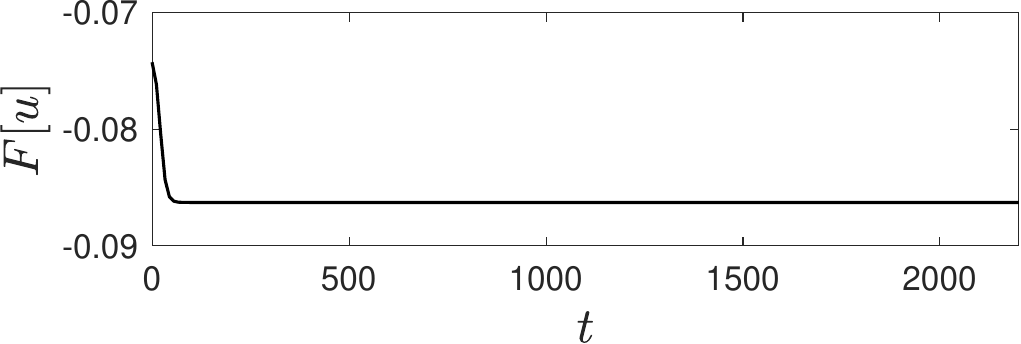}
 \includegraphics [width=5.1cm]{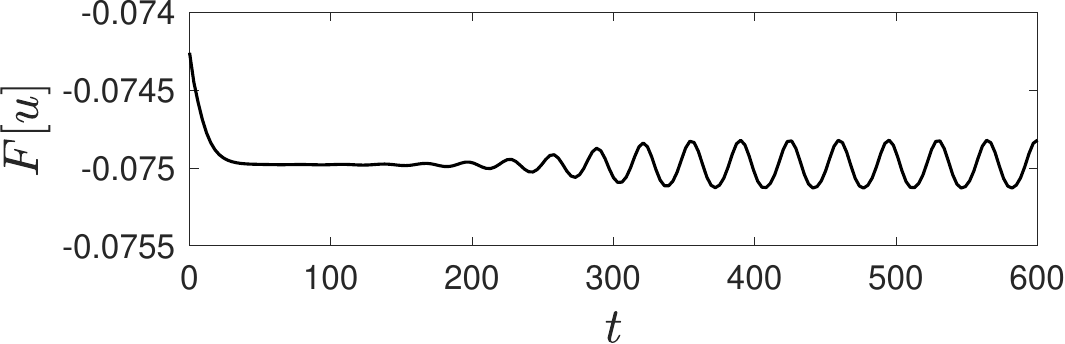}
 
 \vspace{-0.3cm}
 \caption{Numerical solution of the cCH equation (\ref{Eq.2}) on the periodic domain of length $L=25$ for $\bar u=0.4$ and (a) $D=0.3$, (b) $D=2$ and (c) $D=5$, with the initial condition $u(x,0)=\bar u-0.1\cos(2\pi x/L)+0.001\cos(\pi x/L)$. The top panels show space-time plots of the time evolution in moving frames. The bottom panels show the time evolutions of the corresponding energies $F[u]$.
 }
 \label{fig:time_D_5_0_bu_0_4_L_25}
%
\vspace{0.3cm}
 \includegraphics [width=6cm]{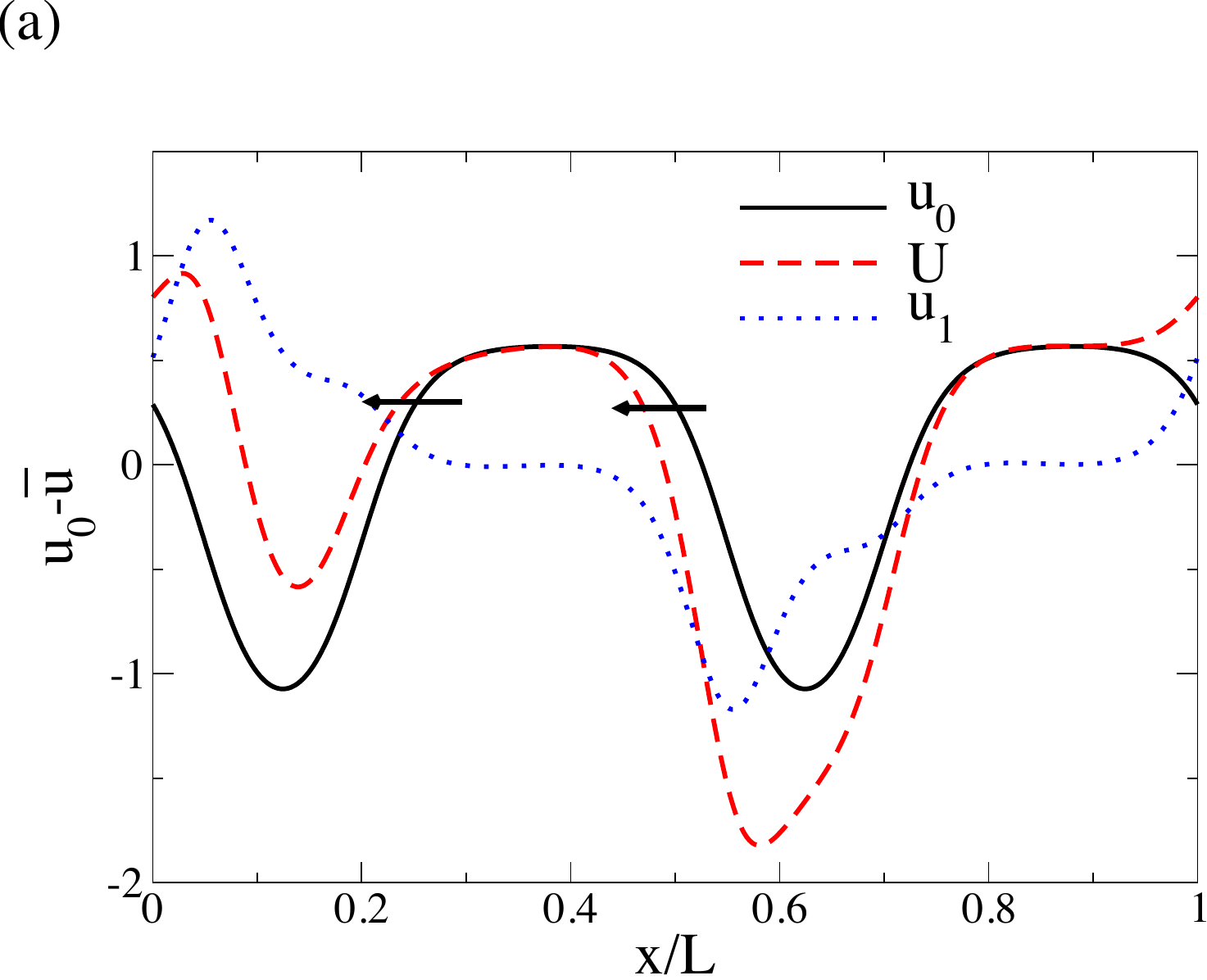}\hspace{0.5cm}
 \includegraphics [width=6cm]{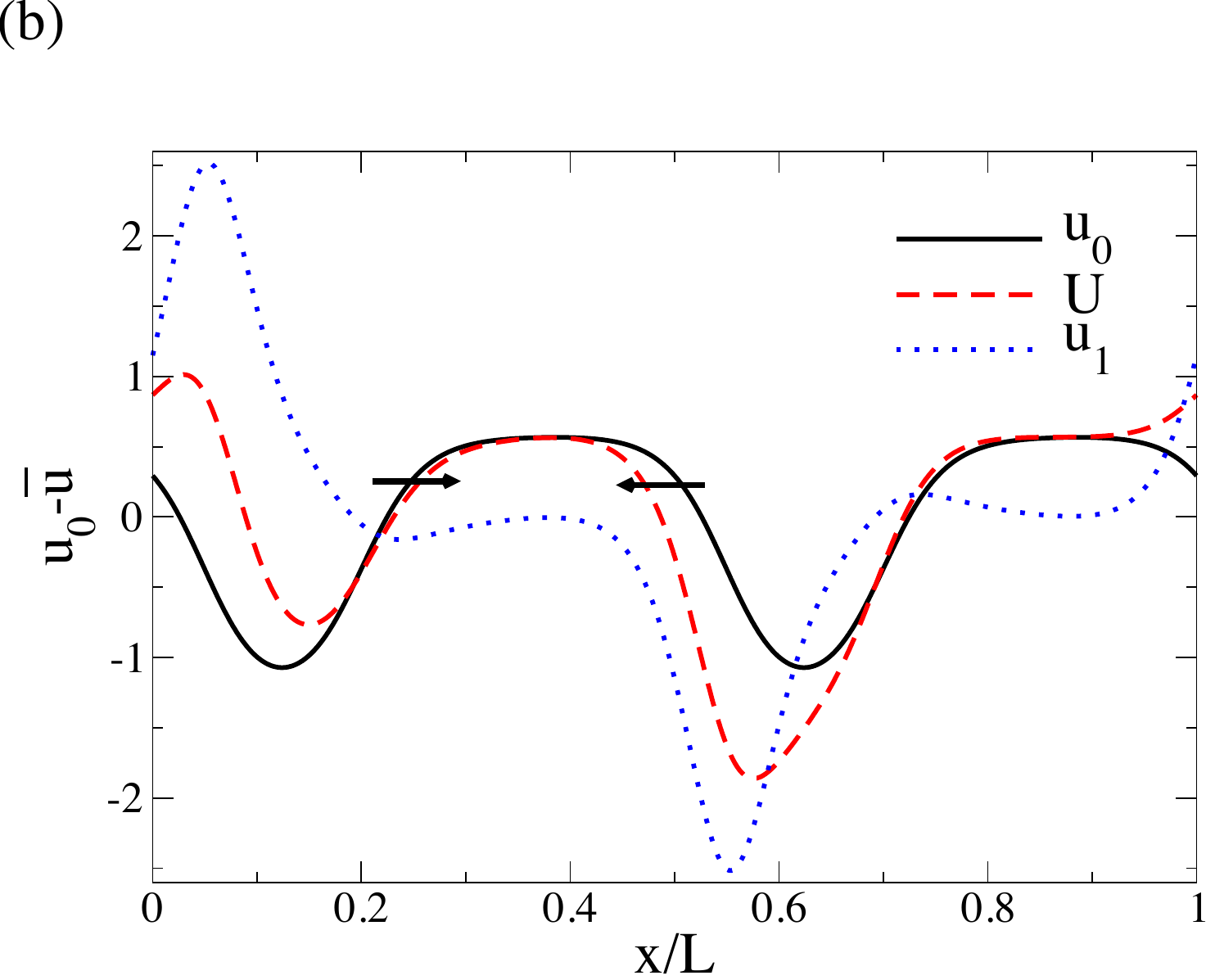}
 
 \vspace{-0.3cm}
 \caption{Shown are the most unstable eigenfunctions for two-drop solutions at $\bar u=0.4$ and $L=25$ for (a) $D=0.005$ and (b) $D=0.1$. The solid lines correspond to the solutions $u_0$.
The dotted lines correspond to the eigenfunctions $u_1$. The dashed lines correspond to the solutions $u_0$ superimposed with eigenfunction $u_1$ multiplied by a small coefficient $\epsilon$, i.e., $U=u_0+\epsilon u_1$. 
}
 \label{fig:c2}   
%
\vspace{0.3cm}
\includegraphics [width=6cm]{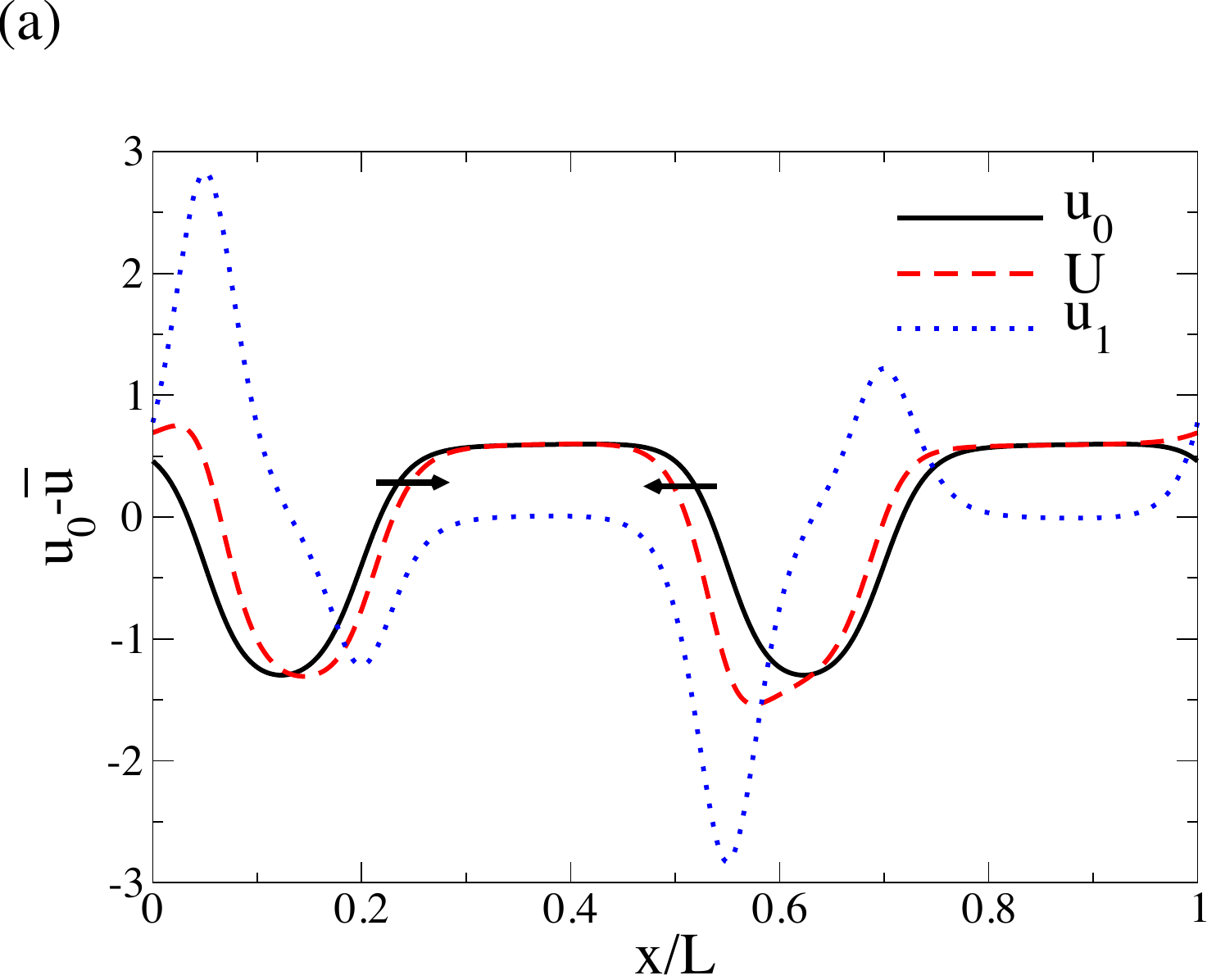}\hspace{0.5cm}
\includegraphics [width=6cm]{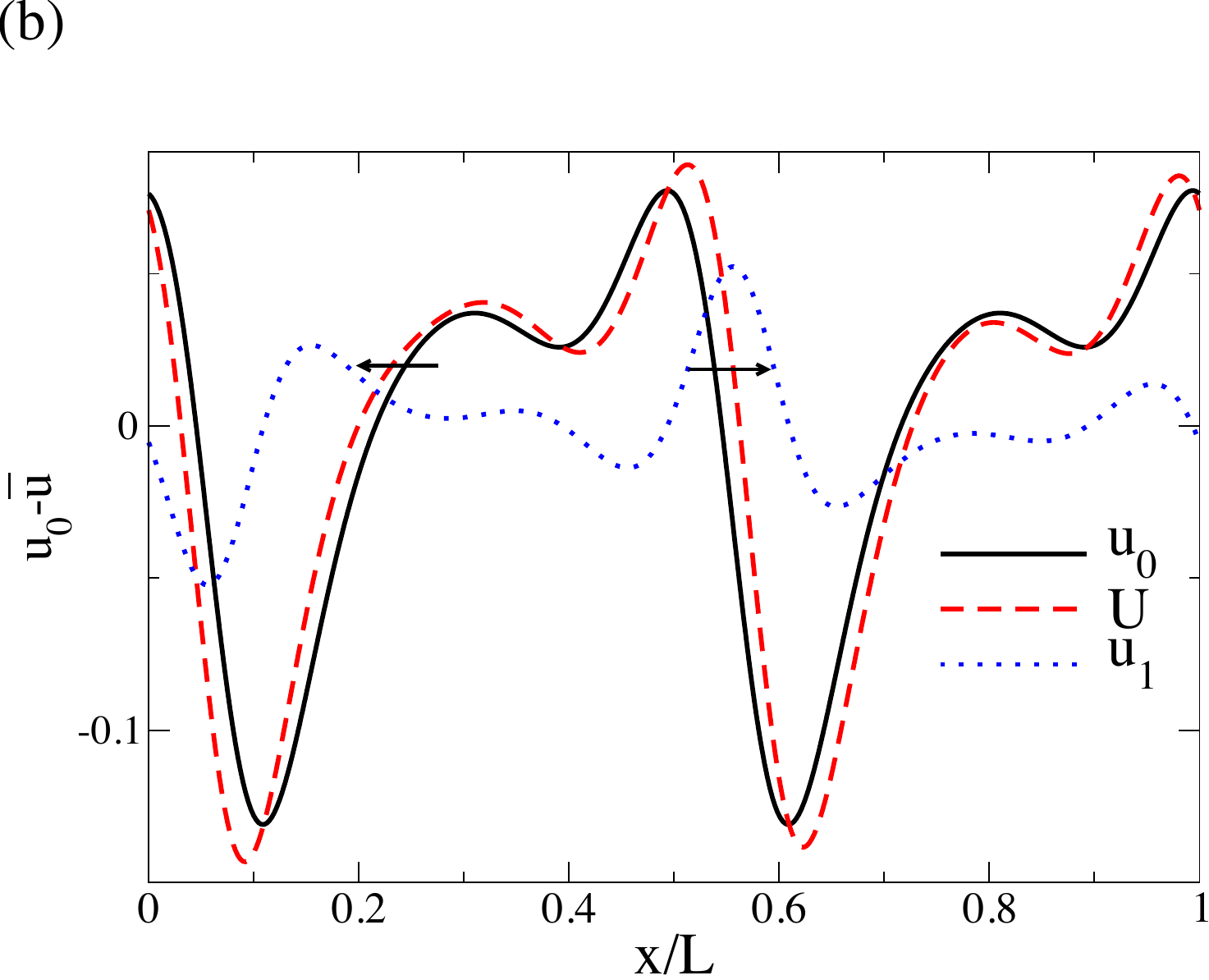}

\vspace{-0.3cm}
   \caption{Shown are the most unstable eigenfunctions for two-drop solutions at $\bar u=0.4$ and $L=35$ for (a) $D=0.1$ and (b) $D=9$. The solid lines correspond to the solutions $u_0$. The dotted lines correspond to the eigenfunctions $u_1$. The dashed lines correspond to the solutions $u_0$ superimposed with eigenfunction $u_1$ multiplied by a small coefficient $\epsilon$, i.e., $U=u_0+\epsilon u_1$. 
}
 \label{fig:c4}   
\end{figure}

Figures~\ref{fig:A1.4}(a) and (b) show the real parts of the dominant eigenvalues along the two-drop primary branches presented in Figs.~\ref{fig:A1.1}(a) and (b). Figure~\ref{fig:A1.4}(a) shows that for $L=25$ there are two pitchfork bifurcation points to side branches, one Hopf bifurcation to a branch of time-periodic solutions, and there is a stable interval between the second bifurcation point to a side branch and the Hopf bifurcation point, i.e., between $D\approx 1.41$ and  $D\approx 2.21$. Interestingly, this means that sufficiently strong driving $D$ can prevent coarsening, resulting in a stable two-drop traveling-wave solution. However, increasing $D$ further may again destabilize such a solution resulting in two drops periodically interacting with each other (note that coarsening is still prevented). These observations are corroborated by the time-dependent simulations shown in Fig.~\ref{fig:time_D_5_0_bu_0_4_L_25} for $\bar u=0.4$, $L=25$ and $D=0.3$, $2$ and $5$ (panels, (a), (b) and (c), respectively). The initial conditions are $u(x,0)=\bar u-0.1\cos(2\pi x/L)$ for panel (a) and (b) and $u(x,0)=\bar u-0.1\cos(2\pi x/L)+0.001\cos(\pi x/L)$ for panel (c). 
The top row shows the time evolutions of the solutions and the bottom row shows the time evolution of the energies of the solutions. (We use the same energy functional $F[u]$ as for the standard CH equation, although it should be pointed out that for $D\neq 0$ this functional is not anymore a Lyapunov functional and should not necessarily be minimized in the time evolution.) 
It can be observed that for $D=0.3$, the solution initially evolves into a two-drop solution, but around $t=1500$ the drops coarsen and a one-drop solution is obtained (a one-drop solution is linearly stable for this value of $D$, see Fig.~\ref{fig:AA1.4}(b)). In contrast, for $D=3$, a two-drop solution remains stable during the course of evolution, which agrees with the theoretical prediction (a one-drop solution is also linearly stable for this value of $D$, see Fig.~\ref{fig:AA1.4}(b), so the long-time evolution of solutions depends on initial conditions). For $D=5$, the solution again initially tends to evolve into a two-drop solution. But as is evident from the energy and norm plots, around $t=150$, the drops start to oscillate, and the solution eventually evolves into a time-periodic state resembling two drops periodically exchanging mass. We note that a one-drop solution is also linearly stable for this value of $D$, see Fig.~\ref{fig:AA1.4}(b). So we expect that different initial conditions can lead to time-periodic solutions or one-drop traveling-wave solutions.

Figure~\ref{fig:c2} shows the most unstable eigenmode $u_1$ superimposed with the primary two-drop solution $u_0$ for $\bar u=0.4$ and $L=25$. The arrows indicate the directions in which the fronts are shifted (in the same way as in Fig.~\ref{fig:standardCH_eig} for the standard CH equation).  Panels (a) and (b) correspond to $D=0.005$ and $0.1$. An interesting observation is that for the smaller value of $D$ the most unstable mode appears to be translational (in agreement with the $D=0$ case), whereas for the larger value of $D$ the mode seems to change into a volume mode. Thus, the driving force can change the type of coarsening.

In Fig.~\ref{fig:A1.4}(b), we can see that for $L=35$ there are five pitchfork bifurcation points to side branches and no Hopf bifurcations. We also see that there are two stable intervals in $D$, namely, $0.82 \le D\le 1.23$  and $2.32 \le D \le 8.28$. 
Figure~\ref{fig:c4} shows the most unstable eigenmode $u_1$ superimposed with the primary two-drop solution $u_0$ for $\bar u=0.4$ and $L=35$. Panels (a) and (b) correspond to $D=0.1$ and $9$. We observe that both modes are apparently volume modes.

\begin{figure}
 \centering
\includegraphics [width=6cm]{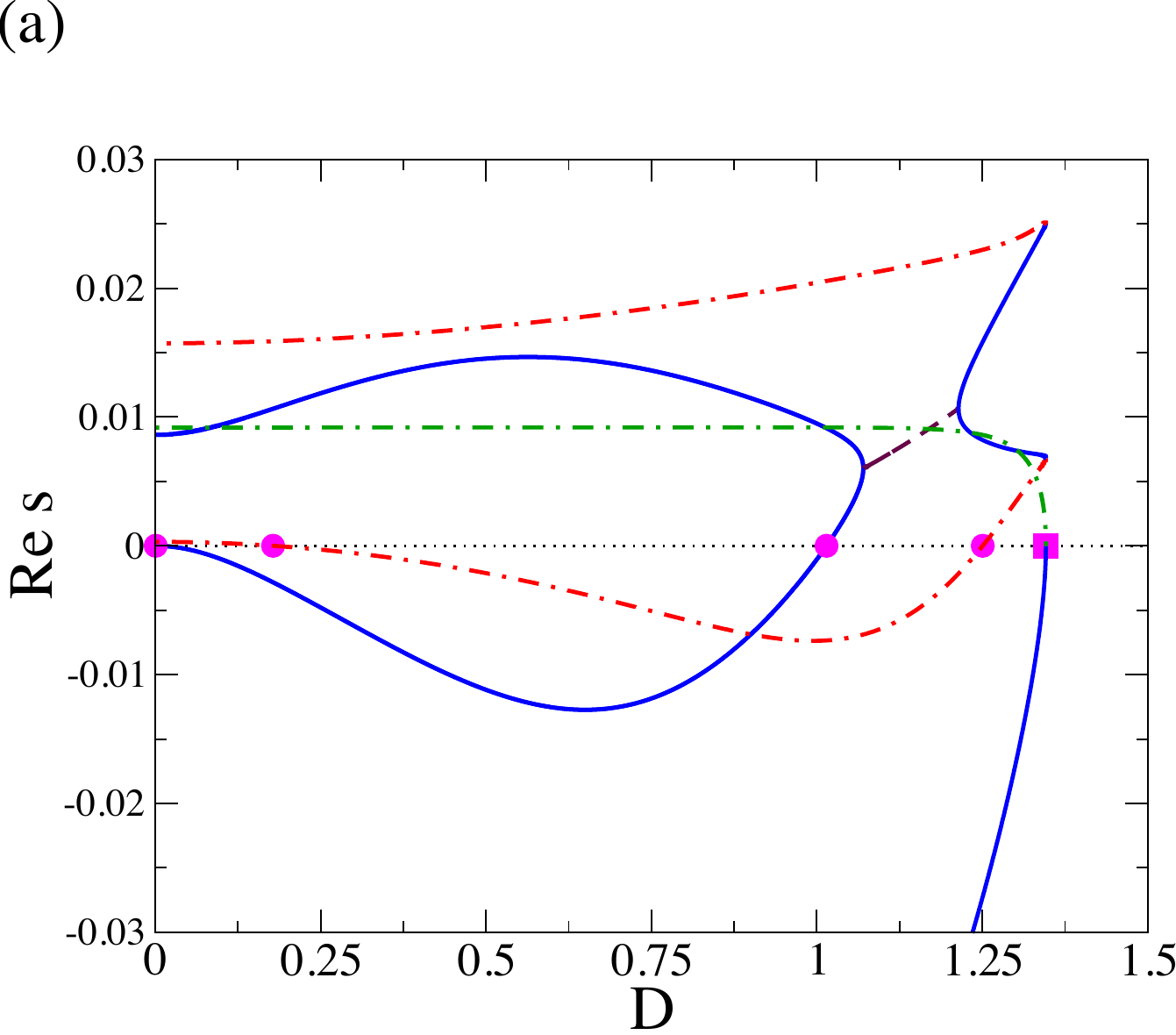} \hspace{0.5cm}
\includegraphics [width=6cm]{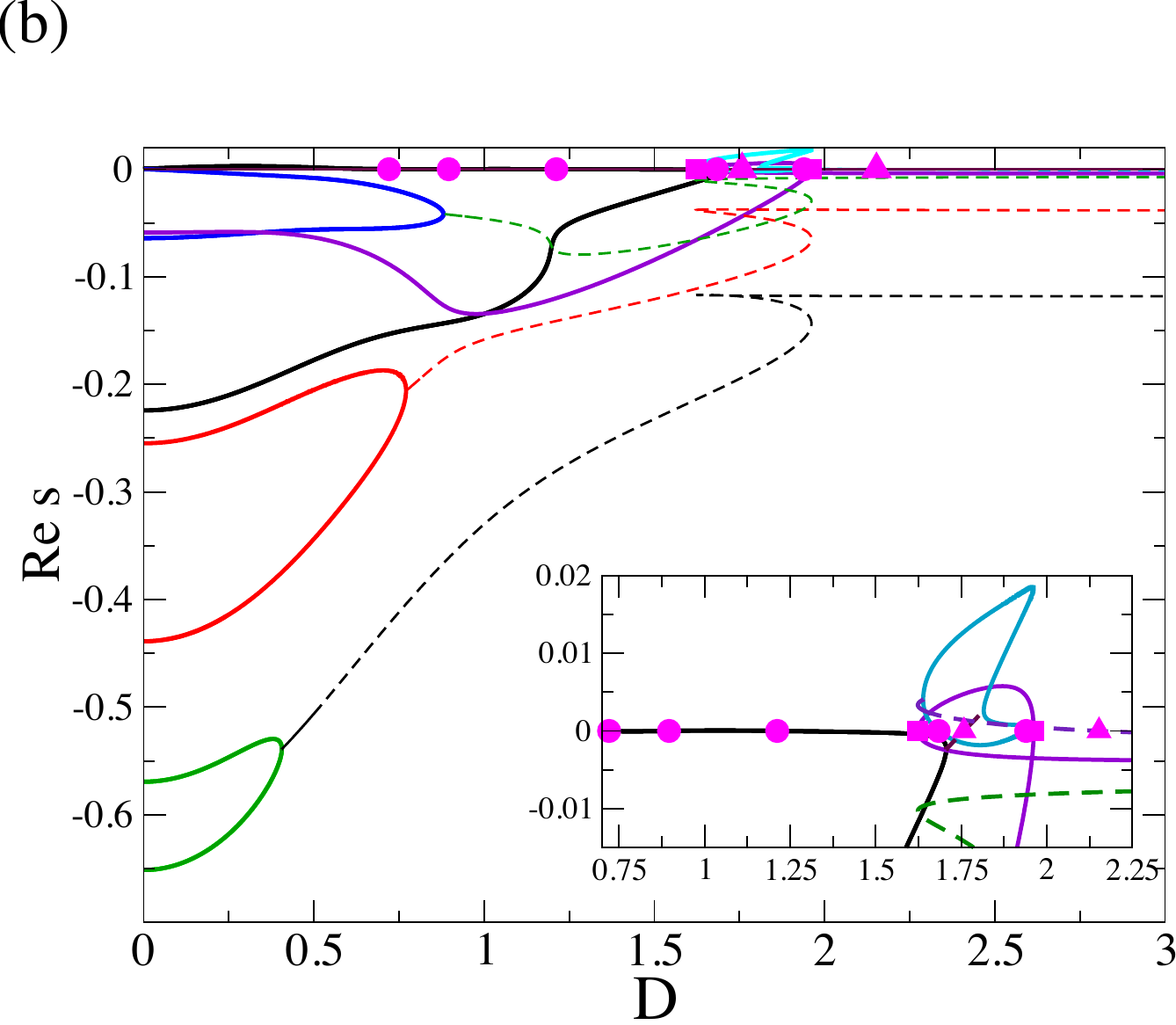} 

\vspace{-0.3cm}
 \caption{The dependence of the real parts of the dominant eigenvalues $s$ on $D$ along the two-drop primary branch when $\bar u=0.55$ and (a) $L=35$ and (b) $L=50$ (cf. Fig.~\ref{fig:A1.2}). 
  }
 \label{fig:A1.5}
  \vspace{-0.3cm}
\end{figure}

Figures~\ref{fig:A1.5}(a) and (b) correspond to $L=35$ and 50, respectively, at $\bar u=0.55$ (cf. Figs.~\ref{fig:A1.2} (a) and (b)). In Fig.~\ref{fig:A1.5}(a), the solid and dashed lines correspond to the real and complex (having nonzero imaginary parts) eigenvalues, respectively, for part a (the upper part) of the bifurcation curve shown in Fig.~\ref{fig:A1.2}(a). However, we additionally introduce the dot-dashed lines that correspond to the real eigenvalues for part b (the lower part) of the bifurcation curve shown in Fig.~\ref{fig:A1.2}(a). Note that for part b, the eigenvalues are real in the shown range. Note that the green dot-dashed line corresponds to the unstable eigenvalue that is inherited from the one-drop primary branch (that is unstable). 
We can see that for $L=35$ there are no stable intervals for the driving force $D$, and, therefore, in this case coarsening cannot be stabilized by sufficiently strong driving. 

Figure~\ref{fig:A1.5}(b) shows that for $L=50$ there are two stable intervals on part a (the upper part) of the bifurcation diagram shown in Fig.~\ref{fig:A1.2}(b), namely, $0.72 \lesssim D \lesssim 0.90$ and $1.21\lesssim D\lesssim 1.76$. Part b (the middle part of the bifurcation diagram) is unstable, and there is a stable interval on part c (the lower part) of the bifurcation diagram, namely, $D\gtrsim 2.15$.  

\subsubsection{Linear stability of secondary branches}

\begin{figure}
 \centering
\includegraphics [width=6cm]{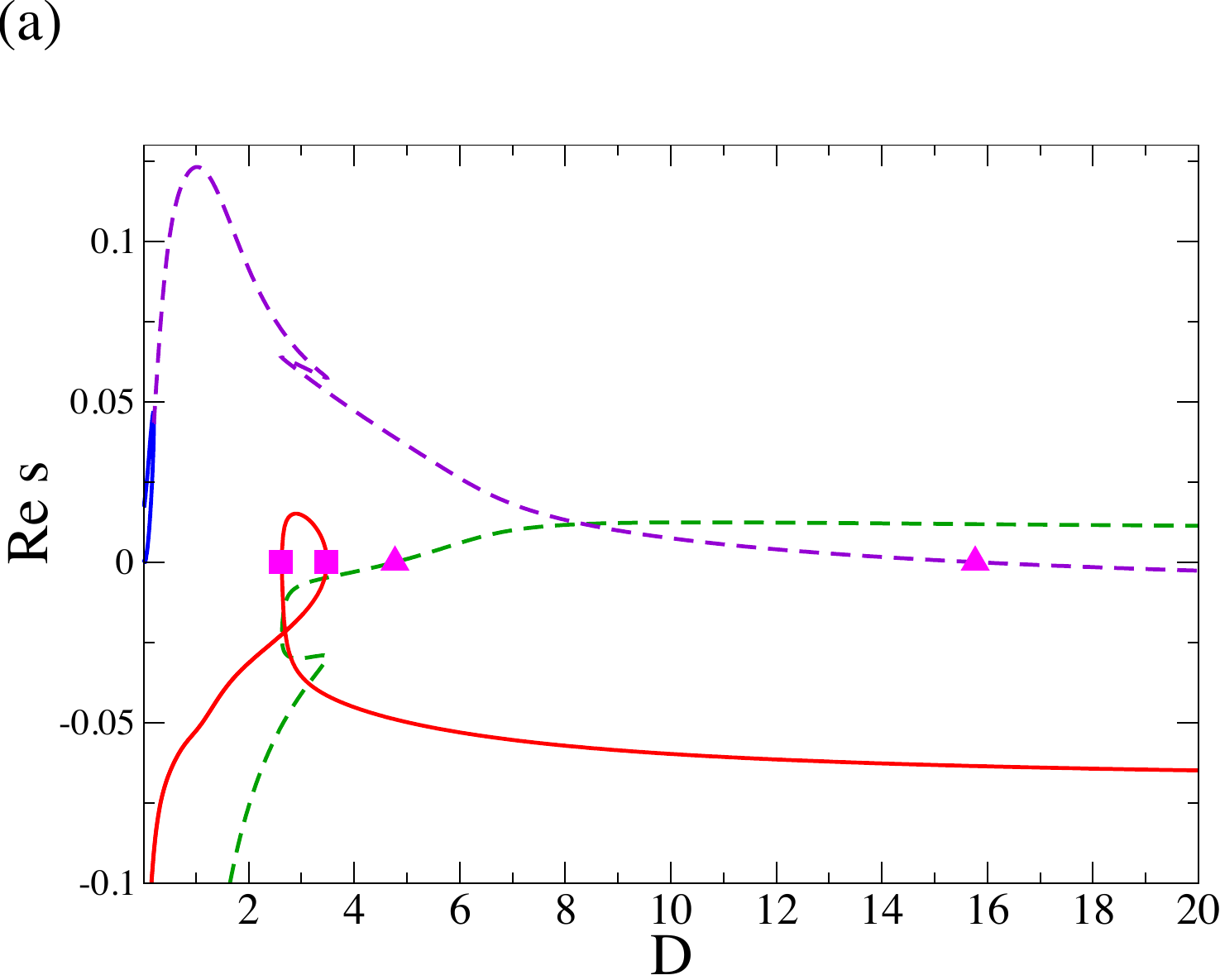}\hspace{0.5cm}
\includegraphics [width=6cm]{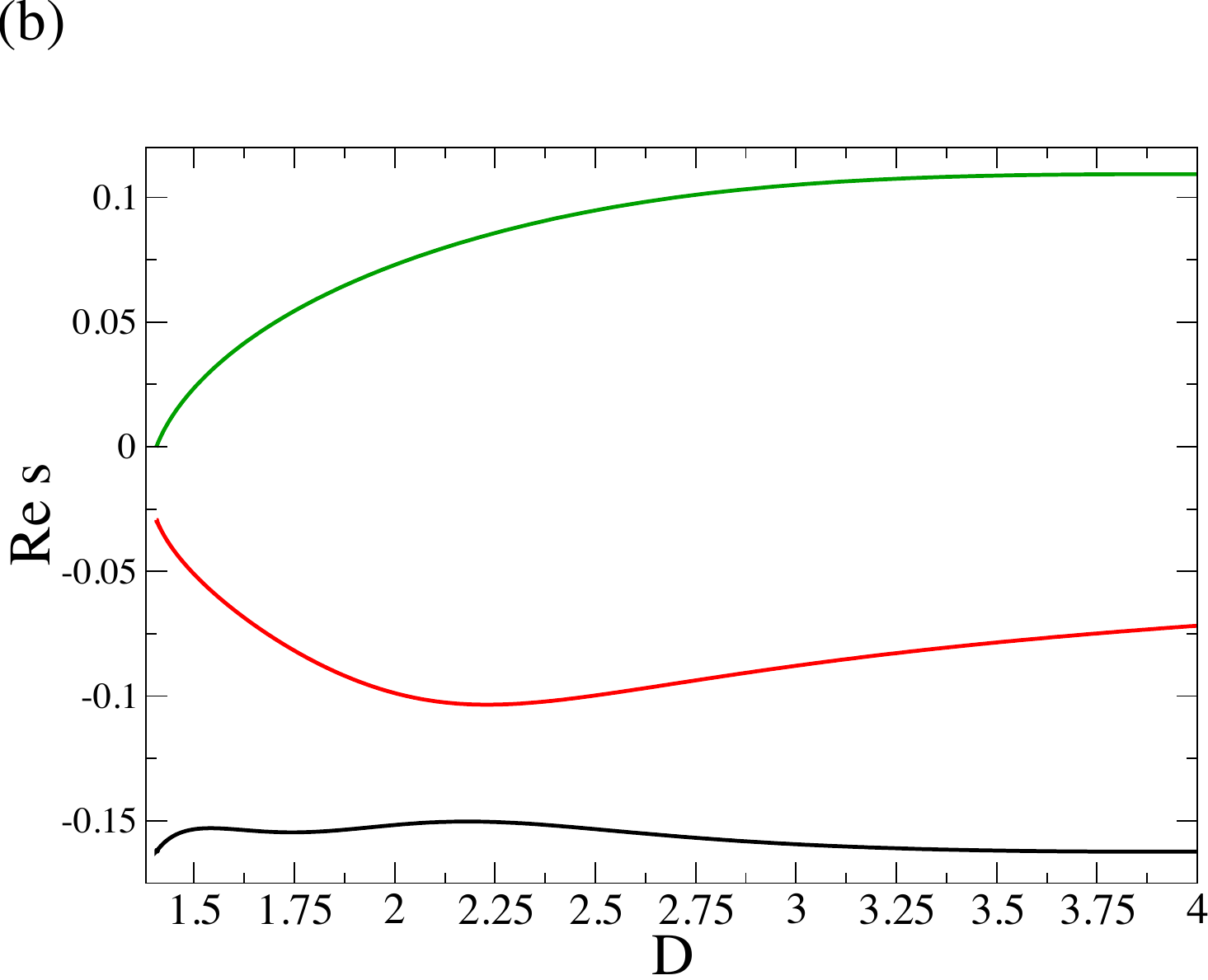}

\vspace{-0.3cm}
   \caption{
The dependence of the real parts of the dominant eigenvalues $s$ on $D$ for $\bar u=0.4$ and $L=25$ along the secondary branches
starting (a) at point 1 and (b) at point 2 in Fig.~\ref{fig:A1.1}(a).   
}
 \label{fig:B.2}
 \vspace{-0.3cm}
\end{figure}

In this section, we analyze the linear stability of the secondary branches. Figures~\ref{fig:B.2}(a) and (b) correspond to $L=25$ at $\bar u=0.4$ (cf. Fig.~\ref{fig:A1.1}(a)). Panels (a) and (b) correspond to the first and second secondary branches shown by the red and green dashed lines, respectively, in Fig.~\ref{fig:A1.1}(a). For the first secondary branch, there are two saddle-node bifurcations, while for the second secondary branch there are no saddle-node bifurcations. We can observe that for both secondary branches there is at least one eigenvalue with a positive real part for all the values of $D$. Therefore, both branches are unstable for all $D$ values. So, in a time evolution, a solution does not evolve into  a solution on the secondary branch. Instead, it can evolve into a two-drop solution (if $D$ belongs to the stable interval), or a one-drop solution, or a time-periodic solution -- such time evolutions are shown in Fig.~\ref{fig:time_D_5_0_bu_0_4_L_25}.

\begin{figure}
 \centering
\includegraphics [width=6cm]{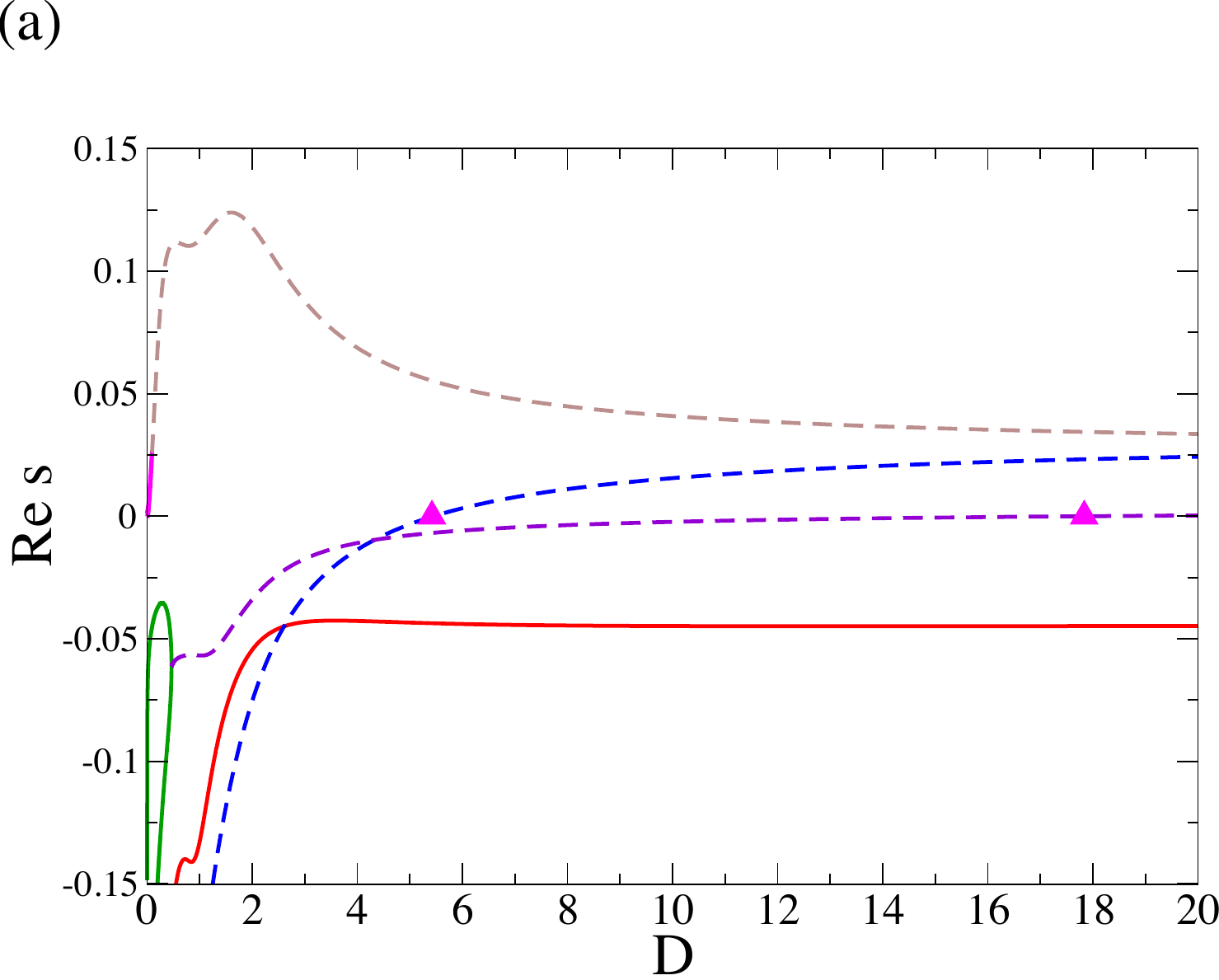}\hspace{0.5cm}
\includegraphics [width=6cm]{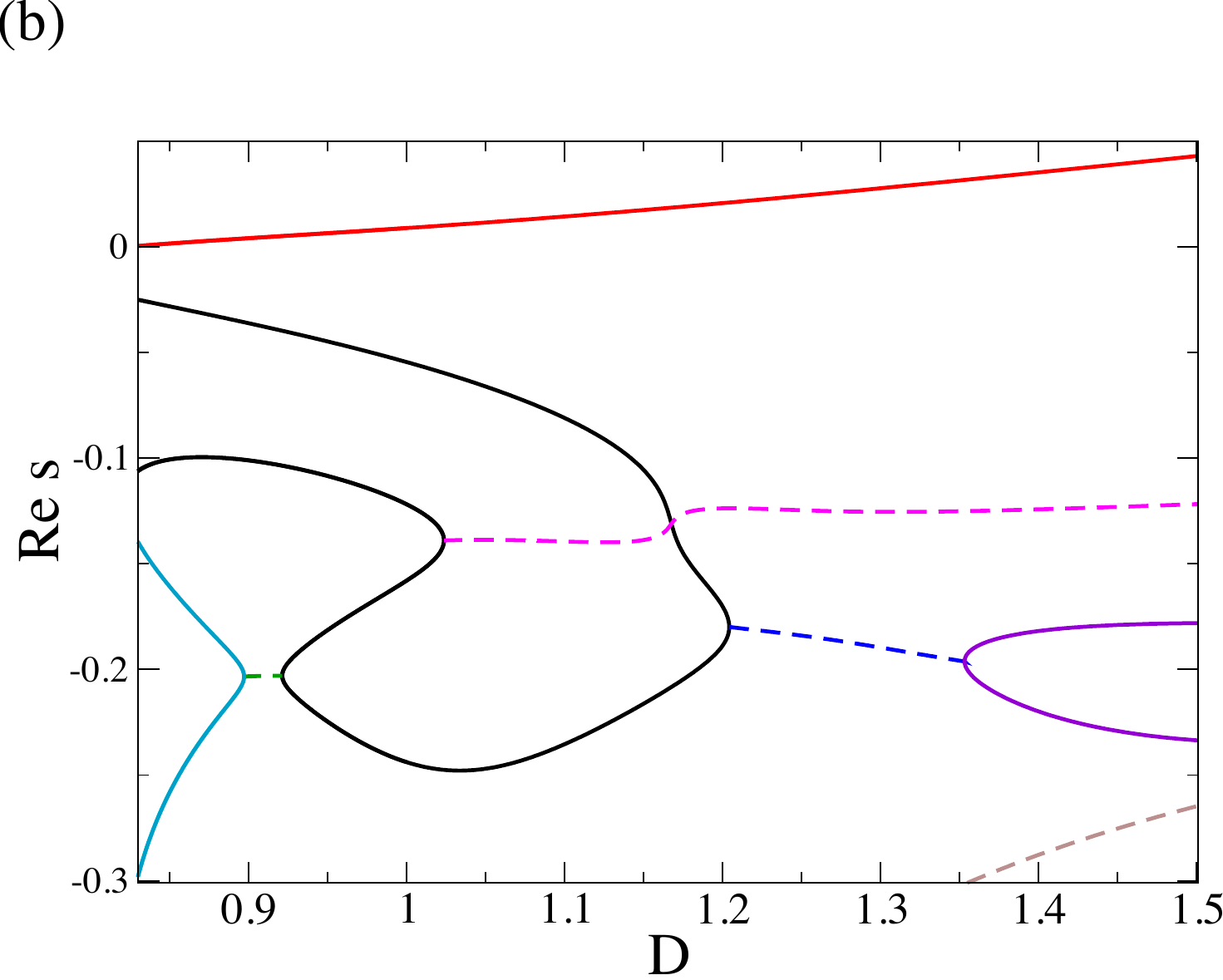}
\includegraphics [width=6cm]{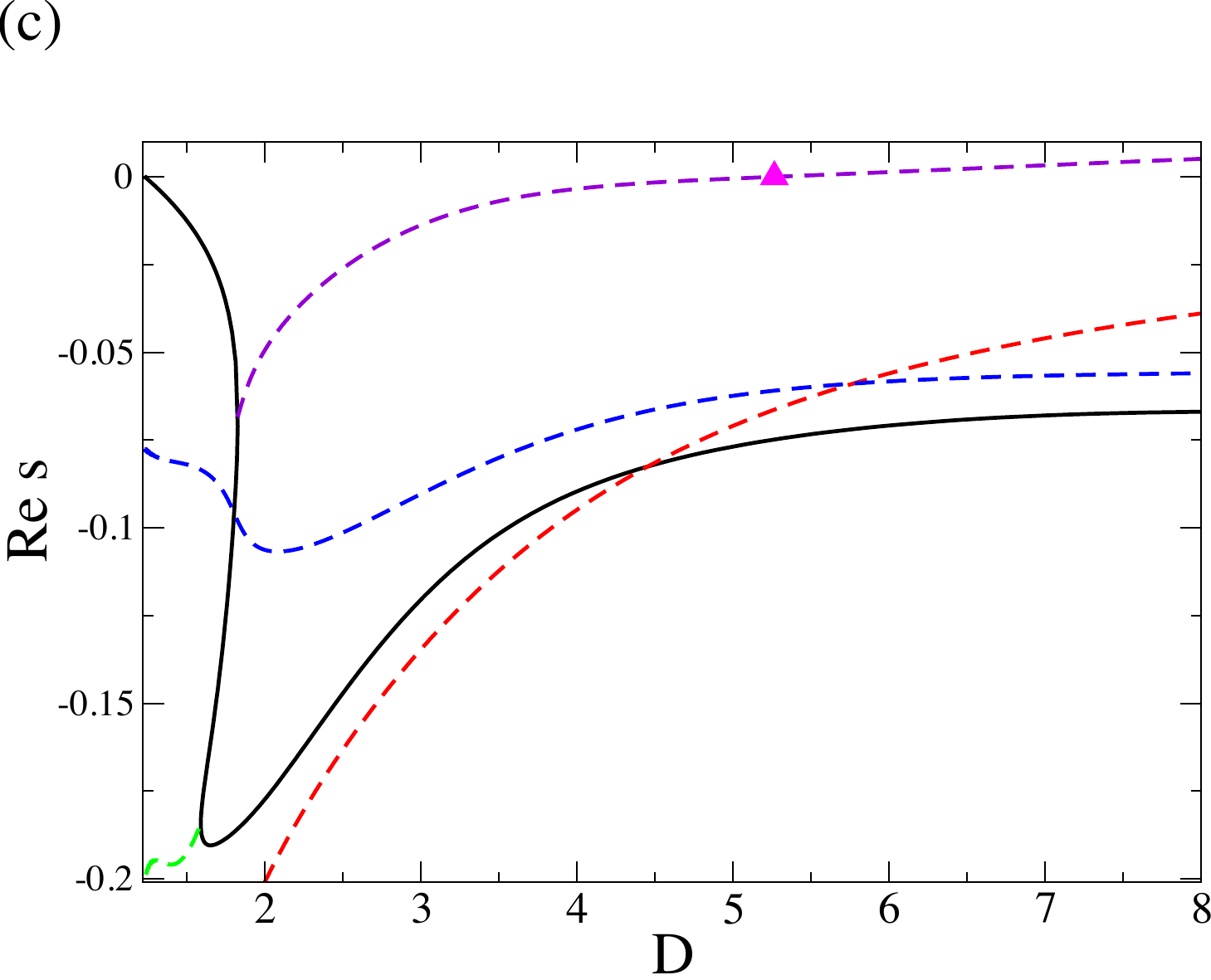}\hspace{0.5cm}
\includegraphics [width=6cm]{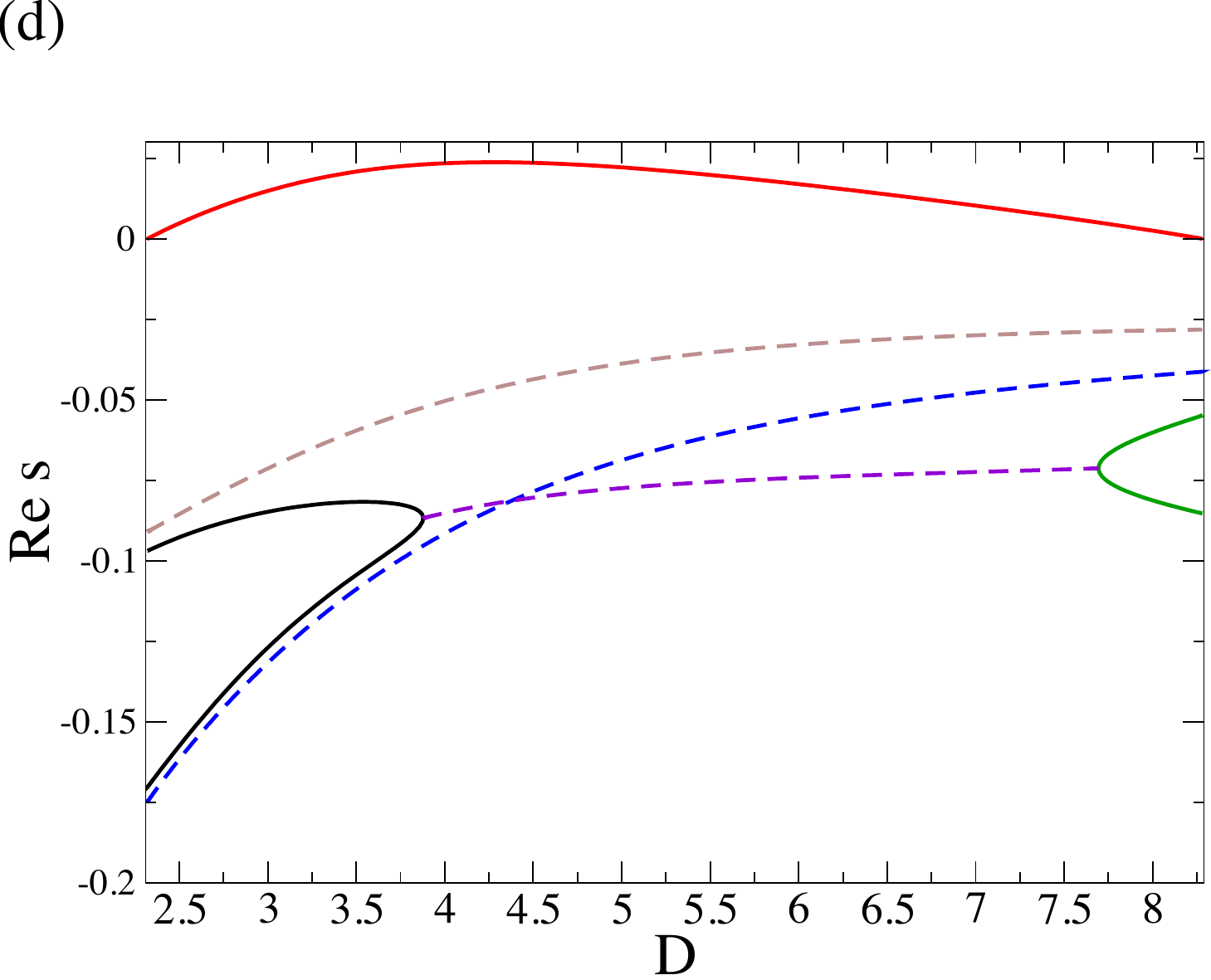}

\vspace{-0.3cm}
   \caption{The dependence of the real parts of the dominant eigenvalues $s$ on $D$ for $\bar u=0.4$ and $L=35$ along the secondary branches starting (a) at point 1, (b) at point 2, (c) at point 3 and (d) at point 4 (and ending at point 5) in Fig.~\ref{fig:A1.1}(b).   
   }
 \label{fig:B.4}
\end{figure}

\begin{figure}
 \centering
 \includegraphics [width=5.1cm]{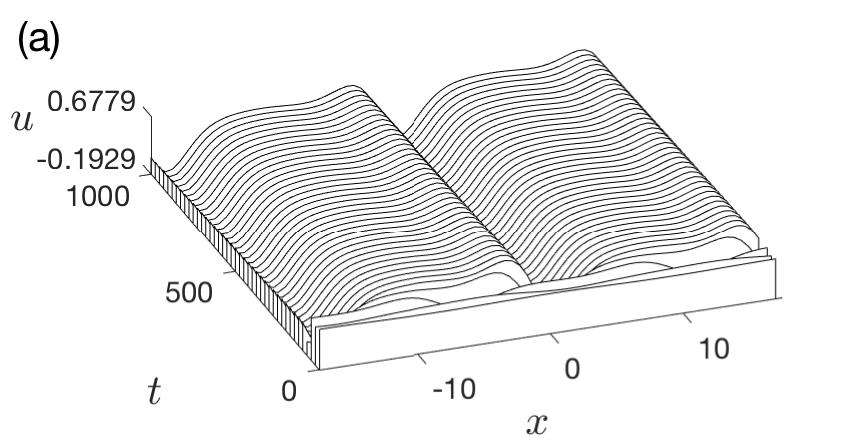}
 \includegraphics [width=5.1cm]{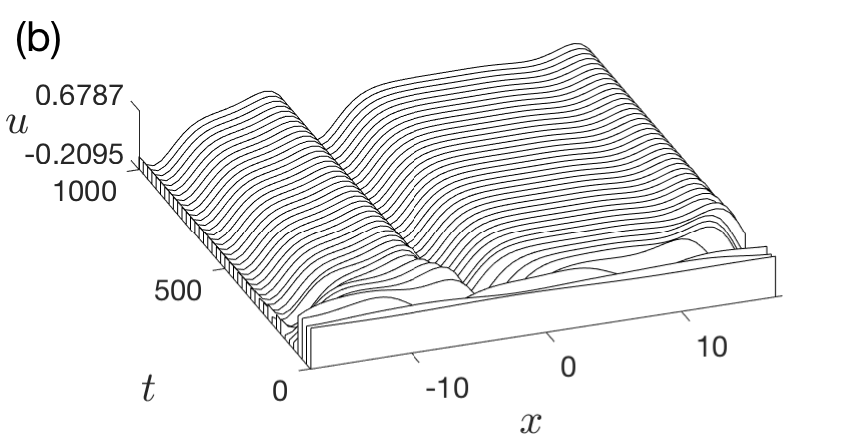}
 \includegraphics [width=5.1cm]{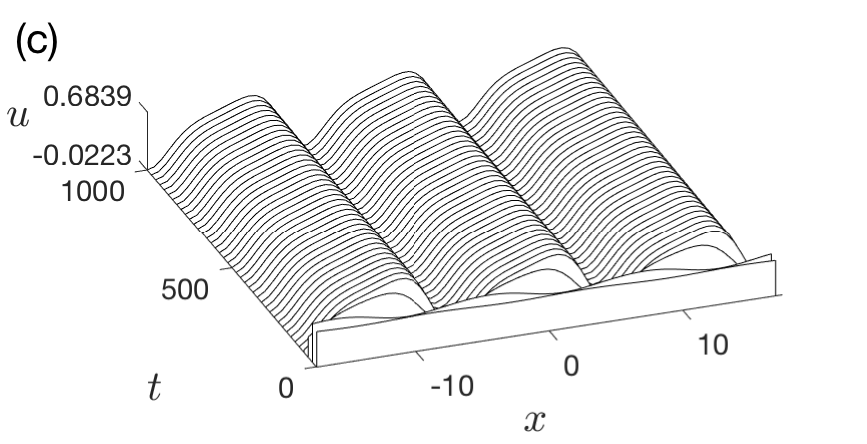}
 \includegraphics [width=5.1cm]{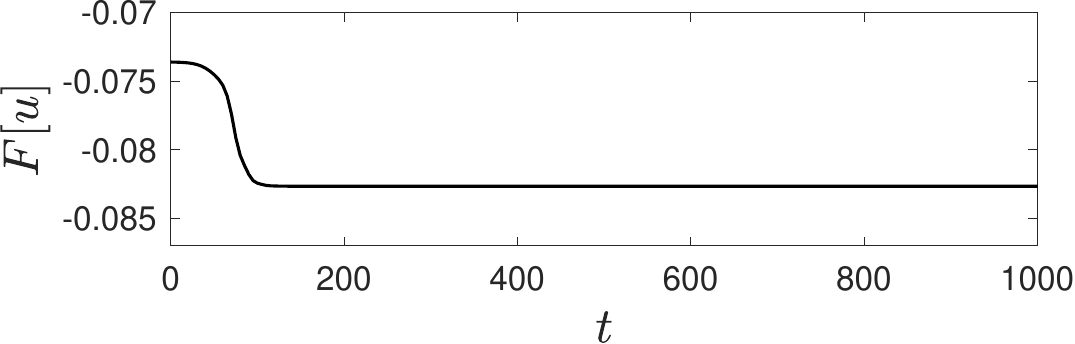}
 \includegraphics [width=5.1cm]{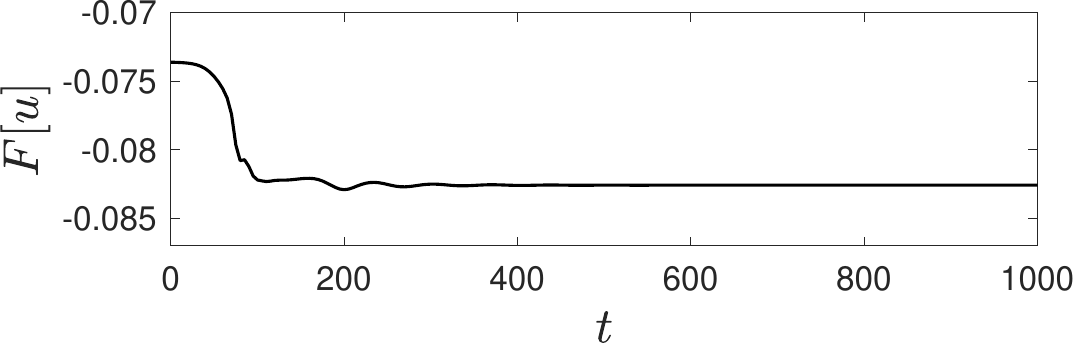}
 \includegraphics [width=5.1cm]{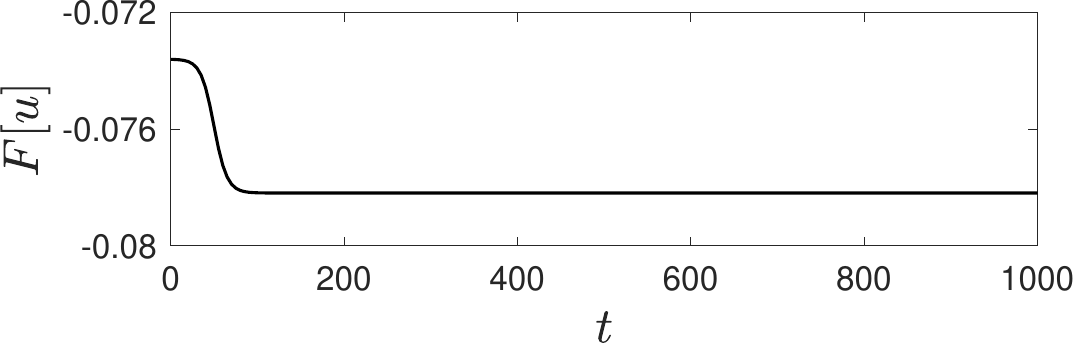}

\vspace{-0.3cm}
 \caption{Numerical solution of the cCH equation (\ref{Eq.2}) on the periodic domain of length $L=35$ for $\bar u=0.4$ and $D=3$, with three different initial conditions (as explained in the main text). 
 }
\label{fig:time_D_3_0_bu_0_4_L_35abc}
\vspace{-0.3cm}
\end{figure}

Figure~\ref{fig:B.4} corresponds to $L=35$ at $\bar u=0.4$ (cf. Fig.~\ref{fig:A1.1}(b)). Panels (a), (b), (c) and (d) correspond to the  secondary branches starting at the bifurcation points 1, 2, 3 and 4, respectively, in Fig.~\ref{fig:A1.1}(b). 
Figures~\ref{fig:B.4} (a), (b) and (d) imply that there are no stable intervals for the first, second and fourth secondary branches, while  Fig.~\ref{fig:B.4}(c) implies that there is a stable interval for the third secondary branch between $D\approx 1.23$ and $D\approx 5.26$. Taking into account the fact that for the two-drop primary branch the stable intervals are $0.82\lesssim D\lesssim1.23$ and $2.32\lesssim D\lesssim 8.28$, we can conclude that for $D\in (0.82,1.23)$ a two-drop solution is stable, for $D\in(1.23,2.32)$ a symmetry-broken solution is stable, for $D\in(2.32,5.26)$ both a two-drop solution and a symmetry-broken solution are stable, for $D\in(5.26,8.28)$ a two-drop solution is stable. Of course, there may exist other branches of solutions that are stable for these values of $D$, e.g., solutions of the one-drop primary branch with $L=35$ are stable for $D\lesssim7.28$, or there may exist some time-periodic solutions (or even quasi-periodic or chaotic solutions). For relatively large values of $L$ there can also exist $n$-drop branches with $n>2$,  
and there can of course also exist other symmetry-broken solutions bifurcating from such $n$-drop branches. For example, for $L=35$, the first four modes are linearly unstable for $\bar u=0.4$, and, therefore, in time-dependent simulations we may also observe three- and four-drop solutions. 
For other values of $D$, one-drop, two-drop and corresponding symmetry-broken solutions are unstable. Then, a time-dependent solution can evolve, for example, into a time-periodic solution or a multi-drop solution (or even a quasi-periodic or chaotic solution).

\begin{figure}
 \centering
\includegraphics [width=6cm]{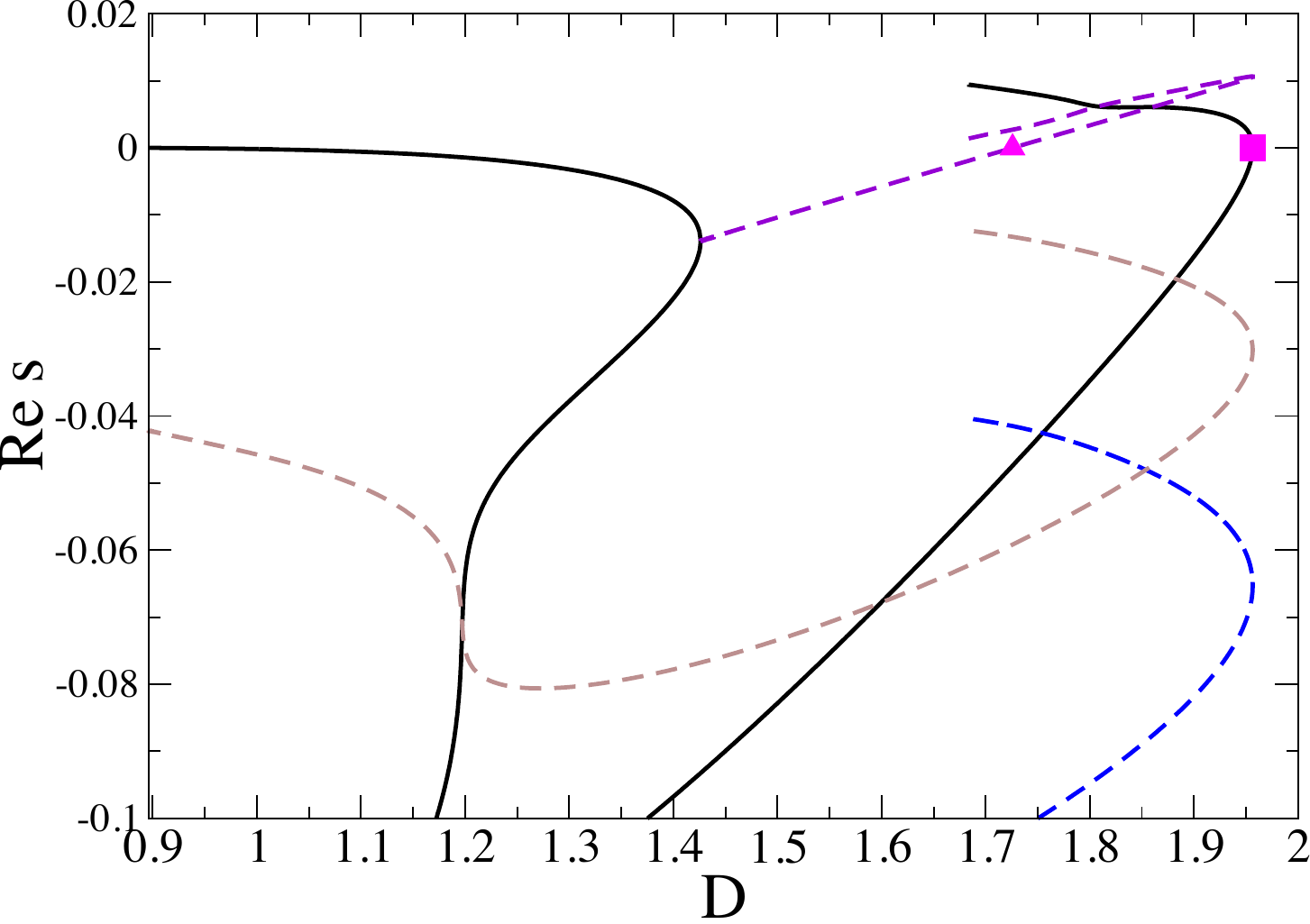}

\vspace{-0.3cm}
   \caption{The dependence of the real parts of the dominant eigenvalues $s$ on $D$ for $\bar u=0.55$ and $L=50$ along the secondary branch connecting points 2 and 5 in Fig.~\ref{fig:A1.2}.
}\label{fig:B.6}
\vspace{-0.3cm}
\end{figure}

Some of these predictions are confirmed in the time-dependent simulations presented in Fig.~\ref{fig:time_D_3_0_bu_0_4_L_35abc} for $\bar u=0.4$, $L=35$ and $D=3$. Panels (a), (b) and (c) correspond to initial conditions $u(x,0) = \bar u+0.01\cos(2\pi x/L)$, $u(x,0) = \bar u+0.01\cos(2\pi (x+2)/L)$ and $u(x,0) = \bar u+0.01\cos(3\pi (x-3)/L)$,  respectively. For this value of $D$, we expect both a two-drop solution and a symmetry-broken solution to be stable. Indeed, the simulation in panel (a) converges to a two-drop solution, whereas the simulation in panel (b) evolves into a symmetry-broken solution consisting of two drops of different sizes. It is interesting to note that there may exist other stable solutions, and, in particular, for the initial condition chosen for panel (c), we observe that the solution evolves into a three-drop solution (that appears to be stable, at least in the time interval presented in Fig.~\ref{fig:time_D_3_0_bu_0_4_L_35abc}(c)). In this work, we do not investigate in detail branches of $n$-drop solutions with $n>2$.

For $\bar u=0.55$ and $L=35$, we have verified that both secondary branches (shown by the red and green dashed lines in Fig.~\ref{fig:A1.2}(a)) are unstable for all the values of $D$. For  $\bar u=0.55$ and $L=50$, we have verified that the only secondary branch that has a stable interval in $D$ is the one connecting points 2 and 5 in Fig.~\ref{fig:A1.2}(b). The dominant eigenvalue for this branch are shown in Fig.~\ref{fig:B.6}. This branch has one Hopf bifurcation and there is a stable interval between $D\approx 0.90$ and $D\approx 1.68$. Taking into account the fact that for the two-drop primary branch the stable intervals are $0.72\lesssim D\lesssim0.90$ and $D\gtrsim 2.12$, we can conclude that for $D\in (0.72,0.90)$ a two-drop solution is stable, for $D\in(0.9,1.68)$ a symmetry-broken solution is stable, for $D\in(1.68,2.12)$ both a two-drop solution and a symmetry-broken solution are stable, for $D\gtrsim2.12$ a two-drop solution is stable. For other values of $D$, neither a two-drop solution nor a symmetry-broken solution are stable. Then, as also discussed above for other cases, a time-dependent solution can, for example, evolve into a one-drop solution (that is stable for $D\lesssim 7.13$), a time-periodic or multi-drop or quasi-periodic or chaotic solution. These observations can be corroborated by time-dependent simulations, however, we decided not to present such calculations here, as the results agree with the expectations and are generally qualitatively similar to the already presented time-dependent simulations.

\subsubsection{Linear stability regions for two-drop solutions in the $(D, L)$- and $(D,\bar u)$-planes}

In the previous section, we found that for fixed $\bar u$ and $L$, there can exist stability intervals for the driving force $D$, i.e., a carefully chosen driving force can be used to prevent coarsening. We also found that there can exist intervals for the driving force $D$ where solutions evolve into time-periodic solutions of two drops of different sizes that periodically exchange mass. In this section, we construct detailed stability and state diagrams showing the locations of the bifurcation points on the two-drop primary branches in the 
$(D, L)$- and $(D,\bar u)$-planes to obtain deeper insight into various possible behaviours of the solutions of the cCH equation at different parameter values. The solid and dashed lines in the diagrams correspond to zero crossings of the real part of real and complex eigenvalues, respectively.

\begin{figure}
 \centering
  \includegraphics [width=5.1cm]{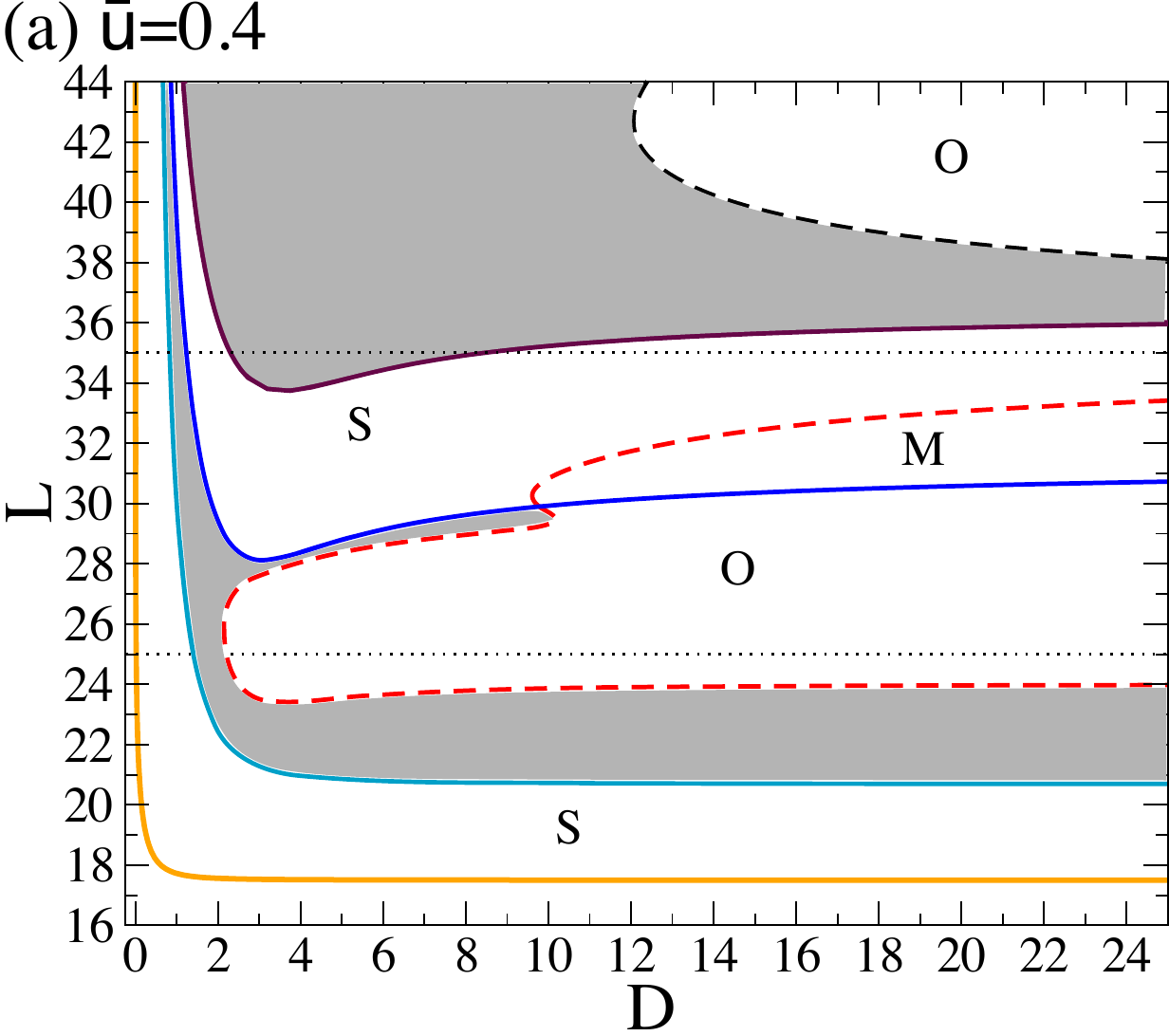}
   \includegraphics [width=5.1cm]{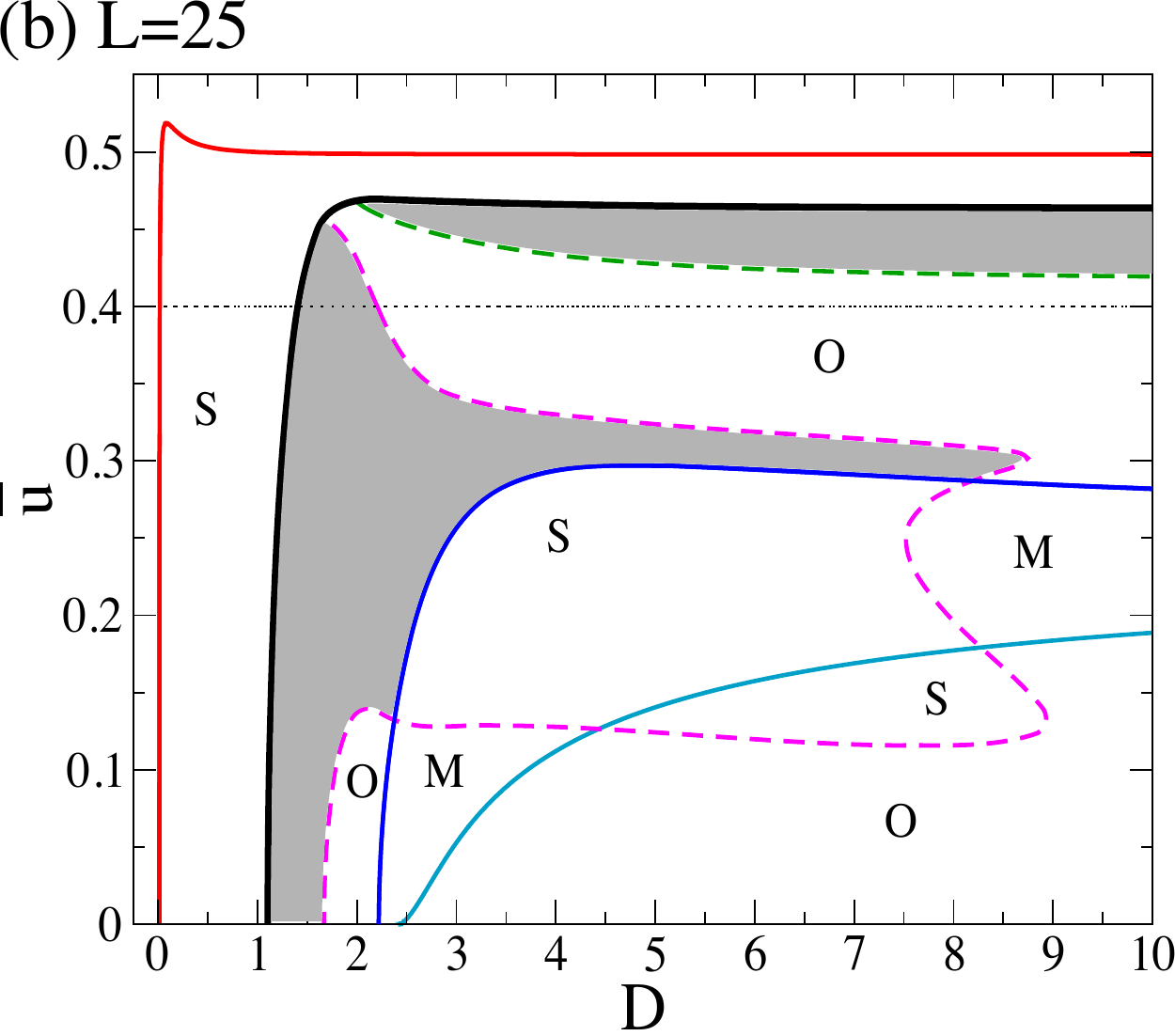}
   \includegraphics [width=5.1cm]{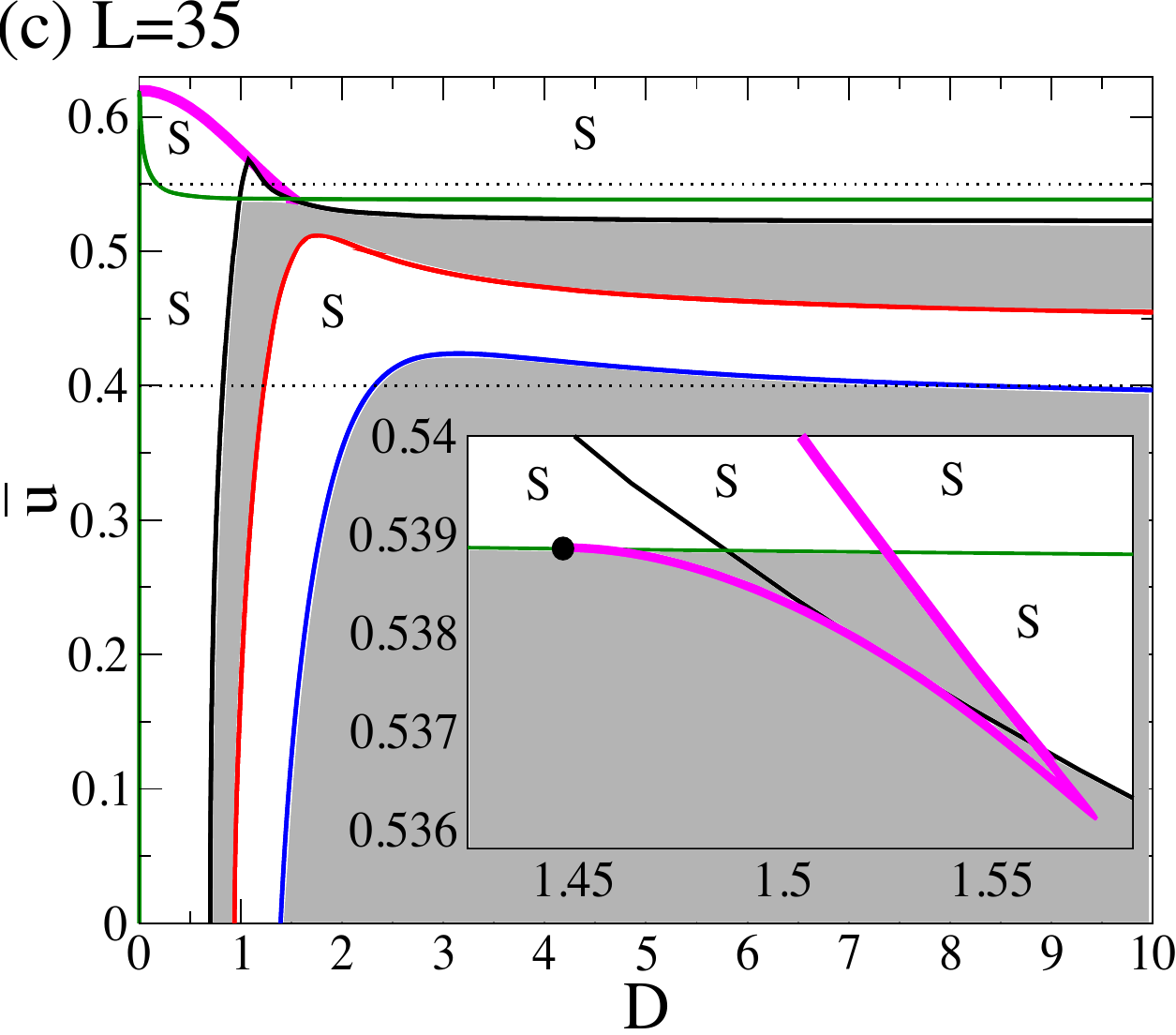}

   \vspace{-0.3cm}
   \caption{Loci of the bifurcation points on the two-drop primary branch (a) in the $(D,L)$-plane for $\bar u=0.4$ and in the $(D,\bar u)$-plane for (b) $L=25$ and (c) $L=35$. The solid and dashed lines correspond to real eigenvalues and the eigenvalues with nonzero imaginary parts, respectively. The linear stability regions are shown in grey. Labels S, O and M correspond to regions of different instability types (as explained in the text).
}
 \label{fig:2.1}
 \vspace{-0.3cm}
\end{figure}

Figure~\ref{fig:2.1}(a) shows the loci of the bifurcation points on the two-drop primary branch in the $(D,L)$-plane for $\bar u=0.4$. The thin horizontal dotted lines mark the values $L=25$ and $35$. These are the values that were chosen in Figs.~\ref{fig:A1.1} and \ref{fig:A1.4}.
As expected, for $L=25$ we have two bifurcation points to secondary branches and one Hopf bifurcation to a time-periodic branch, and for $L=35$ we have five bifurcation points to secondary branches. We can now clearly see how the various bifurcation points move as either $D$ or $L$ changes, and we can also obtain stability regions (shown in grey). In this diagram and in the other diagrams in this section, the instability types in various instability regions are indicated by letters~S (for regions where there exist other stable steady traveling-wave solutions), O (for regions where there exist stable oscillatory solutions) and~M (for regions where there exist stable steady traveling-wave  and oscillatory solutions, i.e., multistability of solutions of different types).

Figure~\ref{fig:2.1}(b) shows the loci of the bifurcation points on the two-drop primary branch in the $(D,\bar u)$-plane for $L=25$. The thin dotted horizontal line corresponds to $\bar u=0.4$. Note that we can have various numbers of bifurcation points to side branches and time-periodic solutions for smaller values of $\bar u$. However, further increasing $\bar u$, we first loose bifurcations to time-periodic states, and then, we loose bifurcations to side branches. Note the Takens-Bogdanov bifurcations occuring where Hopf bifurcations meet pitchfork bifurcations, i.e., where red and green dashed lines end on the black solid line.

Figure~\ref{fig:2.1}(c) shows the loci of the bifurcation points on the two-drop primary branch in the $(D,\bar u)$-plane for $L=35$. The horizontal dotted lines correspond to $\bar u=0.4$ and $0.55$. Note that the thick solid line in this figure shows the locations of the saddle-node bifurcations. For $\bar u=0.4$ we have five bifurcation points to side branches, while for $\bar u=0.55$ we have four bifurcation points to side branches and one saddle-node bifurcation, in agreement with Fig.~\ref{fig:A1.4}(b). Note also that the line showing the locations of the saddle-node bifurcations emerges from a certain point in the $(D,\bar u)$ (see the black circle in the inset in the figure). This point can be obtained using the weakly nonlinear analysis, see the Appendix. Indeed, for a given domain size $L$ for a two-drop solution, using (\ref{Eq.13}) we find that the value of $\bar u$ at which the spatially-uniform solution changes its stability and a nonuniform solution emerges is
$\bar u_c=\sqrt{{(1-k^2)}/{3}}$, 
where $k=4\pi/L$ (the wavenumber is equal to $4\pi/L$ but not to $2\pi/L$, since the value of $L$ that we consider corresponds to a two-drop solution). For this value of $\bar u$, we can then find the value $D_c$ of $D$ using (\ref{eq:Dc}) at which the nature of the primary bifurcation changes (between subcritical and supercritical). Thus, we expect (and, in fact, observe in our numerical results, that we decided not to show here) that when $L$ is fixed and $D$ is used as the principal continuation parameter, for $\bar u$ slightly greater than $\bar u_c$ the primary branch has a single saddle-node bifurcation and returns to $D=0$, for $\bar u=\bar u_c$ the primary branch has a single saddle-node bifurcation but it does not return to $D=0$ and instead hits the $D$-axis at $D=D_c$, and  for $\bar u$ slightly smaller than $\bar u_c$ there appears one more saddle-node bifurcation out of $(D_c,\bar u_c)$, and the branch extends to large values of $D$. For $L=35$, we find that $k\approx 0.3590$, $\bar u_c\approx 0.5389$ and $D_c\approx 1.4480$. This is in agreement with the results presented in Fig.~\ref{fig:2.1}(c) (see the inset showing point (1.4480, 0.5389) by a black circle -- the branch showing the locations of saddle-node bifurcations appears exactly from this point). 

\begin{figure}
\centering
\includegraphics [width=5.1cm]{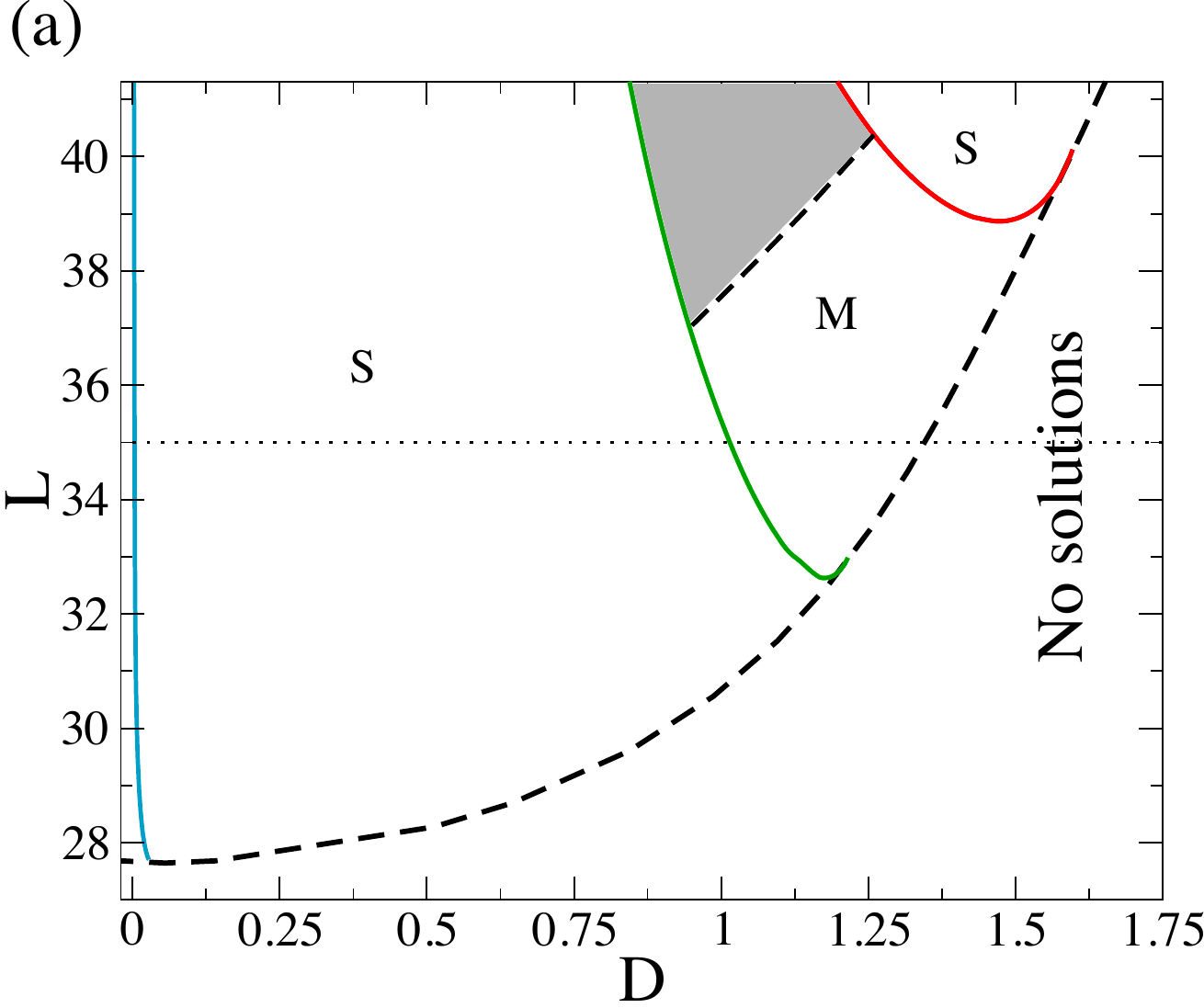}\hspace{0.5cm}
\includegraphics [width=5.1cm]{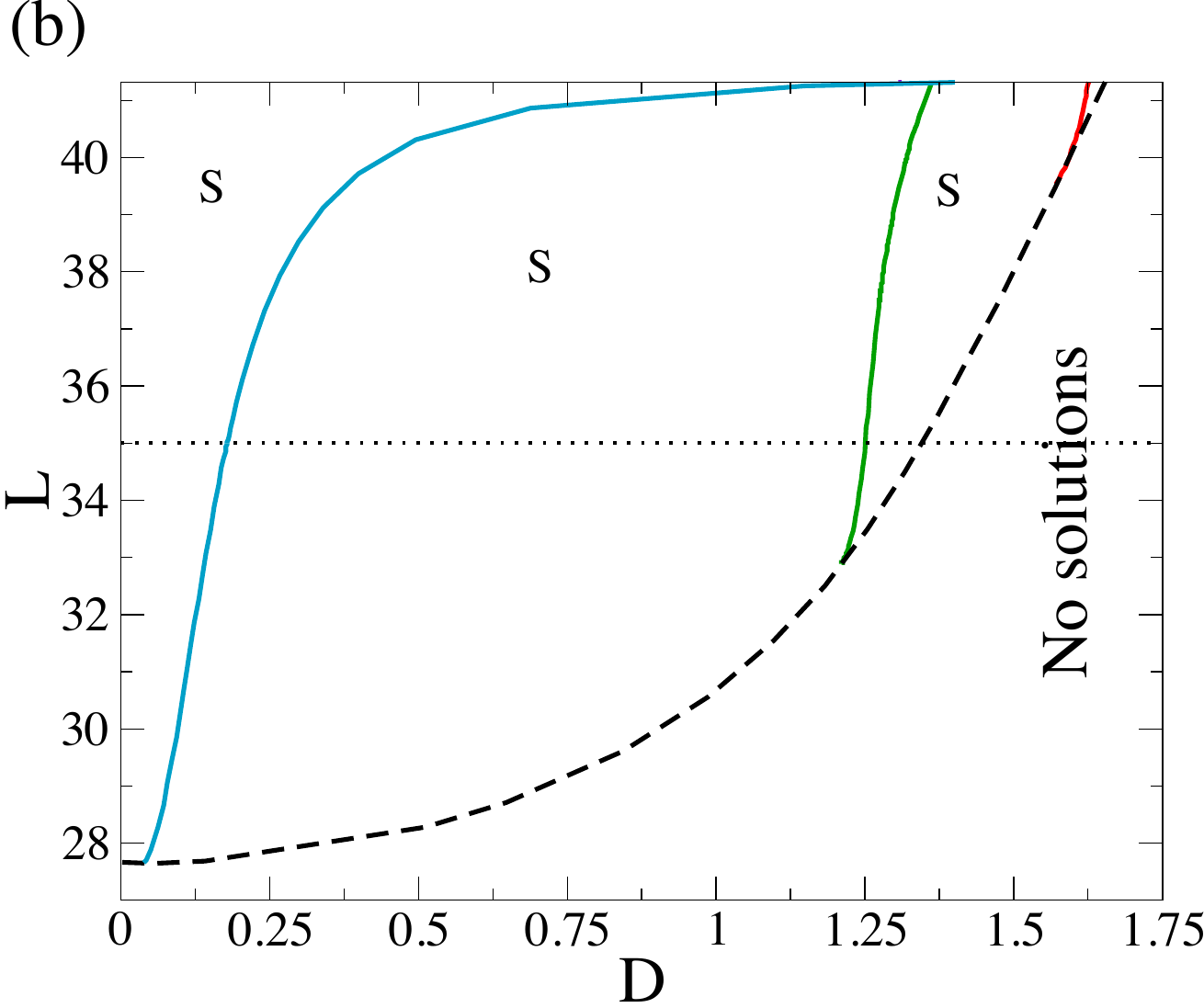}\\[0.3cm]
\includegraphics [width=5.1cm]{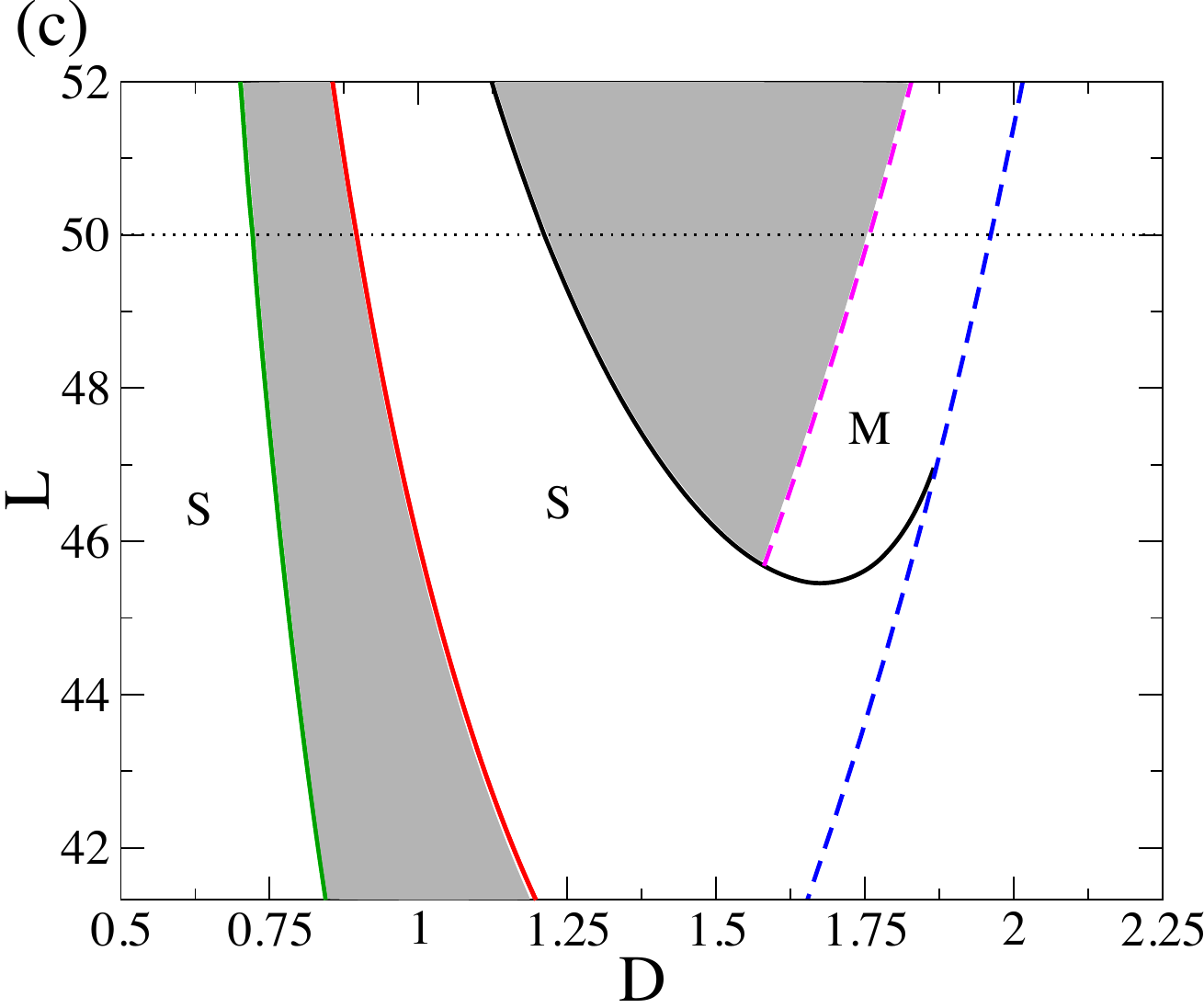}
\includegraphics [width=5.1cm]{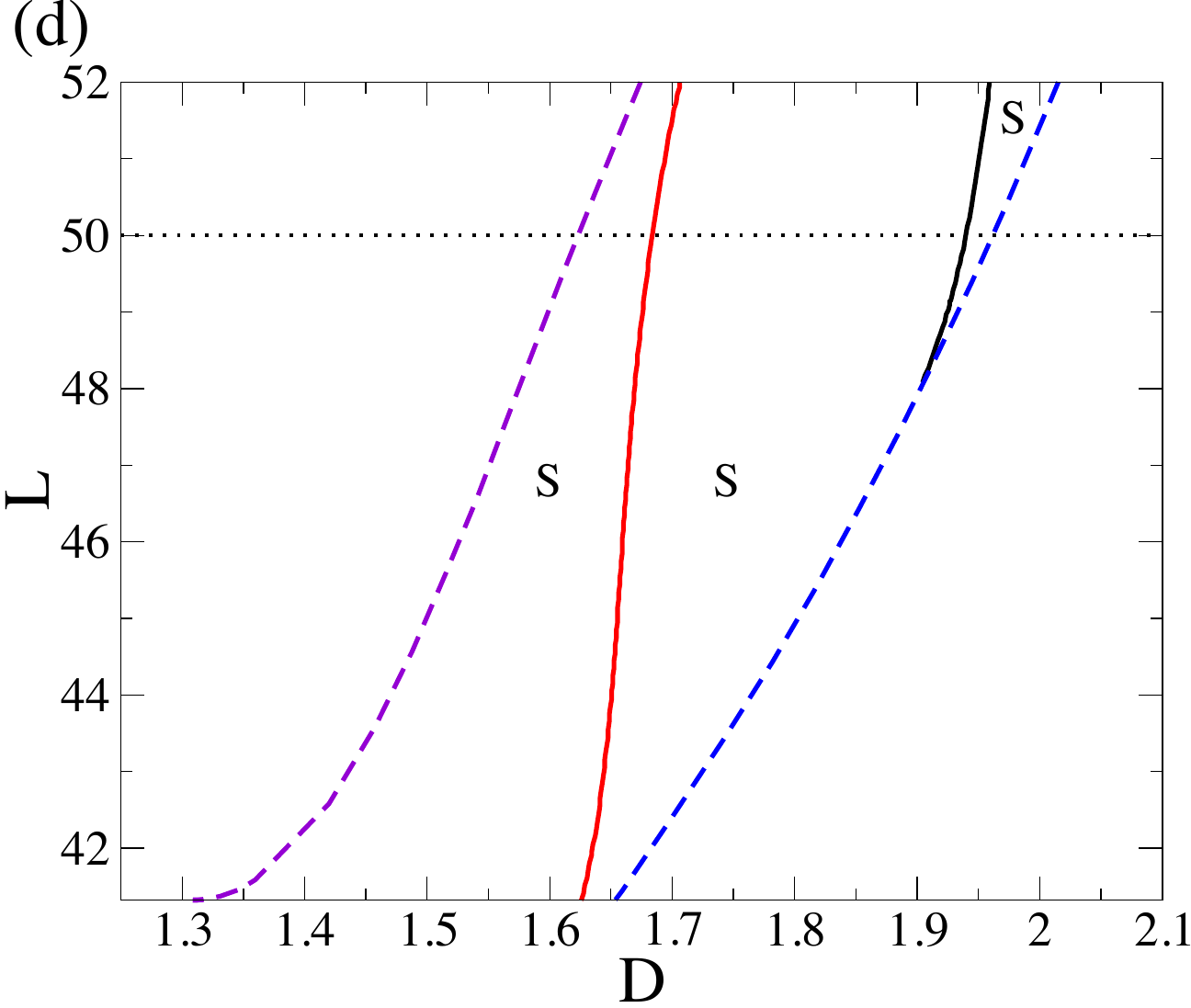}
\includegraphics [width=5.0cm]{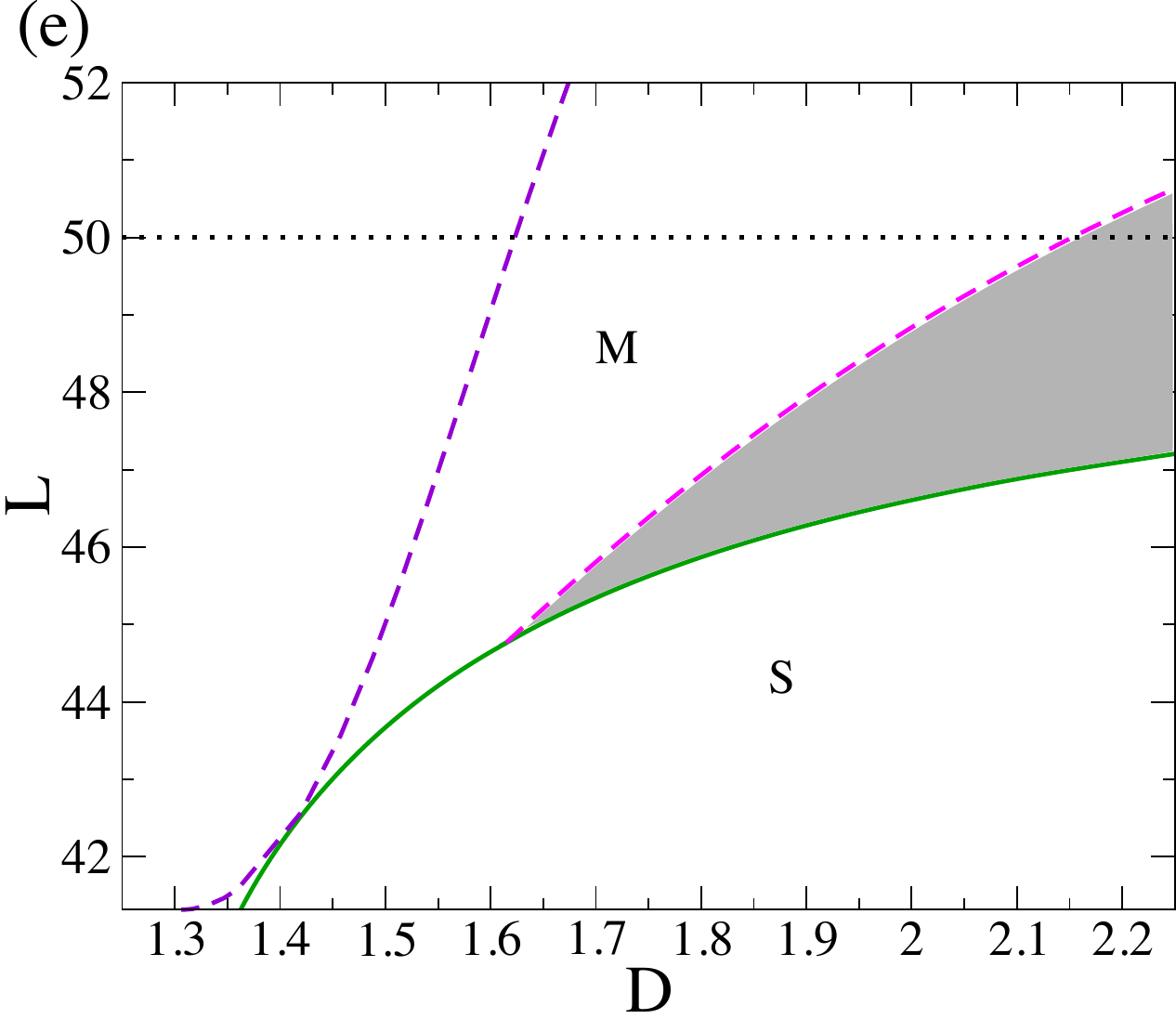}

\vspace{-0.3cm}
\caption{Loci of the bifurcation points on the two-drop primary branch in the $(D,L)$-plane for $\bar u=0.55$. The solid and dashed lines correspond to real eigenvalues and the eigenvalues with nonzero imaginary parts, respectively. The linear stability regions are shown in grey. Labels S and M correspond to regions of different instability types (as explained in the text). In panels (a) and (b)  $L<L_{c}= 41.32$ (so that the primary branch has a single saddle-node bifurcation, see Fig.~\ref{fig:A1.2}(a)), 
and these panels correspond to parts (a) and (b), respectively, of the primary branch shown in Fig.~\ref{fig:A1.2}(a). In (c), (d) and (e) $L>L_{c}= 41.32$ (so that the primary branch has two saddle-node bifurcations, see Fig.~\ref{fig:A1.2}(b)), and these branches correspond to parts (a), (b) and (c), respectively, of the primary branch shown in Fig.~\ref{fig:A1.2}(b).
}
 \label{fig:2.3}
 \vspace{-0.3cm}
\end{figure}

Figure~\ref{fig:2.3} shows the loci of the bifurcation points on the two-drop primary branch in the $(D,L)$-plane for $\bar u=0.55$.  We have split this figure into several parts. Panels (a) and (b) correspond to $L< L_c\approx 41.32$. For these values of $L$, the primary branch has one saddle-node bifurcation (when $D$ is used as the principle continuation parameter, see Fig.~\ref{fig:A1.2}(a)), and the branch returns to $D=0$. Thus, the branch consists of an upper and a lower part, in Fig.~\ref{fig:A1.2}(a) denoted by ``$\alpha$'' and ``$\beta$'', respectively. Panels (a) and (b) of Fig.~\ref{fig:2.3} correspond to parts $\alpha$ and $\beta$, respectively. Panels (c), (d) and (e) of Fig.~\ref{fig:2.3} belong to $L>L_{c}\approx 41.32$. For these values of $L$, the primary branch has a pair of saddle-node bifurcations (when $D$ is used as the principle continuation parameter, see Fig.~\ref{fig:A1.2}(b)), and consists of three parts, the upper one denoted by ``$\alpha$'', the middle one (connecting the two saddle-nodes) denoted by ``$\beta$'', and the lower one (starting from the second saddle-node and extending to infinity) denoted by ``$\gamma$''. Panels (c), (d) and (e) of Fig.~\ref{fig:2.3} correspond to these parts $\alpha$, $\beta$ and $\gamma$, respectively.

We note that for a more complete picture, it would be beneficial to more precisely indicate which solutions (e.g., one-drop, symmetry-broken or time-periodic solutions) are stable in the various regions where two-drop solutions are unstable. However, we do not present such a detailed `morphological phase diagram' here and leave this as a topic for future investigation.

{Finally, we would like to point out that linear stability of periodic one-drop solutions of the cCH equation was previously analysed by Zaks et al.\ \cite{Ref.51} but only for $\bar{u}=0$ and on the infinite domain, using a Floquet-Bloch-type analysis. We study stability also for nonzero values of $\bar{u}$ focusing on the analysis of coarsening modes, i.e., we consider stability on finite domains with lengths equal to twice the period of the one-drop solutions. This implies that for $\bar{u}=0$  the stability regions computed by Zaks et al.\ must be subsets of those computed here. Therefore, direct comparison is only appropriate for $\bar{u}=0$. In particular, we find good agreement with the stability results of Zaks et al.~\cite{Ref.51}  given in their Table 1 on p.\ 715, where stability intervals (in terms of solution wavenumbers) for $D=0.5$, $0.8$, $1$, $2$, $5$ and $D\rightarrow\infty$ are presented. Consider, for example, our stability diagram Fig.~\ref{fig:2.1}(b) for $L=25$. This value of $L$ corresponds to the wavenumber of the one-drop solution $K=2\pi/(L/2)\approx 0.503$. The results of Zaks et al.\ \cite{Ref.51} indicate that for $D=0.5$, $0.8$, $1$, $2$, $5$ the solution with $K=0.503$ must be linearly unstable. This fully agrees with the results presented in Fig.~\ref{fig:2.1}(b). Moreover, an interpolation of results of Zaks et al.\  indicates that for $\bar u=0$ and $K=0.503$ there must exist a stability interval in $D$ between $D=1$ and $D=2$, which agrees with the results in Fig.~\ref{fig:2.1}(b). Similarly,  the results of Zaks et al.\ imply for $\bar u=0$ and $L=35$ (corresponding to $K\approx0.359$) the existence of a stability interval $D\in(D_1,D_2)$, where $D_1\in(0.5,0.8)$ and $D_2\in(0.8,1)$. This agrees with the results given in Fig.~\ref{fig:2.1}(c). An important difference for $L=35$ is that for larger values of $D$ (say 1, 2, 5) we predict stability whereas the results of Zaks et al.\ imply instability. This is not a contradiction, since the stability regions of Zaks et al.\ must only be subsets of those computed here, as mentioned above.
  Finally note that our results also show good agreement with related studies for thin-film equations, in particular, when comparing the respective regimes of moderately strong driving. For instance, the sequence of instabilities and their dependence on driving strength for $D\gtrsim 3$ in our Fig.~\ref{fig:2.1}(a) is very similar to the corresponding behaviour in Fig.~22(b) of Ref.~\cite{Ref.11}. However, the regimes of weak driving notably differ as then the different underlying energies have a crucial influence.  }

\section{Conclusions}
\label{Sect:conclusions}

We have analysed the effect of the driving force on the solutions of the cCH equation. Initial insight was obtained by temporal and spatial linear stability analyses of homogenous solutions and we concluded that for the driving force parameter $D$ in the interval $[0,\sqrt{2}/3)$ the ``horizontal" parts of the fronts and drops/holes are expected to be monotonic, while for $D\in(\sqrt{2}/3,\sqrt{2})$ spatial, decaying oscillations are expected. For $D>\sqrt{2}$, we do not expect to see proper drop or hole solutions. Instead, we expect to observe, for example, localized positive/negative-pulse solutions. In addition, for $D\in(2\sqrt{2}/3,\sqrt{2})$, the horizontal parts of front- and drop/hole-solutions are linearly unstable, and thus, the solutions on large spatial domains are expected to break up into smaller structures.

Next, we presented the results of numerical continuation of single- and double-interface solutions (i.e., fronts and drops/holes). We first discussed the results of numerical continuation with respect to the domain size $L$ for the standard CH equation for several values of the mean solution thickness $\bar u$ and showed that for smaller values of $\bar u$ the primary bifurcation from the branch of homogeneous solutions is supercritical, whereas at some value of $\bar u$ it changes to subcritical. The value of $\bar u$ at which the type of the primary bifurcation switches can be found by the weakly nonlinear analysis. At some even larger value of $\bar u$ (that, in fact, follows from the linear stability analysis), the primary bifurcation disappears, and beyond a certain value of the domain size, linearly stable homogeneous and inhomogeneous solutions and a linearly unstable inhomogeneous solution coexist. After that, we studied the effect of the driving force on inhomogeneous solutions of the CH equation. For smaller values of $\bar u$, we found that when continuation is performed in the driving force parameter $D$, branches of solutions extend to infinity for all sufficiently large values of the domain size. Whereas for larger values of $\bar u$ the branches of solutions exhibit saddle-nodes and return to $D=0$, if $L$ is sufficiently small. For larger values of $L$, the branches exhibit an additional saddle-node and extend to infinity. The transition from one type of the bifurcation diagram to the other type of the bifurcation diagram happens at $L=L_c$, where $L_c$ is the wavelength of a small-amplitude neutrally stable sinusoidal wave. For this value of $L$, the branch of solutions terminates at the horizontal axis at $D=D_c$, where $D_c$ can be found by the weakly nonlinear analysis. So, for $L$ just beyond $L_c$, there is a range of $D$ values for which two different stable spatially inhomogeneous solutions and one unstable inhomogeneous solution coexist. For even larger values of $L$, the saddle-nodes annihilate each other, and the branches extend to infinity. Also, if $\bar u$ becomes sufficiently large, the branches of inhomogeneous solutions exhibit a saddle-node and return to $D=0$ for all sufficiently large values of $L$. 

Finally, we studied in detail the linear stability properties of the various possible spatially periodic traveling solutions of the cCH equation by performing numerical continuation of inhomogeneous solutions along with the dominant eigenvalues. To obtain more complete bifurcation diagrams, we also implemented a numerical procedure for continuation of time-periodic solutions. Our primary interest was in the study of the stability of symmetric two-drop solutions, and coarsening of such solutions in particular. Without driving force, the two-drop solutions have two real positive (unstable) eigenvalues that correspond to two different coarsening modes -- volume and translation modes. For the volume mode, the corresponding eigenfunction tends to increase the volume of one of the drops and decrease the volume of the other one. For the translation mode, the corresponding eigenfunction tends to shift both drops in the opposite directions, so that they move towards each other. When driving is introduced, we found that one of the coarsening modes is stabilized at relatively small values of $D$. In addition, our results indicate that the type of a coarsening mode can change as $D$ increases. We also found that  there may be intervals in the driving force $D$, where there are no unstable eigenvalues, and, therefore, driving can be used to prevent coarsening. We, in addition, computed side branches of symmetry-broken solutions and analysed the stability of such solutions, and also branches of time-periodic solutions, and presented detailed stability diagrams in the $(D,L)$- and $(D,\bar u)$-planes. The predictions from the numerical continuation results have been confirmed by time simulations for the cCH equation. In the future, it will be of interest to undertake similar studies for related equations, such as, for example, the various variants of the Kuramoto-Sivashinsky equation and related thin-film models and to extend the study to two-dimensional and three-dimensional solutions.

\section*{Appendix: Weakly nonlinear analysis for the general cCH-type equation}
\label{sec:weakly_nonlin1}

The aim of this section is to analyze the primary bifurcation for the cCH equation when the domain size is used as the control parameter.
In particular, we perform a Stuart-Landau-type analysis to derive an amplitude equation for the first linearly unstable mode in the vicinity of the bifurcation point. We consider the general cCH-type equation that in the frame moving at constant velocity $v$ in the $x$-direction has the form
\begin{equation} \label{Eq.B1} 
u_t=vu_{x}-D[\chi(u)]_{x}+ \left[Q(u) \left( \frac{\delta F(u)}{\delta u} \right)_{x}\right]_{x},
\end{equation}
where $D\chi(u)$ is the driving force term with $D$ being the driving force strength (for the cCH equation considered above, $\chi(u)=u^2/2$), $Q(u)$ is the mobility (that will be assumed to be nonnegative for any $u$) and $F[u]=\int \varphi(u, u_{x}) dx$ is the free energy functional with $\varphi(u, u_{x})=\frac{1}{2} u_{x}^{2}+f(u)$ denoting the free energy density. Here, $f(u)$ is the local free energy that for the standard CH equation is $f(u)= u^{4}/4 - u^{2}/2.$

Next, let us consider a uniform solution $\bar u$. The linear stability analysis implies that the cutoff wavenumber is $k_c= \sqrt{-f''(\bar u)}$ so that the period of neutral small-amplitude sinusoidal waves is $L_c=2\pi/k_c$, and the phase speed of small-amplitude sinusoidal waves is $v=D\chi'(\bar u)$. We consider the equation in a frame moving at this speed and we set $k = k_c - \epsilon^2$, where $k=2\pi/L$ (with $L$ denoting the domain size) and $\epsilon\ll 1$. For convenience, we rescale the independent variables by writing $x = \xi/k$ (so that $\xi\in[0,2\pi]$) and $t = \epsilon^{-2}\tau/k$ (the slow time scale follows from the linear stability analysis and the fact that we are close to the neutral stability point).

Next, we use a regular asymptotic expansion for $u$:  
\begin{equation} \label{Eq.B6} 
u= \bar u+\epsilon w_{1}(\xi, \tau)+\epsilon^2 w_{2}(\xi, \tau)+ \epsilon^3 w_{3}(\xi, \tau)+\cdots.
\end{equation}
Substituting (\ref{Eq.B6}) in the rescaled general cCH equation, we obtain at order $O(\epsilon)$:
\begin{equation} \label {Eq.B8} 
w_{1 \xi \xi} + w_{1 \xi \xi \xi \xi} =0.
\end{equation}
It can be readily found that the general periodic solution of zero mean to this equation~is
\begin{equation} \label {Eq.B9} 
w_{1}= A_{1} e^{i \xi} +\mathrm{c.c.}, 
\end{equation}
where $A_{1}= A_{1} (\tau)$ is the amplitude of the unstable mode $e^{i \xi}$, and c.c.\ denotes the complex conjugate of the right-hand side. At order $O(\epsilon^2)$, we obtain: 
\begin{equation} 
w_{2\xi \xi \xi \xi}+ w_{2 \xi \xi}= \left[-\frac{2f'''(\bar u)}{ k_{c}^{2}}-\frac {iD\chi''(\bar u)}{Q(u_0)  k_{c}^{3}}\right] A_{1}^{2} e^{2i \xi} +\mathrm{c.c.}
\end{equation}
The general solution is
\begin{equation}  
w_{2}= A_{2} e^{i \xi} +B_2 e^{2 i \xi}+\mathrm{c.c.},\label {Eq.B14}
\end{equation}
where $A_{2}= A_{2} (\tau)$ and $B_2=\left[-{f'''(\bar u)}/{6 k_{c}^{2}}- {iD\chi''(\bar u)}/{12Q(\bar u)  k_{c}^{3}}\right] A_{1}^{2}$.

At order $O (\epsilon^3)$, we find: 
\begin{equation}
w_{3 \xi \xi \xi \xi}+ w_{3\xi \xi}= r_1 e^{i \xi} +r_2 e^{2 i \xi}+\mathrm{c.c.},
\label {Eq.B15} 
\end{equation}
where
\begin{eqnarray} 
r_1&=&\frac{1}{Q(\bar u)  k_{c}^{3}}\biggl[A_{1\tau}-2  k_{c}^{2} Q(\bar{u})A_1+\biggl( \frac{1}{2} k_{c} Q(\bar{u}) f''''(\bar{u})- \frac{1}{6} \frac{Q(\bar{u}) (f'''(\bar{u}))^2}{ k_c}
\nonumber\\
&&+\frac{1}{12}  \frac{D^2 (\chi''(\bar{u}))^2}{ k_{c}^{3} Q(\bar{u})} +\frac{1}{2} iD \chi'''(\bar{u})- \frac{1}{4} i \frac{D\chi''(\bar{u})f'''(\bar{u})}{ k_{c}^{2}}
\nonumber\\
&&-\frac{1}{2} i \frac{D\chi''(\bar{u}) Q'(\bar{u})}{Q(\bar{u})} \biggr)A_1^2 A_{1}^{*} \biggr],
\end{eqnarray}
and $r_2$ is a lengthy coefficient whose particular form is not important for our purposes and, therefore, not shown.

To exclude secular terms, we must have $r_1=0$. We, therefore, obtain the following amplitude (or Stuart-Landau) equation:
\begin{eqnarray} 
\frac{dA_1}{d\tau}=2  k_{c}^{2} Q(\bar{u})A_1 - h A_1^2 A_{1}^{*},
\label {Eq.B16c} 
\end{eqnarray}
where 
\begin{eqnarray} 
h=\frac{1}{2} k_{c} Q(\bar{u}) f''''(\bar{u})-  \frac{Q(\bar{u}) (f'''(\bar{u}))^2}{6 k_c} + \frac{D^2 (\chi''(\bar{u}))^2}{12 k_{c}^{3} Q(\bar{u})}\nonumber
\\
\qquad +\frac{1}{2} iD \chi'''(\bar{u})- \frac{1}{4} i \frac{D\chi''(\bar{u})f'''(\bar{u})}{ k_{c}^{2}} - \frac{1}{2} i \frac{D\chi''(\bar{u}) Q'(\bar{u})}{Q(\bar{u})},
\end{eqnarray}
and we can ultimately arrive at the following equation for $|A_1|$:
\begin{equation}
\frac{d(|A_1|)}{d\tau}=\bigl(2  k_{c}^{2} Q(\bar{u})-\mathrm{Re}(h)|A_1|^2\bigr)|A_1|.
\label{eq:ODE_abs_A1}
\end{equation}
When $\mathrm{Re}(h)<0$, this equation for $|A_1|$ has only one fixed point, namely, $|A_1|=0$. Therefore, for $\mathrm{Re}(h)<0$ there do not exist small-amplitude sinusoidal solutions beyond the primary bifurcation point. Therefore, the primary bifurcation is  subcritical in this case. On the other hand, when $\mathrm{Re}(h)>0$, equation (\ref{eq:ODE_abs_A1}) for $|A_1|$ has two fixed points, namely, an unstable fixed point $|A_1|=0$ and a stable fixed point $|A_1|=(2  k_{c}^{2} Q(\bar{u})/\mathrm{Re}(h))^{1/2}$. Therefore, for $\mathrm{Re}(h)>0$ there exists a small-amplitude sinusoidal solutions beyond the primary bifurcation point. 
Therefore, the primary bifurcation is supercritical when $\mathrm{Re}(h)>0$. Thus, we find that the change from supercritical to subcritical bifurcation happens when $\mathrm{Re}(h)=0$. For the cCH equation~(\ref{Eq.2}), this condition becomes 
\begin{equation}
\mathrm{Re}(h)=3 k_c - 6 \frac{\bar u^{2}}{ k_c}+ \frac{D^2}{12 k_{c}^{3}}=0.\label{eq:Sub_Sup_cond}
\end{equation}

\section*{Acknowledgments}
We thank the Center of Nonlinear Science (CeNoS) of the University of M{\"u}nster for support of the author's collaboration. DT and TS acknowledge support by the EPSRC under grant No.\ EP/J001740/1. TS acknowledges support by Ministry of Science and Technology of Taiwan under research grant MOST-107-2115-M-009-008-MY2. UT acknowledges support by the Deutsche Forschungsgemeinschaft (DFG; Grant No.~TH781/8-1) and the German-Israeli Foundation for Scientific Research and Development (GIF, Grant No. I-1361-401.10/2016).


\end{document}